\documentclass[11pt,onecolumn]{article}
\usepackage{geometry,setspace}
\usepackage[font=small,format=plain,labelfont=bf,justification=raggedright,singlelinecheck=false]{caption}
\usepackage{amsmath,amsfonts,amssymb,amsthm,upgreek,mathrsfs,bm}
\usepackage{hyperref,cite}
\usepackage{graphicx,subfig,tikz}
\usepackage{physics}
\usepackage{extarrows}
\usepackage{orcidlink}

\hypersetup{colorlinks=true,linkcolor=blue,anchorcolor=blue,citecolor=blue,urlcolor=blue}
\allowdisplaybreaks[3]

\numberwithin{equation}{section}

\definecolor{darkgreen}{RGB}{40,150,60}

\title{Supersymmetric BMS$_4$ Algebras Revisited: Electric/Magnetic Superalgebras and Free Field Realization}
\author{Yu-fan Zheng\orcidlink{0000-0001-7405-582X}$^{a}$\footnote{\href{zhengyufan@bimsa.cn}{zhengyufan@bimsa.cn}}}
\date{\today}

\begin{document}

\maketitle
\begin{center}
    {\it
        $^{a}$ Beijing Institute of Mathematical Sciences and Applications (BIMSA), Huaibei Town, Huairou District, Beijing 101408, China \\
    }
    \vspace{10mm}
\end{center}

\begin{abstract}
    \vspace{5mm}
    \begin{spacing}{1.5}
        In this work, we present a systematic classification of supersymmetric extensions of the BMS$_4$ algebra and their realizations in free field theories. By requiring that supercharges admit finite-dimensional subsectors, we identify ten distinct electric super BMS$_4$ algebras and six magnetic ones. The electric case is characterized by supercharge anticommutators closing on supertranslations, while the magnetic case necessarily involves superrotations. To realize these algebras in free field theories, we follow a constructive procedure: first identify the modes bilinears of which generate the symmetry algebra, then determine the fields with appropriate transformation properties under the BMS$_4$ algebra, and finally construct consistent theories whose equations of motion admit the desired supersymmetry. Notably, $R$-symmetry with nonvanishing spin is essential for the Type II-II, Type I-II, and Type II-I magnetic super BMS$_4$ algebras, shedding new light on spacetime structure of string theory for flat holography. Moreover, in the Type I-I theory, $R$-symmetry relates electric and magnetic scalars, indicating their equal significance. In addition, $R$-symmetry maps the electric scalar to a spin-$1$ field and vice versa, offering a novel perspective on supersymmetric extensions of soft theorems.
    \end{spacing}
\end{abstract}

\setcounter{tocdepth}{2}
\tableofcontents

\section{Introduction}\label{sec:Introduction}

    The study of asymptotic symmetries in general relativity originates from the seminal work of Bondi, van der Burg, Metzner, and Sachs \cite{Bondi:1962px,Sachs:1962wk,Sachs:1962zza,Penrose:1962ij}. They demonstrated that the symmetry group of $4$D asymptotically flat spacetime at null infinity extends beyond the Poincar\'e group. This enlarged symmetry is now referred to as the BMS$_4$ algebra. A central feature of this algebra is the infinite-dimensional family of generators known as supertranslations, which extend the usual translations in Minkowski space. Beyond supertranslations, additional structure was subsequently identified. A systematic formulation of these ``superrotations" was developed by Barnich and Troessaert \cite{Banks:2003vp,Barnich:2009se,Barnich:2010eb,Barnich:2010ojg}, who showed that they can be interpreted as local conformal transformations of the celestial sphere. In this sense, superrotations generalize global Lorentz transformations in the same way that supertranslations generalize translations. The resulting structure, often called the extended BMS$_4$ algebra, takes the form
    \begin{equation}\label{eq:BMS4Algebra}
        \begin{aligned}
            &[L_{m},L_{n}]=(m-n)L_{m+n}, \quad [\bar{L}_{\bar{m}},\bar{L}_{\bar{n}}]=(\bar{m}-\bar{n})\bar{L}_{\bar{m}+\bar{n}}, \\
            &[L_{m}, M_{r,\bar{r}}] = \left(\frac{m}{2}-r\right)M_{m+r,\bar{r}}, \quad [\bar{L}_{\bar{m}}, M_{r,\bar{r}}] = \left(\frac{\bar{m}}{2}-\bar{r}\right)M_{r,\bar{m}+\bar{r}},
        \end{aligned}
    \end{equation}
    where the superrotations $L, \bar{L}$ and the supertranslations $M$ together generate an infinite-dimensional symmetry. This extension has significant implications for gravitational scattering, and later work established its relation to subleading soft graviton theorems \cite{Kapec:2014opa,Campiglia:2014yka}. For recent reviews, see \cite{Alessio:2017lps,Ashtekar:2018lor,Strominger:2017zoo}. \par

    For several decades the BMS algebra remained primarily a mathematical curiosity, until renewed interest arose from its role in infrared physics. In particular, Strominger and collaborators demonstrated that the Ward identities associated with BMS supertranslations reproduce Weinberg’s soft graviton theorem \cite{Strominger:2013jfa,He:2014laa, He:2014cra, Strominger:2014pwa, Pasterski:2015tva, Conde:2016rom, Nande:2017dba}. This correspondence established a precise relation among asymptotic symmetries, soft theorems, and memory effects, often referred to as the infrared triangle \cite{Strominger:2017zoo}. In recent years, BMS$_4$ symmetries have become central to the framework of celestial holography. Within this approach, scattering amplitudes in asymptotically flat spacetime are reinterpreted as correlation functions of conformal primaries defined on the celestial sphere \cite{Adamo:2014yya, Pasterski:2016qvg, Fotopoulos:2020bqj, Raclariu:2021zjz, Pasterski:2021rjz, Prema:2021sjp, Donnay:2023mrd}. In this setting, the BMS$_4$ algebra provides the organizing principle of the theory, much like the Virasoro and affine Kac–Moody algebras in $2$D conformal field theory. Current research continues to develop its representation theory \cite{Barnich:2011mi, Barnich:2014kra, Barnich:2015uva, Compere:2020lrt} and investigate its significance for flat holography \cite{Barnich:2010eb, Pasterski:2016qvg, Donnay:2020fof, Donnay:2023mrd, Chen:2025gaz}, quantum gravity \cite{Strominger:2013jfa, Kapec:2014opa}, and even black hole physics \cite{Donnay:2015abr, Hawking:2016msc, Hawking:2016sgy}. \par

    While the BMS$_4$ algebra encodes the asymptotic symmetries of flat spacetime, it is natural to ask whether these structures admit supersymmetric extensions. Supersymmetry plays a central role in modern theoretical physics, appearing in supergravity, string theory, and proposals for ultraviolet-complete quantum gravity theories. Extending BMS$_4$ by incorporating fermionic generators enables the study of asymptotic symmetries in supergravity and the analysis of constraints imposed by supersymmetry at asymptotic infinities \cite{Awada:1985by, Barnich:2015sca, Fuentealba:2020aax, Henneaux:2020ekh}. Moreover, such extensions provide a systematic framework for examining celestial holography in supersymmetric regimes, offering insights into scattering amplitudes involving gravitinos and their superpartners \cite{Fotopoulos:2020bqj, Narayanan:2020amh, Pano:2021ewd, Fuentealba:2023hzq, Tropper:2024evi}. The simplest extension introduces spin-$\frac{1}{2}$ generators $Q_\alpha$ alongside the bosonic supertranslation and superrotation generators \cite{Awada:1985by, Teitelboim:1977fs, Tabensky:1977ic, Avery:2015iix, Barnich:2015sca, Banerjee:2015kcx, Fuentealba:2021xhn, Bagchi:2022owq}. These generators satisfy anticommutation relations consistent with the Poincar\'e supersymmetry algebra, while commuting appropriately with the extended BMS generators. Schematically, the supersymmetric BMS$_4$ algebra contains generators $\{L_{n}, \bar{L}_{\bar{n}}, M_{r,\bar{r}}, Q^{(\frac{1}{2},0)}_{r,\bar{r}}, Q^{(0,\frac{1}{2})}_{r,\bar{r}}\}$\footnote{
        $(l,\bar{l})$ for the super charge label the $\mathfrak{so}(3)$ spins in Lorentzian group $\mathfrak{so}(3)\times\mathfrak{so}(3)$. See Section \ref{subsec:ChargeRepresentations} for more details. 
    }, and the commutation relation besides \eqref{eq:BMS4Algebra} is
    \begin{equation}\label{eq:SimplestSUSY}
        \{Q^{(\frac{1}{2},0)}_{r,\bar{r}}, Q^{(0,\frac{1}{2})}_{s,\bar{s}}\} = M_{r+s,\bar{r}+\bar{s}}.
    \end{equation}
    This algebra illustrates how supersymmetry intertwines with the infinite-dimensional asymptotic symmetries, thereby generalizing the infrared structures identified in purely bosonic theories. \par

    In \cite{Lodato:2016alv, Bagchi:2017cte, Chen:2023esw}, the authors proposed the democratic (homogeneous) and despotic (inhomogeneous) limits to obtain two types of super BMS$_3$ algebras. Applying the same method to the BMS$_4$ case yields \eqref{eq:SimplestSUSY} as well as algebra \eqref{eq:MagneticBMS4TypeI} respectively. However, taking limit is not the full story of supersymmetric extension of the BMS$_4$ algebra. In this work, we identify additional supersymmetric extensions, which can be classified as electric and magnetic types. Specifically, the electric super BMS$_4$ algebra refers to the case where anticommutators of supercharges produce only supertranslations, while the magnetic super BMS$_4$ algebra is characterized by anticommutators that also yield superrotations (See \cite{Banerjee:2019lrv, Prabhu:2021bod} for early discussion on magnetic examples). In this classification, the simplest extension \eqref{eq:SimplestSUSY} corresponds to an electric super BMS$_4$ algebra. This paper focuses on solutions admitting finite-dimensional subalgebras, which have potential applications in flat holography \cite{Banks:2014iha, Adamo:2014wea, Fotopoulos:2020bqj, Banerjee:2022lnz}. For the electric case, there are ten distinct unextended superalgebras. It was realized in section \ref{sec:Realization} that only three of them can be realized without introducing spin higher than $\frac{1}{2}$ in the corresponding symmetry representations. The magnetic cases are more intricate. There are two types of chiral super BMS$_4$ algebras: Type I \eqref{eq:MagneticBMS4TypeI} and Type II \eqref{eq:MagneticBMS4TypeII}, along with four nonchiral superalgebras formed by combinations of these two types. Notably, for the nonchiral cases, $R$-symmetry is necessary for the closure of the algebra. The representation of the $R$-charges varies in each case, depending on the types of left- and right-handed supercharges. The appearance of the $R$-charges indicates that these combined super BMS$_4$ algebras should be interpreted as extended supersymmetry. \par

    Beyond the algebraic formulation, this study presents an explicit free field realization of the super BMS$_4$ algebra, by expressing the algebra generators as bilinears of the field modes. Specifically, we construct the supersymmetric extension of the massless electric scalar theory $\mathcal{L} = \frac{1}{2}\partial_0\phi\partial_0\phi$, representing the simplest scalar multiplet. For the electric superalgebras, the theory \eqref{eq:ElectricTheory} admits three distinct electric superalgebras; conversely, for the magnetic superalgebras, each constructed theory is unique. Notably, although theory \eqref{eq:MagneticTypeIITheory} resembles the electric theory \eqref{eq:ElectricTheory}, they are fundamentally different due to the distinct conformal dimensions of the fermionic fields. Interestingly, theory \eqref{eq:MagneticTypeI-ITheory} establishes a relationship between the electric and magnetic scalars via $R$-symmetry, highlighting the essential role of magnetic scalars in supersymmetric flat holography. In theory \eqref{eq:MagneticTypeI-IITheory}, the emergence of a spin-$1$ massless boson suggests potential applications in supersymmetric soft theorems. Moreover, part of the theory \eqref{eq:MagneticTypeI-IITheory} appeared in \cite{Adamo:2014yya} as free perturbative supergravity theory. These realizations establish a direct correspondence between the algebraic structure of super BMS$_4$ and physical observables, thereby facilitating the exploration of supersymmetric celestial conformal field theories \cite{Taylor:2023bzj} and the investigation of how supersymmetry influences the infrared behavior of gravitational scattering processes. Such theories serve as fundamental building blocks for understanding the spectrum and dynamics of supersymmetric theories in asymptotically flat spacetime. Moreover, the free field approach provides a systematic framework for deriving operator product expansions and correlation functions, which are essential for studying holographic dualities and soft theorems in supersymmetric gravity. \par

    The remainder of this paper is organized as follows. In Section \ref{sec:Algebra}, we discuss the possible super BMS$_4$ algebras. After introducing the charge representations in Section \ref{subsec:ChargeRepresentations}, we identify ten electric super BMS$_4$ algebras in Section \ref{subsec:ElectricAlgebra}, along with two chiral magnetic super BMS$_4$ algebras and four nonchiral superalgebras in Section \ref{subsec:MagneticAlgebra}. We then proceed to the free field realizations of these algebras in Section \ref{sec:Realization}. Specifically, in Section \ref{subsec:ElectricRealization}, we demonstrate that theory \eqref{eq:ElectricTheory} is trivial in the sense that it preserves three of the electric superalgebras. In the magnetic sector discussed in Section \ref{subsec:MagneticRealization}, we show that the field theories admitting different superalgebras are distinct. Lastly, we conclude this work in Section \ref{sec:Discussion}. Appendix \ref{app:TensorProductOfVirReprs} contains detailed calculations of the linear difference equations arising from the constraints on the representations. \par

\section{Super \texorpdfstring{BMS$_4$}{BMS4} Algebras}\label{sec:Algebra}

    It is well known that the global part of the BMS$_4$ algebra is the $3$D Carrollian conformal algebra\footnote{
        In this paper, the $d$-dimensional Carrollian conformal algebra refers to the finite-dimensional algebra that is isomorphic to the $(d+1)$-dimensional Poincar\'e algebra, as well as to the global part of the BMS$_d$ algebra.
    }
    \cite{Duval:2014uva}. As established in the literature, every aspect of Carrollian theories, such as the taking limit procedures and types of field theories, has electric and magnetic versions \cite{Duval:2014uoa, Duval:2014uva, Duval:2014lpa, Lodato:2016alv, Bagchi:2017cte, Chen:2023esw}. In this paper, we adopt the terminology from Carrollian theory to classify the super BMS$_4$ algebras. Specifically, we define a super BMS$_4$ algebra as electric if the anticommutators of the supercharges yield only supertranslations. In contrast, we classify a super BMS$_4$ algebra as magnetic if the anticommutators produce both superrotations and supertranslations:
    \begin{equation}
        \begin{aligned}
            &\text{electric:} && \{Q,Q\}\propto M \\
            &\text{magnetic:} && \{Q,Q\}\propto L, \bar{L} \text{ or } M \\
        \end{aligned}
    \end{equation}
    \par

    In this section, we systematically derive the super BMS$_4$ algebras under the following assumptions. First, we assume that supersymmetry is generated by certain currents, implying that the supercharges reside in charge representations, which will be introduced in Section \ref{subsec:ChargeRepresentations}. Second, considering the potential applications in flat space holography, we require that the super BMS$_4$ algebra possesses finite-dimensional subalgebras, which can be viewed as supersymmetric extensions of the Poincaré algebra. Lastly, to respect the spin-statistics theorem, we stipulate that the supercharges must have half-integer spins. With these assumptions, we classify the super BMS$_4$ algebras into electric and magnetic types in Sections \ref{subsec:ElectricAlgebra} and \ref{subsec:MagneticAlgebra} respectively. \par

\subsection{Charge representation of \texorpdfstring{BMS$_4$}{BMS4} algebra} \label{subsec:ChargeRepresentations}

    In the context of supersymmetry, fermionic charges act as modes of supersymmetric currents. This subsection addresses the representation theory of these generators, which are associated with specific currents. For simplicity, we first examine the representation under a single branch of the Virasoro algebra and subsequently extend this to the full BMS$_4$ algebra within this framework. \par

    The symmetric generator $Q_{n}$ of a holomorphic current $J(z)$ with conformal dimension $h$ is expressed as
    \begin{equation}\label{eq:CurrentExpansion}
        J(z) = \sum_{r \in \mathbb{Z} + r_0} Q_r z^{-r-h} .
    \end{equation}
    The action of Virasoro algebra on $Q_{n}$ is given by
    \begin{equation}\label{eq:ChargeRepresentation}
        [L_{m},Q_r] = ((h-1)m - r) Q_{m+r}.
    \end{equation}
    Verification of the Jacobi identity for $\{L_{m},L_{n},Q_r\}$ confirms that this defines a self consistent representation. In this work, we refer to this as the charge representation. Notably, the charge representation admits a finite subrepresentation if there exists a cutoff in the action of $L_{\pm 1}$ on $Q$:
    \begin{equation}
        \exists ~ r>s \in \mathbb{Z} + r_0 \text{ s.t. } [L_1,Q_r] = [L_{-1},Q_s] = 0.
    \end{equation}
    This amounts to solving $(h-1) - r = -(h-1) - s = 0$, which yields the following solutions:
    \begin{equation}
        \left\{~~\begin{aligned}
            &r_0=0,\quad h\in\mathbb{Z},\quad h\ge 1, \\
            &r_0=\frac{1}{2},\quad h\in\mathbb{Z}+\frac{1}{2},\quad h\ge \frac{3}{2}. \\
        \end{aligned}\right.
    \end{equation}
    For these cases, $\{Q_{h-1},\dots,Q_{-h+1}\}$ forms a finite-dimensional representation space of the subalgebra $\{L_{1},L_{0},L_{-1}\} = \mathfrak{so}(3)$ with spin-$l=h-1$. In the remainder of this paper, we use the notation introduced in \cite{Haag:1974qh}, and label the representations with $\mathfrak{so}(3)$ spin-$l$ (and $\bar{l}$ for the anti-holomorphic part), with $l(\bar{l})$ extended to general $\mathbb{R}$. \par

    Considering a (super)algebra that contains multiple such symmetry charges $Q^{i}$, the Jacobi identities of the algebra should be saturated such that the algebra is well-defined. Assume that the commutation relation is
    \begin{equation}
        [Q^i,Q^j\}\propto Q^k, \quad [Q^i,Q^j\} = \begin{cases}
            Q^iQ^j + Q^jQ^i & \text{$Q^i$ and $Q^j$ are fermionic,} \\
            Q^iQ^j - Q^jQ^i & \text{otherwise.} \\
        \end{cases}
    \end{equation}
    The Jacobi identity of $\{L_0, Q^i, Q^j\}$ requires that the index of $Q^k$ equals the sum of the indices of $Q^i$ and $Q^j$:
    \begin{equation}\label{eq:TensorProductOfVirasoro}
        [Q^i_{r_i},Q^j_{r_j}\}= f_{l_i,l_j}^{l_k}(r_i,r_j) ~ Q^k_{r_i+r_j},
    \end{equation}
    then the Jacobi identity of $\{L_{n}, Q^i, Q^j\}$ for generic $n$ further gives a constraint on the function $f$:
    \begin{equation} \label{eq:JacobiEquation}
        (l_k n - (r_i+r_j)) f_{l_i,l_j}^{l_k}(r_i,r_j) - (l_j n - r_j) f_{l_i,l_j}^{l_k}(r_i,r_j + n) - (l_i n - r_i) f_{l_i,l_j}^{l_k}(r_i+n,r_j) = 0.
    \end{equation}  
    This is a linear difference equation which is only defined on discrete $n, r_i, r_j\in \mathbb{Z}$ points. Numerically, we find the following useful solutions\footnote{
        Here we consider solutions with $l\in\mathbb{Z}/2$ and $l>0$, i.e., charge representations that admit finite-dimensional sub-representations. See Appendix \ref{app:TensorProductOfVirReprs} for more details.
    }
    \begin{subequations}\label{eq:TensorProductOfVirReprsSolution}
        \begin{align}
            &f_{l_i,l_j}^{l_k}(r_i,r_j) = 1, && l_k = l_i+l_j, \label{eq:TensorProductOfVirReprsSolution1}\\
            &f_{l_i,l_j}^{l_k}(r_i,r_j) = l_j r_i - l_i r_j, && l_k = l_i+l_j-1, \label{eq:TensorProductOfVirReprsSolution2}\\
            &f_{l_i,l_j}^{l_k}(r_i,r_j) = r_i(l_j r_i + r_j), && l_i = 0, \quad l_k = l_j-2, \text{ or } (i \xleftrightarrow{} j), \label{eq:TensorProductOfVirReprsSolution3}
        \end{align}
    \end{subequations}
    up to overall factors. Especially, \eqref{eq:TensorProductOfVirReprsSolution2} applies to the BMS$_4$ algebra itself. The case that $Q^i=Q^j=Q^k=L$, which corresponds to $l_i=l_j=1$ and $l_k=l_i+l_j-1=1$, gives the commutation relation
    \begin{equation}
        [L_{m},L_{n}] = (1*m-1*n)L_{m+n},
    \end{equation}
    while the case that $Q^i=L, Q^j=Q^k=M$, which corresponds to $l_i=1, l_j=\frac{1}{2}$ and $l_k=l_i+l_j-1=\frac{1}{2}$, gives the commutation relation
    \begin{equation}
        [L_{m}, M_{r,\bar{r}}] = \left(\frac{m}{2}-1*r\right)M_{m+r,\bar{r}}.
    \end{equation}
    This result aligns with the fact that generators in the finite subalgebra $L_{m}, \bar{L}_{\bar{m}}, M_{r,\bar{r}}$ are respectively in $(1,0)$, $(0,1)$ and $\left(\frac{1}{2},\frac{1}{2}\right)$ representations under the Lorentzian algebra $\mathfrak{so}_L(3)\times\mathfrak{so}_R(3)$. \par

    It is worth emphasizing that the tensor product rules in \eqref{eq:TensorProductOfVirReprsSolution} deviate from the familiar decomposition of $\mathfrak{so}(3)$ spin representations. For the latter, one has the standard rule 
    \begin{equation}
        (l_i)\otimes(l_j) = (|l_i-l_j|)\oplus \cdots \oplus (l_i+l_j),
    \end{equation}
    where all intermediate spins appear. In contrast, for the charge representations under consideration, not all such intermediate spins are admissible. Specifically, the condition $l_k < l_i + l_j - 2$ fails to satisfy the Jacobi identity equation \eqref{eq:JacobiEquation} for generic $l_i$ and $l_j$. This indicates that the charge representation enforces a more restrictive structure, reflecting the algebraic consistency required by supersymmetric extensions rather than the purely group theoretic rules of $\mathfrak{so}(3)$. \par

    Building on the above discussion, a charge $Q$ in the BMS$_4$ algebra can be characterized by its holomorphic and anti-holomorphic spins $(l,\bar{l})$, with commutation relations given by
    \begin{equation}
        [L_{m},Q^{(l,\bar{l})}_{r,\bar{r}}] = (lm - r) Q^{(l,\bar{l})}_{m+r,\bar{r}}, \qquad [\bar{L}_{\bar{m}},Q^{(l,\bar{l})}_{r,\bar{r}}] = (\bar{l}\bar{m} - \bar{r}) Q^{(l,\bar{l})}_{r,\bar{m}+\bar{r}}.
    \end{equation}
    When the supertranslation generators $M$ act nontrivially on $Q$, i.e. $[M,Q^{a}] \propto Q^{b} \neq 0$, the spins of $M$ and $Q^{(i)}$ are constrained by the tensor product rule in \eqref{eq:TensorProductOfVirReprsSolution}. In this sense, a super BMS$_4$ algebra can be systematically constructed by enlarging the BMS$_4$ algebra with additional fermionic generators, provided that the extended commutation relations are consistent. The labels $(l,\bar{l})$ already determine the action of $L$ and $\bar{L}$ on these new generators, so the nontrivial part of the construction lies in specifying the commutators of $M$ with the fermionic charges, as well as the (anti-)commutators among the new generators themselves. Ensuring algebraic closure then requires verifying the Jacobi identities case by case. In the following subsections, we will present explicit constructions of the electric and magnetic super BMS$_4$ algebras. \par

\subsection{Electric super \texorpdfstring{BMS$_4$}{BMS4} algebra} \label{subsec:ElectricAlgebra}
    
    In this subsection, we reformulate the electric super BMS$_4$ algebra using the charge representation framework introduced in Section \ref{subsec:ChargeRepresentations}. A standard construction of the electric extension begins with the infinite-dimensional uplift of the super Poincar\'e algebra. The resulting extended algebra is generated by $\{L_{n}, \bar{L}_{\bar{n}}, M_{r,\bar{r}}, Q^{(\frac{1}{2},0)}_r, Q^{(0,\frac{1}{2})}_{\bar{r}}\}$, where $Q^{(\frac{1}{2},0)}_r$ and $Q^{(0,\frac{1}{2})}_{\bar{r}}$ denote the fermionic supercharges. Their defining relations are 
    \begin{equation}\label{eq:ElectricBMS4FromPoincare}
        [M_{r,\bar{r}},Q^{(\frac{1}{2},0)}_s]=[M_{r,\bar{r}}, Q^{(0,\frac{1}{2})}_{\bar{s}}]=0, \qquad \{Q^{(\frac{1}{2},0)}_r, Q^{(0,\frac{1}{2})}_{\bar{r}}\} = M_{r,\bar{r}}.
    \end{equation}
    The largest finite subalgebra of this infinite extension is generated by
    \begin{equation}
        \{L_{\pm 1,0}, \bar{L}_{\pm 1,0}, M_{\pm\frac{1}{2},\pm\frac{1}{2}}, Q^{(\frac{1}{2},0)}_{\pm\frac{1}{2}}, Q^{(0,\frac{1}{2})}_{\pm\frac{1}{2}}\},
    \end{equation}
    which is isomorphic to the standard $\mathcal{N}=1$ super Poincar\'e algebra. This correspondence becomes explicit after relabeling the generators as 
    \begin{equation}
        \begin{aligned}
            & J^{0,1}_\text{P} = \frac{i}{2}(- L_{1} - L_{-1} + \bar{L}_{1} + \bar{L}_{-1} ), ~
            J^{0,2}_\text{P} = \frac{1}{2}(L_{1} - L_{-1} + \bar{L}_{1} - \bar{L}_{-1} ), ~
            J^{0,3}_\text{P} = - L_{0}- \bar{L}_{0}, \\
            & J^{1,2}_\text{P} = -i L_{0} + i \bar{L}_{0},  ~
            J^{1,3}_\text{P} = \frac{i}{2}(- L_{1} + L_{-1} + \bar{L}_{1} - \bar{L}_{-1} ), ~ 
            J^{2,3}_\text{P} = \frac{1}{2}(L_{1} + L_{-1} + \bar{L}_{1} + \bar{L}_{-1} ), \\
            & P^0_\text{P} = \frac{1}{\sqrt{2}}(M_{\frac{1}{2},\frac{1}{2}} + M_{-\frac{1}{2},-\frac{1}{2}}), \qquad
            P^1_\text{P} = \frac{i}{\sqrt{2}}(M_{\frac{1}{2},-\frac{1}{2}} - M_{-\frac{1}{2},\frac{1}{2}}), \\
            &P^2_\text{P} = \frac{1}{\sqrt{2}}(-M_{\frac{1}{2},-\frac{1}{2}} - M_{-\frac{1}{2},\frac{1}{2}}), \qquad
            P^3_\text{P} = \frac{1}{\sqrt{2}}(-M_{\frac{1}{2},\frac{1}{2}} + M_{-\frac{1}{2},-\frac{1}{2}}), \\
            & Q_{\frac{1}{2}}^\text{P} = i~ 2^{\frac{3}{4}} Q^{(0,\frac{1}{2})}_{-\frac{1}{2}}, \quad
            Q_{-\frac{1}{2}}^\text{P} = 2^{\frac{3}{4}} Q^{(0,\frac{1}{2})}_{\frac{1}{2}}, \quad
            \bar{Q}_{\frac{1}{2}}^\text{P} = -i~ 2^{\frac{3}{4}} Q^{(\frac{1}{2},0)}_{-\frac{1}{2}}, \quad
            \bar{Q}_{-\frac{1}{2}}^\text{P} = 2^{\frac{3}{4}} Q^{(\frac{1}{2},0)}_{\frac{1}{2}},
        \end{aligned}
    \end{equation}
    With this identification, the set $\{J^{\mu\nu}_{\text{P}}, P^\mu_{\text{P}}, Q^{\text{P}}_\alpha, \bar{Q}^{\text{P}}_{\dot{\alpha}}\}$ reproduces the standard super Poincar\'e algebra. \par

    However, the algebra obtained in \eqref{eq:ElectricBMS4FromPoincare} does not exhaust all possible constructions of electric super BMS$_4$. To proceed more systematically, we adopt a general strategy based on the following guiding principles. First, the supercharges are required to transform in charge representations, ensuring consistency with the underlying current algebra structure. Second, the labels $(l,\bar{l})$ of the supercharges are taken to lie in $\mathbb{Z}/2$ with $l,\bar{l}\geq 0$, so that the resulting algebra admits a finite-dimensional subalgebra, which can be viewed as a supersymmetric extension of the Poincar\'e algebra. Finally, to respect the spin–statistics theorem, the total spin must satisfy $l+\bar{l}\in \mathbb{Z}+\frac{1}{2}$. These ansatzes provide a systematic framework to classify and construct possible electric super BMS$_4$ algebras beyond the minimal realization. \par

    The cases with a single set of supercharges fail to satisfy the ansatzes outlined above. In fact, there are precisely two possible constructions compatible with the constraint \eqref{eq:TensorProductOfVirReprsSolution}. The first is given by      \eqref{eq:TensorProductOfVirReprsSolution}:
    \begin{equation}
        \{L_{n}, \bar{L}_{\bar{n}}, M_{r,\bar{r}}, Q^{(\frac{1}{4},\frac{1}{4})}_{r,\bar{r}}\}, \text{ with } \{Q^{(\frac{1}{4},\frac{1}{4})}_{r,\bar{r}},Q^{(\frac{1}{4},\frac{1}{4})}_{s,\bar{s}}\} = M_{r+s,\bar{r}+\bar{s}},
    \end{equation}
    which satisfies the condition \eqref{eq:TensorProductOfVirReprsSolution1}. The second possibility reads  
    \begin{equation}
        \{L_{n}, \bar{L}_{\bar{n}}, M_{r,\bar{r}}, Q^{(\frac{3}{4},\frac{3}{4})}_{r,\bar{r}}\}, \text{ with } \{Q^{(\frac{3}{4},\frac{3}{4})}_{r,\bar{r}},Q^{(\frac{3}{4},\frac{3}{4})}_{s,\bar{s}}\} = \left(\frac{3}{4} r - \frac{3}{4} s\right)\left(\frac{3}{4} \bar{r} - \frac{3}{4} \bar{s}\right)M_{r+s,\bar{r}+\bar{s}},
    \end{equation}
    consistent with the constraint \eqref{eq:TensorProductOfVirReprsSolution2}. Although these constructions are algebraically consistent, neither of them admits a finite-dimensional subalgebra. As a result, they fall outside the class of super BMS$_4$ algebras envisioned by our ansatzes and do not yield a physically viable supersymmetric extension. \par

    We now turn to the case with two sets of supercharges. A representative example is provided by the extension of the super Poincar\'e algebra \eqref{eq:ElectricBMS4FromPoincare}, where both the holomorphic and anti-holomorphic labels satisfy the condition \eqref{eq:TensorProductOfVirReprsSolution1}. More generally, the supercharge $Q^{(\frac{1}{2},0)}_{r}$ can be extended to $Q^{(\frac{1}{2},0)}_{r,\bar{r}}$, with the original operator identified as $Q^{(\frac{1}{2},0)}_{r}=Q^{(\frac{1}{2},0)}_{r,0}$. This extension introduces additional super BMS$_4$ charges while leaving the finite-dimensional subalgebra unchanged. In a completely analogous fashion, the supercharge $Q^{(0,\frac{1}{2})}_{\bar{r}}$ can be extended to $Q^{(0,\frac{1}{2})}_{r,\bar{r}}$. Following this logic, we obtain ten distinct realizations of the electric super BMS$_4$ algebra, classified according to the rules \eqref{eq:TensorProductOfVirReprsSolution}, up to exchange of holomorphic and anti-holomorphic labels. \par
    \noindent \textbf{holomorphic \eqref{eq:TensorProductOfVirReprsSolution1}, anti-holomorphic \eqref{eq:TensorProductOfVirReprsSolution1}:}
    \begin{equation}\label{eq:ElectricBMS4FromPoincareFull}
        \{L_{n}, \bar{L}_{\bar{n}}, M_{r,\bar{r}}, Q^{(\frac{1}{2},0)}_{r,\bar{r}}, Q^{(0,\frac{1}{2})}_{r,\bar{r}}\}, \quad \{Q^{(\frac{1}{2},0)}_{r,\bar{r}}, Q^{(0,\frac{1}{2})}_{s,\bar{s}}\} = M_{r+s,\bar{r}+\bar{s}}. 
    \end{equation}
    \textbf{holomorphic \eqref{eq:TensorProductOfVirReprsSolution1}, anti-holomorphic \eqref{eq:TensorProductOfVirReprsSolution2}:}
    \begin{align}
        &\{L_{n}, \bar{L}_{\bar{n}}, M_{r,\bar{r}}, Q^{(\frac{1}{2},1)}_{r,\bar{r}}, Q^{(0,\frac{1}{2})}_{r,\bar{r}}\}, \quad \{Q^{(\frac{1}{2},1)}_{r,\bar{r}}, Q^{(0,\frac{1}{2})}_{s,\bar{s}}\} = \left(\frac{1}{2}\bar{r} - \bar{s}\right)M_{r+s,\bar{r}+\bar{s}}, \label{eq:ElectricBMS4Type12A} \\
        &\{L_{n}, \bar{L}_{\bar{n}}, M_{r,\bar{r}}, Q^{(\frac{1}{2},0)}_{r,\bar{r}}, Q^{(0,\frac{3}{2})}_{r,\bar{r}}\}, \quad \{Q^{(\frac{1}{2},0)}_{r,\bar{r}}, Q^{(0,\frac{3}{2})}_{s,\bar{s}}\} = \frac{3}{2}\bar{r}M_{r+s,\bar{r}+\bar{s}}. \label{eq:ElectricBMS4Type12B}
    \end{align}
    \textbf{holomorphic \eqref{eq:TensorProductOfVirReprsSolution1}, anti-holomorphic \eqref{eq:TensorProductOfVirReprsSolution3}:}
    \begin{equation} \label{eq:ElectricBMS4Type13}
        \{L_{n}, \bar{L}_{\bar{n}}, M_{r,\bar{r}}, Q^{(\frac{1}{2},0)}_{r,\bar{r}}, Q^{(0,\frac{5}{2})}_{r,\bar{r}}\}, \quad \{Q^{(\frac{1}{2},0)}_{r,\bar{r}}, Q^{(0,\frac{5}{2})}_{s,\bar{s}}\} = \bar{r}\left(\frac{5}{2}\bar{r} + \bar{s}\right)M_{r+s,\bar{r}+\bar{s}}.
    \end{equation}
    \textbf{holomorphic \eqref{eq:TensorProductOfVirReprsSolution2}, anti-holomorphic \eqref{eq:TensorProductOfVirReprsSolution2}:}
    \begin{align}
        &\{L_{n}, \bar{L}_{\bar{n}}, M_{r,\bar{r}}, Q^{(\frac{1}{2},1)}_{r,\bar{r}}, Q^{(1,\frac{1}{2})}_{r,\bar{r}}\}, \quad \{Q^{(\frac{1}{2},1)}_{r,\bar{r}}, Q^{(1,\frac{1}{2})}_{s,\bar{s}}\} = \left(r - \frac{1}{2}s\right)\left(\frac{1}{2}\bar{r} - \bar{s}\right)M_{r+s,\bar{r}+\bar{s}}, \label{eq:ElectricBMS4Type22A} \\
        &\{L_{n}, \bar{L}_{\bar{n}}, M_{r,\bar{r}}, Q^{(\frac{1}{2},0)}_{r,\bar{r}}, Q^{(1,\frac{3}{2})}_{r,\bar{r}}\}, \quad \{Q^{(\frac{1}{2},0)}_{r,\bar{r}}, Q^{(1,\frac{3}{2})}_{s,\bar{s}}\} = \left(r - \frac{1}{2}s\right) \frac{3}{2}\bar{r} M_{r+s,\bar{r}+\bar{s}}, \label{eq:ElectricBMS4Type22B} \\
        &\{L_{n}, \bar{L}_{\bar{n}}, M_{r,\bar{r}}, Q^{(\frac{3}{2},0)}_{r,\bar{r}}, Q^{(0,\frac{3}{2})}_{r,\bar{r}}\}, \quad \{Q^{(\frac{3}{2},0)}_{r,\bar{r}}, Q^{(0,\frac{3}{2})}_{s,\bar{s}}\} = \frac{9}{4}s\bar{r} M_{r+s,\bar{r}+\bar{s}}. \label{eq:ElectricBMS4Type22C}
    \end{align}
    \textbf{holomorphic \eqref{eq:TensorProductOfVirReprsSolution2}, anti-holomorphic \eqref{eq:TensorProductOfVirReprsSolution3}:}
    \begin{align}
        &\{L_{n}, \bar{L}_{\bar{n}}, M_{r,\bar{r}}, Q^{(\frac{1}{2},0)}_{r,\bar{r}}, Q^{(1,\frac{5}{2})}_{r,\bar{r}}\}, \quad \{Q^{(\frac{1}{2},0)}_{r,\bar{r}}, Q^{(1,\frac{5}{2})}_{s,\bar{s}}\} = \left(r - \frac{1}{2}s\right)\bar{r}\left(\frac{5}{2}\bar{r} + \bar{s}\right)M_{r+s,\bar{r}+\bar{s}}, \label{eq:ElectricBMS4Type23A} \\
        &\{L_{n}, \bar{L}_{\bar{n}}, M_{r,\bar{r}}, Q^{(\frac{3}{2},0)}_{r,\bar{r}}, Q^{(0,\frac{5}{2})}_{r,\bar{r}}\}, \quad \{Q^{(\frac{3}{2},0)}_{r,\bar{r}}, Q^{(0,\frac{5}{2})}_{s,\bar{s}}\} = \frac{3}{2}s \bar{r}\left(\frac{5}{2}\bar{r} + \bar{s}\right)M_{r+s,\bar{r}+\bar{s}}.\label{eq:ElectricBMS4Type23B}
    \end{align}
    \textbf{holomorphic \eqref{eq:TensorProductOfVirReprsSolution3}, anti-holomorphic \eqref{eq:TensorProductOfVirReprsSolution3}:}
    \begin{equation}\label{eq:ElectricBMS4Type33}
        \{L_{n}, \bar{L}_{\bar{n}}, M_{r,\bar{r}}, Q^{(\frac{5}{2},0)}_{r,\bar{r}}, Q^{(0,\frac{5}{2})}_{r,\bar{r}}\}, \quad \{Q^{(\frac{5}{2},0)}_{r,\bar{r}}, Q^{(0,\frac{5}{2})}_{s,\bar{s}}\} = s\left(r+\frac{5}{2}s\right)\bar{r}\left(\frac{5}{2}\bar{r} + \bar{s}\right)M_{r+s,\bar{r}+\bar{s}}.
    \end{equation}
    \par

    All of the above algebras contain finite-dimensional subalgebras. However, plugging in the coefficients, the finite-dimensional subalgebras are trivial with all $\{Q,Q\}=0$, except for \eqref{eq:ElectricBMS4FromPoincareFull}, \eqref{eq:ElectricBMS4Type12A} and \eqref{eq:ElectricBMS4Type22A}. Furthermore, \eqref{eq:ElectricBMS4Type12A} and \eqref{eq:ElectricBMS4Type22A} necessarily include spin-$\frac{3}{2}$ components. Such higher-spin fermionic generators are not compatible with the Haag–Łopuszański–Sohnius theorem\cite{Haag:1974qh}, which allows only spin-$\frac{1}{2}$ supercharges in a consistent extension of the Poincar\'e algebra. Therefore, the only physically admissible super Poincar\'e algebra is the finite-dimensional subalgebra of \eqref{eq:ElectricBMS4FromPoincareFull}. \par

\subsection{Magnetic chiral super \texorpdfstring{BMS$_4$}{BMS4} algebras} \label{subsec:MagneticAlgebra}

    In the magnetic case, it's required the existence of supercharges $Q^a$ such that $\{Q^a, Q^b\}\propto L$. Since $L$ is purely holomorphic, consistency demands that the supercharges anticommuting into $L$ must also be purely holomorphic. Otherwise, the relation $\{Q^a_{r,\bar{r}}, Q^b_{s,\bar{s}}\} = G_{r+s,\bar{r}+\bar{s}}$ would generate unphysical bosonic generators $L_{n,\bar{n}}$. An analogous argument applies to the anti-holomorphic sector. Therefore, it suffices to consider $\{Q^a_{r}, Q^b_{r}\}\propto L$ with $Q^a_{r}\in (l_a,0)$ and $Q^b_{r}\in (l_b,0)$, together with $\{Q^c_{\bar{r}}, Q^d_{\bar{r}}\}\propto \bar{L}$ where $Q^c_{\bar{r}}\in (0,l_c)$ and $Q^d_{\bar{r}}\in (0,l_d)$. \par

    Assuming a single set of supercharges $Q$ generates $L$, the only consistent choice is $Q\in \left(\frac{1}{2},0\right)$ such that $\{Q^{(\frac{1}{2},0)}_{r},Q^{(\frac{1}{2},0)}_{s}\} = L_{r+s}$. The Jacobi identity for $\{M, Q^{(\frac{1}{2},0)}, Q^{(\frac{1}{2},0)}\}$ then implies a nonvanishing commutator $[M, Q^{(\frac{1}{2},0)}]$. Applying the rules of \eqref{eq:TensorProductOfVirReprsSolution}, one finds two distinct chiral magnetic BMS$_4$ algebras:
    \par\noindent \textbf{Type I} $\{L_{n}, \bar{L}_{\bar{n}}, M_{r,\bar{r}}, Q^{(\frac{1}{2},0)}_{r}, Q^{(0,\frac{1}{2})}_{r,\bar{r}}\}$:
    \begin{equation}\label{eq:MagneticBMS4TypeI}
        [M_{r,\bar{r}}, Q^{(\frac{1}{2},0)}_{s}] = \left( \frac{1}{2}r-\frac{1}{2}s \right) Q^{(0,\frac{1}{2})}_{r+s,\bar{r}}, \quad \{Q^{(\frac{1}{2},0)}_{r}, Q^{(\frac{1}{2},0)}_{s}\} = L_{r+s}, \quad \{Q^{(\frac{1}{2},0)}_{r}, Q^{(0,\frac{1}{2})}_{s,\bar{s}}\} = M_{r+s,\bar{s}},
    \end{equation}
    \textbf{Type II} $\{L_{n}, \bar{L}_{\bar{n}}, M_{r,\bar{r}}, Q^{(\frac{1}{2},0)}_{r}, Q^{(1,\frac{1}{2})}_{r,\bar{r}}\}$:
    \begin{equation}\label{eq:MagneticBMS4TypeII}
        [M_{r,\bar{r}}, Q^{(\frac{1}{2},0)}_{s}] = Q^{(1,\frac{1}{2})}_{r+s,\bar{r}}, \quad \{Q^{(\frac{1}{2},0)}_{r}, Q^{(\frac{1}{2},0)}_{s}\} = L_{r+s}, \quad \{Q^{(\frac{1}{2},0)}_{r}, Q^{(1,\frac{1}{2})}_{s,\bar{s}}\} = -\left(r-\frac{1}{2}s\right) M_{r+s,\bar{s}},
    \end{equation}
    In Type I, $[M, Q^{(\frac{1}{2},0)}] \propto Q^{(0,\frac{1}{2})}$ satisfies \eqref{eq:TensorProductOfVirReprsSolution2}, whereas in Type II, $[M, Q^{(\frac{1}{2},0)}] \propto Q^{(1,\frac{1}{2})}$ satisfies \eqref{eq:TensorProductOfVirReprsSolution1}. Unlike the electric case, the supercharges $Q^{(\frac{1}{2},0)}_{r}$ in \eqref{eq:MagneticBMS4TypeI} and \eqref{eq:MagneticBMS4TypeII} are not extended to $Q^{(\frac{1}{2},0)}_{r,\bar{r}}$, thereby avoiding the introduction of nonphysical generators $L_{m,\bar{m}}$. In contrast, the holomorphic index $r$ of $Q^{(0,\frac{1}{2})}_{r,\bar{r}}$ in \eqref{eq:MagneticBMS4TypeI} is required for the nonvanishing commutator $[M_{r,\bar{r}}, Q^{(\frac{1}{2},0)}_{s}] = Q^{(1,\frac{1}{2})}_{r+s,\bar{r}}$. \par

    \sloppy However, a mixture of the above two types does not yield a physically consistent superalgebra. Specifically, if one assumes that $\{Q^{(\frac{1}{2},0)}_{r}, Q^{(0,\frac{1}{2})}_{r,\bar{r}}, Q^{(1,\frac{1}{2})}_{r,\bar{r}}\}$ are the supercharges, the Jacobi identities for $\{Q^{(\frac{1}{2},0)}, Q^{(\frac{1}{2},0)}, Q^{(0,\frac{1}{2})}_{r,\bar{r}}\}$ and for $\{Q^{(\frac{1}{2},0)}, Q^{(\frac{1}{2},0)}, Q^{(1,\frac{1}{2})}_{r,\bar{r}}\}$ become mutually contradictory, unless one introduces an additional bosonic generator $\Tilde{M}^{(\frac{1}{2},\frac{1}{2})}$. The resulting extended algebra takes the form:
    \begin{equation}
        \begin{aligned}
            &\{L_{n}, \bar{L}_{\bar{n}}, M_{r,\bar{r}}, \Tilde{M}_{r,\bar{r}}, Q^{(\frac{1}{2},0)}_{r}, Q^{(0,\frac{1}{2})}_{r,\bar{r}}, Q^{(1,\frac{1}{2})}_{r,\bar{r}}\}, \\[0.5em]
            &[M_{r,\bar{r}}, Q^{(\frac{1}{2},0)}_{s}] = \frac{a}{2}(r-s) Q^{(0,\frac{1}{2})}_{r+s,\bar{r}} + (1-a)Q^{(1,\frac{1}{2})}_{r+s,\bar{r}}, \\[0.5em]
            &[\Tilde{M}_{r,\bar{r}}, Q^{(\frac{1}{2},0)}_{s}] = \frac{1-ab}{2c}(r-s) Q^{(0,\frac{1}{2})}_{r+s,\bar{r}} + \frac{(a-1)b}{c}Q^{(1,\frac{1}{2})}_{r+s,\bar{r}} \\[0.5em]
            &\{Q^{(\frac{1}{2},0)}_{r}, Q^{(\frac{1}{2},0)}_{s}\} = L_{r+s}, \quad \{Q^{(\frac{1}{2},0)}_{r}, Q^{(0,\frac{1}{2})}_{s,\bar{s}}\} = b M_{r+s,\bar{s}} + c\Tilde{M}_{r+s,\bar{s}}, \\[0.5em]
            &\{Q^{(\frac{1}{2},0)}_{r}, Q^{(1,\frac{1}{2})}_{s,\bar{s}}\} = \left(r-\frac{1}{2}s\right) \left(\frac{1-ab}{a-1}M_{r+s,\bar{s}} + \frac{ac}{a-1}\Tilde{M}_{r+s,\bar{s}}\right). \\
        \end{aligned}
    \end{equation}
    where $a,b,c$ are arbitrary constants. Importantly, the additional generator $\Tilde{M}$ lacks a clear physical interpretation. Therefore, such a mixed construction does not correspond to a physically meaningful symmetry. \par

    Observing \eqref{eq:MagneticBMS4TypeI}, we note that $Q^{(\frac{1}{2},0)}$ and $Q^{(0,\frac{1}{2})}$ are not related by conjugation. This makes it clear that both \eqref{eq:MagneticBMS4TypeI} and \eqref{eq:MagneticBMS4TypeII} describe chiral superalgebras, as only the holomorphic Virasoro algebra $L_{n}$ appears in anticommutators of super charges. For this reason, \eqref{eq:MagneticBMS4TypeI} and \eqref{eq:MagneticBMS4TypeII} are considered as holomorphic in this paper. Anti-holomorphic counterparts $\{\bar{Q}^{(0,\frac{1}{2})}_{\bar{r}}, \bar{Q}^{(\frac{1}{2},0)}_{r,\bar{r}}\}$ and $\{\bar{Q}^{(0,\frac{1}{2})}_{\bar{r}}, \bar{Q}^{(\frac{1}{2},1)}_{r,\bar{r}}\}$ can be defined through conjugation, yielding super BMS$_4$ algebra structures analogous to \eqref{eq:MagneticBMS4TypeI} and \eqref{eq:MagneticBMS4TypeII}. \par

    For a super BMS$_4$ algebra containing both holomorphic and anti-holomorphic sectors, simply including the corresponding supercharges is insufficient. For instance, consider holomorphic supercharges $\{Q^{(\frac{1}{2},0)}_{r}, Q^{(0,\frac{1}{2})}_{r,\bar{r}}\}$ and anti-holomorphic supercharges $\{\bar{Q}^{(0,\frac{1}{2})}_{\bar{r}}, \bar{Q}^{(\frac{1}{2},0)}_{r,\bar{r}}\}$, both of Type I as in \eqref{eq:MagneticBMS4TypeI}. The Jacobi identity for $\{Q^{(\frac{1}{2},0)}, Q^{(\frac{1}{2},0)}, \bar{Q}^{(\frac{1}{2},0)}\}$ is satisfied only if a bosonic generator $R^{(0,0)}$ is introduced, such that 
    \begin{equation}
        \{Q^{(\frac{1}{2},0)}, \bar{Q}^{(\frac{1}{2},0)}\}\propto R^{(0,0)}.
    \end{equation}
    Within the finite-dimensional subalgebra, $R = R^{(0,0)}_{0,0}$ naturally functions as an $R$-symmetry. Consequently, the super BMS$_4$ algebra requires the presence of $R$-symmetry even when it is not an extended supersymmetry algebra. This motivates the construction of magnetic super BMS$_4$ algebras combining different types of supercharges, and we label the resulting algebras Type A-B for the holomorphic part being Type A and anti-holomorphic part being Type B. The resulting superalgebras are:
    \par\noindent \textbf{Type I-I} $\{L_{n}, \bar{L}_{\bar{n}}, M_{r,\bar{r}}, Q^{(\frac{1}{2},0)}_{r}, Q^{(0,\frac{1}{2})}_{r,\bar{r}}, \bar{Q}^{(0,\frac{1}{2})}_{\bar{r}}, \bar{Q}^{(\frac{1}{2},0)}_{r,\bar{r}}, R^{(0,0)}_{r,\bar{r}}\}$:
    \begin{equation}\label{eq:MagneticBMS4TypeI-I}
        \begin{aligned}
            &[M_{r,\bar{r}}, Q^{(\frac{1}{2},0)}_{s}] = \left(\frac{1}{2}r-\frac{1}{2}s\right) Q^{(0,\frac{1}{2})}_{r+s,\bar{r}}, \quad [M_{r,\bar{r}}, \bar{Q}^{(0,\frac{1}{2})}_{\bar{s}}] = \left(\frac{1}{2}\bar{r}-\frac{1}{2}\bar{s}\right) \bar{Q}^{(\frac{1}{2},0)}_{r,\bar{r}+\bar{s}}, \\[0.5em]
            &\{Q^{(\frac{1}{2},0)}_{r}, Q^{(\frac{1}{2},0)}_{s}\} = L_{r+s}, \quad \{Q^{(\frac{1}{2},0)}_{r}, Q^{(0,\frac{1}{2})}_{s,\bar{s}}\} = M_{r+s,\bar{s}} , \\[0.5em]
            &\{\bar{Q}^{(0,\frac{1}{2})}_{\bar{r}}, \bar{Q}^{(0,\frac{1}{2})}_{\bar{s}}\} = \bar{L}_{\bar{r}+\bar{s}}, \quad \{\bar{Q}^{(0,\frac{1}{2})}_{\bar{r}}, \bar{Q}^{(\frac{1}{2},0)}_{s,\bar{s}}\} = M_{s,\bar{r}+\bar{s}} , \\[0.5em]
            &\{Q^{(\frac{1}{2},0)}_{r}, \bar{Q}^{(\frac{1}{2},0)}_{s,\bar{s}} \} = \left(\frac{1}{2}r-\frac{1}{2}s\right)R^{(0,0)}_{r+s,\bar{s}}, \quad \{ \bar{Q}^{(0,\frac{1}{2})}_{\bar{r}}, Q^{(0,\frac{1}{2})}_{s,\bar{s}} \} = -\left(\frac{1}{2}\bar{r}-\frac{1}{2}\bar{s}\right)R^{(0,0)}_{s,\bar{r} + \bar{s}}, \\[0.5em]
            &[R^{(0,0)}_{r,\bar{r}}, Q^{(\frac{1}{2},0)}_{s}] = -\bar{Q}^{(\frac{1}{2},0)}_{r+s,\bar{s}}, \quad [R^{(0,0)}_{r,\bar{r}}, \bar{Q}^{(0,\frac{1}{2})}_{\bar{s}}] = Q^{(0,\frac{1}{2})}_{s,\bar{r}+\bar{s}}, 
        \end{aligned}
    \end{equation}
    \textbf{Type II-II} $\{L_{n}, \bar{L}_{\bar{n}}, M_{r,\bar{r}}, Q^{(\frac{1}{2},0)}_{r}, Q^{(1,\frac{1}{2})}_{r,\bar{r}}, \bar{Q}^{(0,\frac{1}{2})}_{\bar{r}}, \bar{Q}^{(\frac{1}{2},1)}_{r,\bar{r}}, R^{(1,1)}_{r,\bar{r}}\}$:
    \begin{equation}\label{eq:MagneticBMS4TypeII-II}
        \begin{aligned}
            &[M_{r,\bar{r}}, Q^{(\frac{1}{2},0)}_{s}] = Q^{(1,\frac{1}{2})}_{r+s,\bar{r}}, \quad [M_{r,\bar{r}}, \bar{Q}^{(0,\frac{1}{2})}_{\bar{s}}] = \bar{Q}^{(\frac{1}{2},1)}_{r,\bar{r}+\bar{s}}, \\[0.5em]
            &\{Q^{(\frac{1}{2},0)}_{r}, Q^{(\frac{1}{2},0)}_{s}\} = L_{r+s}, \quad \{Q^{(\frac{1}{2},0)}_{r}, Q^{(1,\frac{1}{2})}_{s,\bar{s}}\} = -\left(r-\frac{1}{2}s \right)M_{r+s,\bar{s}} , \\[0.5em]
            &\{\bar{Q}^{(0,\frac{1}{2})}_{\bar{r}}, \bar{Q}^{(0,\frac{1}{2})}_{\bar{s}}\} = \bar{L}_{\bar{r}+\bar{s}}, \quad \{\bar{Q}^{(0,\frac{1}{2})}_{\bar{r}}, \bar{Q}^{(\frac{1}{2},1)}_{s,\bar{s}}\} = -\left(\bar{r}-\frac{1}{2}\bar{s}\right)M_{s,\bar{r}+\bar{s}} , \\[0.5em]
            &\{Q^{(\frac{1}{2},0)}_{r}, \bar{Q}^{(\frac{1}{2},1)}_{s,\bar{s}} \} = R^{(1,1)}_{r+s,\bar{s}}, \quad \{\bar{Q}^{(0,\frac{1}{2})}_{\bar{r}}, Q^{(1,\frac{1}{2})}_{s,\bar{s}} \} = -R^{(1,1)}_{s,\bar{r} + \bar{s}}, \\[0.5em]
            &[R^{(1,1)}_{r,\bar{r}}, Q^{(\frac{1}{2},0)}_{s}] = \left(\frac{1}{2}r - s \right)\bar{Q}^{(\frac{1}{2},1)}_{r+s,\bar{s}}, \quad [R^{(1,1)}_{r,\bar{r}}, \bar{Q}^{(0,\frac{1}{2})}_{\bar{s}}] = -\left(\frac{1}{2}\bar{r} - \bar{s}\right)Q^{(0,\frac{1}{2})}_{s,\bar{r}+\bar{s}}.
        \end{aligned}
    \end{equation}
    \textbf{Type I-II} $\{L_{n}, \bar{L}_{\bar{n}}, M_{r,\bar{r}}, Q^{(\frac{1}{2},0)}_{r}, Q^{(0,\frac{1}{2})}_{r,\bar{r}}, \bar{Q}^{(0,\frac{1}{2})}_{\bar{r}}, \bar{Q}^{(\frac{1}{2},1)}_{r,\bar{r}}, R^{(0,1)}_{r,\bar{r}}\}$:
    \begin{equation}\label{eq:MagneticBMS4TypeI-II}
        \begin{aligned}
            &[M_{r,\bar{r}}, Q^{(\frac{1}{2},0)}_{s}] = \left(\frac{1}{2}r-\frac{1}{2}s\right) Q^{(0,\frac{1}{2})}_{r+s,\bar{r}}, \quad [M_{r,\bar{r}}, \bar{Q}^{(0,\frac{1}{2})}_{\bar{s}}] = \bar{Q}^{(\frac{1}{2},1)}_{r,\bar{r}+\bar{s}}, \\[0.5em]
            &\{Q^{(\frac{1}{2},0)}_{r}, Q^{(\frac{1}{2},0)}_{s}\} = L_{r+s}, \quad \{Q^{(\frac{1}{2},0)}_{r}, Q^{(0,\frac{1}{2})}_{s,\bar{s}}\} = M_{r+s,\bar{s}} , \\[0.5em]
            &\{\bar{Q}^{(0,\frac{1}{2})}_{\bar{r}}, \bar{Q}^{(0,\frac{1}{2})}_{\bar{s}}\} = \bar{L}_{\bar{r}+\bar{s}}, \quad \{\bar{Q}^{(0,\frac{1}{2})}_{\bar{r}}, \bar{Q}^{(\frac{1}{2},1)}_{s,\bar{s}}\} = -\left(\bar{r}-\frac{1}{2}\bar{s}\right)M_{s,\bar{r}+\bar{s}} , \\[0.5em]
            &\{Q^{(\frac{1}{2},0)}_{r}, \bar{Q}^{(\frac{1}{2},1)}_{s,\bar{s}} \} = \left(\frac{1}{2}r-\frac{1}{2}s\right)R^{(0,1)}_{r+s,\bar{s}}, \quad \{\bar{Q}^{(0,\frac{1}{2})}_{\bar{r}}, Q^{(0,\frac{1}{2})}_{s,\bar{s}} \} = R^{(0,1)}_{s,\bar{r} + \bar{s}}, \\[0.5em]
            &[R^{(0,1)}_{r,\bar{r}}, Q^{(\frac{1}{2},0)}_{s}] = -\bar{Q}^{(\frac{1}{2},1)}_{r+s,\bar{s}}, \quad [R^{(0,1)}_{r,\bar{r}}, \bar{Q}^{(0,\frac{1}{2})}_{\bar{s}}] = \left(\frac{1}{2}\bar{r} - \bar{s}\right)Q^{(0,\frac{1}{2})}_{s,\bar{r}+\bar{s}}, 
        \end{aligned}
    \end{equation}
    as well as Type II-I, which is the conjugate of Type I-II in \eqref{eq:MagneticBMS4TypeI-II}. Owing to this conjugation relation, we will omit the discussion of Type II-I, as it follows analogously from the Type I-II case.  \par 

    For each case, we introduce distinct $R$-symmetry charges. In particular, $R^{(0,1)}$, $R^{(1,0)}$, and $R^{(1,1)}$ carry nonvanishing spin indices. This implies that, unlike the conventional $R$-symmetries which commute with spacetime symmetries, these $R$-symmetries transform covariantly under Lorentz rotations, distinguishing them from the usual extended super Poincar\'e algebras. \par

    Finally, let us consider the case where two sets of $Q$s anticommute to $L$. To avoid introducing nonphysical $L_{n,\bar{n}}$ generators, the anti-holomorphic spins of these $Q$s must also vanish. Additionally, the spin-statistics requirement restricts us to two possibilities:
    \begin{equation}
        \begin{aligned}
            &\{Q^{(\frac{1}{2},0)}_{r}, \Tilde{Q}^{(\frac{1}{2},0)}_{s}\} = L_{r+s}, && \text{satisfying \eqref{eq:TensorProductOfVirReprsSolution2}}, \\
            &\{Q^{(\frac{1}{2},0)}_{r}, \Tilde{Q}^{(\frac{3}{2},0)}_{s}\} = \left(\frac{3}{2}r-\frac{1}{2}s\right)L_{r+s}, && \text{satisfying \eqref{eq:TensorProductOfVirReprsSolution2}}. \\
        \end{aligned}
    \end{equation}
    The first case reduces to the extended super BMS$_4$ algebras of Type I \eqref{eq:MagneticBMS4TypeI}, Type II \eqref{eq:MagneticBMS4TypeII}, or their combinations, via appropriate rotations of $Q^{(\frac{1}{2},0)}_{r}$ and $\Tilde{Q}^{(\frac{1}{2},0)}_{s}$. In the second case, an additional bosonic generator arising from $\{\Tilde{Q}^{(\frac{3}{2},0)},\Tilde{Q}^{(\frac{3}{2},0)}\}$ would be required to satisfy the Jacobi identity of $\{Q^{(\frac{1}{2},0)}_{r},\Tilde{Q}^{(\frac{3}{2},0)},\Tilde{Q}^{(\frac{3}{2},0)}\}$. However, this generator lacks a well-defined physical motivation. Hence, no unextended super BMS$_4$ algebra exists with two sets of $Q$s anticommuting to $L$. The same argument applies to the anti-holomorphic sector. \par

    To conclude, there are six unextended magnetic super BMS$_4$ algebras: chiral Type I and Type II, as well as nonchiral Type I-I, Type I-II, Type II-I, and Type II-II. All of these algebras admit finite-dimensional subalgebras. However, only in Type I-I and Type II-II we can define conjugation compatible with conjugation of Poincar\'e algebra, and the Gram matrices $\{Q,Q^\dagger\}$ for both cases are indefinite, which means the Hilbert spaces contain negative norm states. Therefore, none of the magnetic superalgebras are connected to the super Poincar\'e algebra. \par

\section{Free Field Realization}\label{sec:Realization}

    In this section, we present free field realizations of the various super BMS$_4$ algebras. \par

\subsection{Symmetry generators in terms of field modes} \label{subsec:ModesInRealization}

    It is standard in QFT, particularly in 2D free CFTs and in BMS$_3$ and BMS$_4$ free theories, to express symmetry generators in terms of field modes \cite{DiFrancesco:1997nk, Green:1987sp, Polchinski:1998rq, Bagchi:2022eav, Hao:2022xhq, Yu:2022bcp, Chen:2023esw, Bagchi:2024qsb}. The conventional approach to obtain the symmetry generators of a given free theory involves first computing the energy-momentum tensor in terms of the fields, then performing a Laurent or Fourier expansion of the fields according to their equations of motion (EOMs), and finally substituting this expansion to read off the generators. Here, since the symmetry algebra is already known, we adopt an inverse approach: we construct the symmetry generators using a specific ansatz for the field modes, and then build the corresponding field theory consistent with these mode expansions. \par

    The expansion of fields is analogous to the expansion \eqref{eq:CurrentExpansion} of conserved currents. For example, a time independent field is expanded as
    \begin{equation}
        \phi(t, z, \bar{z}) = \sum_{r,\bar{r}}\phi^{(h,\bar{h})}_{r,\bar{r}} z^{-r-h} \bar{z}^{-\bar{r}-\bar{h}}.
    \end{equation}
    Consequently, the modes of free fields $\phi^{(h,\bar{h})}_{r,\bar{r}}$ also fall into the charge representations discussed in Section \ref{subsec:ChargeRepresentations}. For simplicity, we show only the holomorphic part:
    \begin{equation}
        [L_{m},\phi^{(h)}_{r}] = ((h-1)m - r) \phi^{(h)}_{m+r},
    \end{equation}
    Similarly, we label the representations by $l=h-1$. Assuming that bilinears of modes $\phi^i$ form a set of charges $Q$ in the $(l_Q)$ representation,
    \begin{equation}
        Q^{(l_Q)}_{m} = \sum_{r} c^{l_Q}_{l_i,l_j}(r,n) \phi^i_{r} \phi^j_{m-r},
    \end{equation}
    the consistency condition from the commutator $[L_{m}, Q]$ imposes a constraint on the function $c$:
    \begin{equation}\label{eq:ModeCombinationEquation}
        ((l_i+1)m-r)c(r-m,t)+(l_j m-t+r)c(r,t)-(l_Q m-t)c(r,t+m)=0.
    \end{equation}
    Note that this equation for $c$ differs from the equation for $f$ in \eqref{eq:JacobiEquation}. For constructing free theories, it suffices to consider the following solutions:
    \begin{subequations}\label{eq:CombinedModesOfVirReprsSolution}
        \begin{align}
            &c_{l_i,l_j}^{l_Q}(r,n) = 1, && l_Q = l_i+l_j+1, \label{eq:CombinedModesOfVirReprsSolution1}\\
            &c_{l_i,l_j}^{l_Q}(r,n) = l_Q r - (l_i+1) n, && l_Q = l_i+l_j+2, \label{eq:CombinedModesOfVirReprsSolution2}
        \end{align}
    \end{subequations}
    Other solutions can be found in Appendix \ref{app:TensorProductOfVirReprs}. \par

    In the free theory, another crucial point is the canonical quantization. Suppose $\varphi$ and $\phi$ are two sets of canonically conjugate modes. Their commutation relation is expected to take the form
    \begin{equation}\label{eq:GeneralCanonicalCommutator}
        [\varphi_r,\phi_s\}= g^{l_I}_{l_\varphi, l_\phi}(r,s)\delta_{r+s} I
    \end{equation}
    where $\delta$ denotes the Kronecker delta and $I$ is the identity operator. In particular, the identity generator $I$ is regarded as a charge in the representation with (anti-)holomorphic labels $l_I = \bar{l}_I = 0$ and (anti-)holomorphic indices $m_I = \bar{m}_I = 0$. Indeed, if we formally extend $I$ to $I_{m}$, we have
    \begin{equation}
        [L_{m},I_{n=0}] = (l_I*m-1*n)I_{m} = (0*m-1*0)I_{m}=0,
    \end{equation}
    showing that additional generators $I_{m\neq 0}$ do not enter the theory as symmetry generators. Therefore, $\varphi$, $\phi$, and $I$ can all be viewed as charges in the theory, and equation \eqref{eq:GeneralCanonicalCommutator} implies that $g(r,s)\delta_{r+s}$ satisfies the constraint \eqref{eq:JacobiEquation}. The discrete solution relevant for this situation is
    \begin{equation}\label{eq:DistreteJacobiSolution}
        f^{l_I}_{l_\varphi, l_\phi}(r,s) = \delta_{r+s}, \quad l_\varphi+l_\phi=-1, \quad l_I=0.
    \end{equation}
    \par

    As an example, consider the BMS$_4$ algebra. Suppose the theory contains bosonic modes $\varphi^{(-\frac{3}{4},-\frac{3}{4})}$ and $\phi^{(-\frac{1}{4},-\frac{1}{4})}$, which are canonically conjugate to each other. Their commutation relation reads
    \begin{equation}\label{eq:BosonicCommutationCondition}
        [\varphi^{(-\frac{3}{4},-\frac{3}{4})}_{r,\bar{r}},\phi^{(-\frac{1}{4},-\frac{1}{4})}_{s,\bar{s}}] = \frac{i}{4\pi^2}\delta_{r+s} \delta_{\bar{r}+\bar{s}},
    \end{equation}
    where the identity operator is implicit. Using these modes, we can construct the BMS$_4$ algebra as
    \begin{equation}
        \begin{aligned}
            &L_{m} =-4\pi^2 i \sum_{s,\bar{s}}\left(s-\frac{1}{4}m\right)\varphi_{s,\bar{s}} \phi_{m-s,-\bar{s}}, \\
            &\bar{L}_{\bar{m}} = -4\pi^2 i \sum_{s,\bar{s}}\left(\bar{s}-\frac{1}{4}\bar{m}\right)\varphi_{s,\bar{s}} \phi_{-s,\bar{m}-\bar{s}}, \\
            &M_{r,\bar{r}} = 2\pi^2 i \sum_{s,\bar{s}}\phi_{s,\bar{s}} \phi_{r-s,\bar{r}-\bar{s}}. \\
        \end{aligned}
    \end{equation}
    In this construction, rule \eqref{eq:CombinedModesOfVirReprsSolution1} is satisfied by the anti-holomorphic index of $L$, the holomorphic index of $\bar{L}$, and both indices of $M$, while rule \eqref{eq:CombinedModesOfVirReprsSolution2} is satisfied by the holomorphic index of $L$ and the anti-holomorphic index of $\bar{L}$. In defining $L$ and $\bar{L}$, we consistently order the modes such that $\phi$ is always to the right of $\varphi$. This reproduces the BMS$_4$ algebra \eqref{eq:BMS4Algebra} without introducing central terms. The ordering convention will be implicitly applied throughout the rest of this section. Furthermore, using the commutation relation \eqref{eq:BosonicCommutationCondition}, one can verify that the transformation of the modes under the BMS$_4$ algebra:
    \begin{equation}
        \begin{aligned}\relax
            &[L_{m}, \varphi^{(-\frac{3}{4},-\frac{3}{4})}_{s,\bar{s}}] = \left(-\frac{3}{4}m -s\right)\varphi^{(-\frac{3}{4},-\frac{3}{4})}_{m+s,\bar{s}}, 
            &&[L_{m}, \phi^{(-\frac{1}{4},-\frac{1}{4})}_{s,\bar{s}}] = \left(-\frac{1}{4}m -s\right)\phi^{(-\frac{1}{4},-\frac{1}{4})}_{m+s,\bar{s}}, \\[0.5em]
            &[\bar{L}_{\bar{m}}, \varphi^{(-\frac{3}{4},-\frac{3}{4})}_{s,\bar{s}}] = \left(-\frac{3}{4}\bar{m} -\bar{s}\right)\varphi^{(-\frac{3}{4},-\frac{3}{4})}_{s,\bar{m}+\bar{s}}, 
            &&[\bar{L}_{\bar{m}}, \phi^{(-\frac{1}{4},-\frac{1}{4})}_{s,\bar{s}}] = \left(-\frac{1}{4}\bar{m} -\bar{s}\right)\phi^{(-\frac{1}{4},-\frac{1}{4})}_{s,\bar{m}+\bar{s}}, \\[0.5em]
            &[M_{r,\bar{r}}, \varphi^{(-\frac{3}{4},-\frac{3}{4})}_{s,\bar{s}}] = \phi^{(-\frac{1}{4},-\frac{1}{4})}_{r+s,\bar{r}+\bar{s}}, 
            &&[M_{r,\bar{r}}, \phi^{(-\frac{1}{4},-\frac{1}{4})}_{s,\bar{s}}] = 0. \\
        \end{aligned}
    \end{equation}
    These commutation relations satisfy \eqref{eq:TensorProductOfVirReprsSolution1} and \eqref{eq:TensorProductOfVirReprsSolution2} up to overall factors respectively. In this example, negative spin indices are allowed because the existence of finite-dimensional subsectors is not essential for field modes.  \par

\begin{figure}
        \centering
        \captionsetup[]{width= 0.8\linewidth} 
        \includegraphics[width=0.3\linewidth]{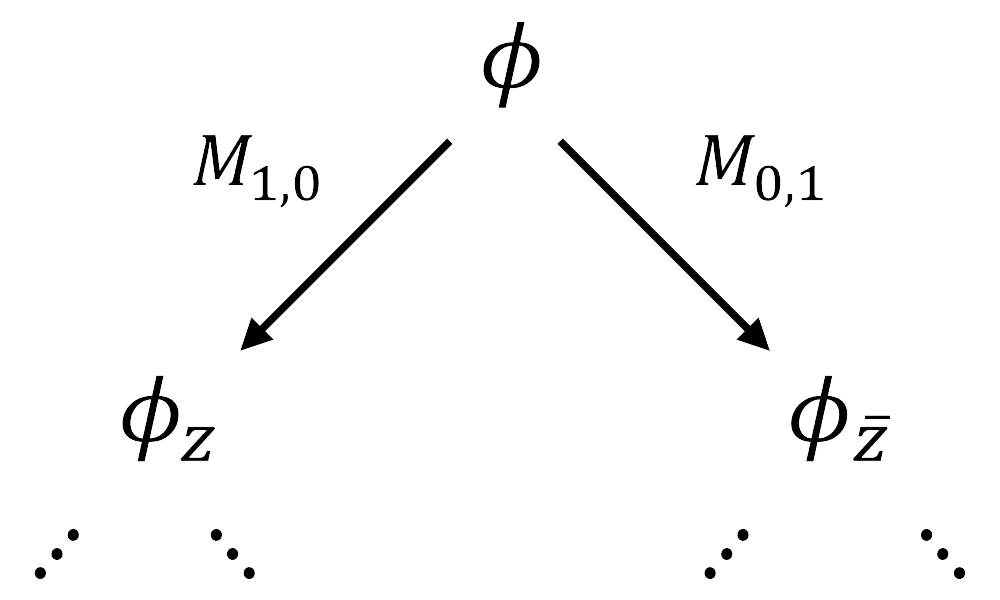}
        \caption{\centering Structure of Carrollian multiplet of $\phi$. \\ The ellipses denote higher-order multiplet fields.}
        \label{fig:CarrollianMultiplet}
    \end{figure}

    Consider the BMS$_4$ theory on $\mathbb{R}\times\mathbb{T}^2$, where the coordinates $(t,\sigma,\bar{\sigma})$ are reparametrized as $(t,z,\bar{z})$ with $z=e^{i\sigma}$ and $\bar{z}=e^{i\bar{\sigma}}$. The BMS$_4$ algebra is generated by the following vector fields:
    \begin{equation}
        \begin{aligned}
            & \xi^{(L)}_{n} = \left(-\frac{n+1}{2}tz^n, -z^{(n+1)},0\right),\\
            & \xi^{(\bar{L})}_{\bar{n}} = \left(-\frac{\bar{n}+1}{2}t\bar{z}^n, 0, -\bar{z}^{(\bar{n}+1)}\right),\\
            & \xi^{(M)}_{r,\bar{r}} = \left(-z^r\bar{z}^{\bar{r}}, 0,0\right).\\
        \end{aligned}
    \end{equation}
    For a BMS$_4$ field $\phi$, the action of the symmetry transformation $G_\xi$ generated by $\xi^\mu$ is given by \cite{Chen:2024voz}
    \begin{equation}\label{eq:BMSFieldTransformation}
        \begin{aligned}
            & [L_{n}, \phi] = -(\xi^{(L)}_{n})^\mu\partial_\mu\phi - 2h \partial_t(\xi^{(L)}_{n})^t \phi + \frac{1}{2}\partial_{z}(\xi^{(L)}_{n})^t \phi_{z} + \frac{1}{2}\partial_{\bar{z}}(\xi^{(L)}_{n})^t \phi_{\bar{z}} + \cdots, \\
            & [\bar{L}_{\bar{n}}, \phi] = -(\xi^{(\bar{L})}_{\bar{n}})^\mu\partial_\mu\phi - 2\bar{h} \partial_t(\xi^{(\bar{L})}_{\bar{n}})^t \phi + \frac{1}{2}\partial_{z}(\xi^{(\bar{L})}_{\bar{n}})^t \phi_{z} + \frac{1}{2}\partial_{\bar{z}}(\xi^{(\bar{L})}_{\bar{n}})^t \phi_{\bar{z}} + \cdots, \\
            & [M_{r,\bar{r}}, \phi] = -(\xi^{(M)}_{r,\bar{r}})^\mu\partial_\mu\phi + \frac{1}{2}\partial_{z}(\xi^{(M)}_{r,\bar{r}})^t \phi_{z} + \frac{1}{2}\partial_{\bar{z}}(\xi^{(M)}_{r,\bar{r}})^t \phi_{\bar{z}} + \cdots. \\
        \end{aligned}
    \end{equation}
    Here, the fields $\phi_{z}$ and $\phi_{\bar{z}}$ belong to the same Carrollian multiplet as the field $\phi$ \cite{Chen:2021xkw}, while the ellipses denote contributions from higher-order multiplet fields. The structure of this relation is illustrated in Figure \ref{fig:CarrollianMultiplet}. \par

    To meet equations \eqref{eq:BMSFieldTransformation}, the example discussed here can be identified as the free electric BMS$_4$ scalar theory:
    \begin{equation}\label{eq:BosonicFieldTheory}
        \mathcal{L} = \frac{1}{2}\partial_0\phi\partial_0\phi.
    \end{equation}
    The EOM of the field $\phi$ is $\partial_0^2\phi=0$, and $\phi$ can be expanded as
    \begin{equation}
        \phi(t,z,\bar{z}) = \sum_{r,\bar{r}} \varphi_{r,\bar{r}} z^{-r-\frac{1}{4}}\bar{z}^{-\bar{r}-\frac{1}{4}} + t \phi_{r,\bar{r}} z^{-r-\frac{3}{4}}\bar{z}^{-\bar{r}-\frac{3}{4}}.
    \end{equation}
    In this construction, the conformal dimensions of the field $\phi$ are $h=\bar{h}=\frac{1}{4}$, giving $\Delta=\frac{1}{2}$ and spin $j=0$. The commutation relation \eqref{eq:BosonicCommutationCondition} is consistent with the canonical commutation relation of the theory:
    \begin{equation}
        [\phi(t,z,\bar{z}), \partial_0\phi(t,z^\prime,\bar{z}^\prime )] = i\delta^2(z-z^\prime).
    \end{equation}
    In the following subsections, we use the same method to construct supersymmetric BMS$_4$ theories with the symmetries discussed in Section \ref{sec:Algebra}. For simplicity, we focus on scalar multiplets of different supersymmetries, with the scalar theory taken as \eqref{eq:BosonicFieldTheory}. \par

\subsection{Electric super \texorpdfstring{BMS$_4$}{BMS4} algebra} \label{subsec:ElectricRealization}
    In Section \ref{subsec:ElectricAlgebra}, we identified ten electric super BMS$_4$ algebras \eqref{eq:ElectricBMS4FromPoincareFull}--\eqref{eq:ElectricBMS4Type33}. However, in the free field mode construction, only the algebras \eqref{eq:ElectricBMS4FromPoincareFull}, \eqref{eq:ElectricBMS4Type12A}, and \eqref{eq:ElectricBMS4Type12B} admit supersymmetry between spin-$0$ scalars and spin-$\frac{1}{2}$ fermions. Realizing the other algebras requires introducing higher-spin fermion fields. Since, in the bulk 4D unextended super Poincar\'e theory, the superpartner of a scalar is a spin-$\frac{1}{2}$ spinor, we focus here only on the free field realizations of \eqref{eq:ElectricBMS4FromPoincareFull}, \eqref{eq:ElectricBMS4Type12A}, and \eqref{eq:ElectricBMS4Type12B}. \par

    \subsubsection*{Realization of electric algebra \eqref{eq:ElectricBMS4FromPoincareFull}} 
    
    To realize \eqref{eq:ElectricBMS4FromPoincareFull}, we introduce bosonic field modes
    \begin{equation}\label{eq:ElectricBosonModesCommutator}
        \varphi^{(-\frac{3}{4},-\frac{3}{4})}, \quad \phi^{(-\frac{1}{4},-\frac{1}{4})}, \quad \text{with }[\varphi^{(-\frac{3}{4},-\frac{3}{4})}_{r,\bar{r}},\phi^{(-\frac{1}{4},-\frac{1}{4})}_{s,\bar{s}}] = \frac{i}{4\pi^2}\delta_{r+s} \delta_{\bar{r}+\bar{s}},
    \end{equation}
    and fermionic field modes
    \begin{equation}\label{eq:ElectricFermionModesCommutator}
        \psi^{(-\frac{1}{4},-\frac{3}{4})}, \quad \bar{\psi}^{(-\frac{3}{4},-\frac{1}{4})}, \quad \text{with }\{\psi^{(-\frac{1}{4},-\frac{3}{4})}_{r,\bar{r}},\bar{\psi}^{(-\frac{3}{4},-\frac{1}{4})}_{s,\bar{s}}\} = \frac{i}{4\pi^2}\delta_{r+s} \delta_{\bar{r}+\bar{s}}.
    \end{equation}
    The super BMS$_4$ algebra \eqref{eq:ElectricBMS4FromPoincareFull} can then be realized as
    \begin{equation}\label{eq:ElectricBMS4ModesConstruction}
        \begin{aligned}
            &L_{m} = -4\pi^2 i \sum_{s,\bar{s}}\left(s-\frac{1}{4}m\right)\varphi_{s,\bar{s}} \phi_{m-s,-\bar{s}} 
                - \left(s-\frac{3}{4}m\right)\psi_{s,\bar{s}} \bar{\psi}_{m-s,-\bar{s}}, \\
            &\bar{L}_{\bar{m}} = -4\pi^2 i \sum_{s,\bar{s}}\left(\bar{s}-\frac{1}{4}\bar{m}\right)\varphi_{s,\bar{s}} \phi_{-s,\bar{m}-\bar{s}}
                - \left(\bar{s}-\frac{1}{4}\bar{m}\right)\psi_{s,\bar{s}} \bar{\psi}_{-s,\bar{m}-\bar{s}}, \\
            &M_{r,\bar{r}} = 2\pi^2 i \sum_{s,\bar{s}}\phi_{s,\bar{s}} \phi_{r-s,\bar{r}-\bar{s}}, \\
        \end{aligned}
    \end{equation}
    together with the supercharges
    \begin{equation}
        Q^{(\frac{1}{2},0)}_{r,\bar{r}} = 2\sqrt{2}\pi^2 \sum_{s,\bar{s}}\phi_{s,\bar{s}} \psi_{r-s,\bar{r}-\bar{s}}, \quad Q^{(0,\frac{1}{2})}_{r,\bar{r}} = 2\sqrt{2}\pi^2 \sum_{s,\bar{s}}\phi_{s,\bar{s}} \bar{\psi}_{r-s,\bar{r}-\bar{s}}.
    \end{equation}
    It is straightforward for the interested reader to check that these constructions satisfy the commutation relations \eqref{eq:ElectricBMS4FromPoincareFull} using \eqref{eq:ElectricBosonModesCommutator} and \eqref{eq:ElectricFermionModesCommutator}. The transformations of the field modes under the super BMS$_4$ algebra are summarized in Table \ref{tab:ModesTransformationElectricFull}. For simplicity, the actions of $L$ and $\bar{L}$ are omitted in the table, as the (anti-)holomorphic indices already encode their transformations. \par

    \begin{table}[htpb]
        \renewcommand\arraystretch{2}
        \centering
        \caption{\centering Field modes transformation under super BMS$_4$ algebra \eqref{eq:ElectricBMS4FromPoincareFull}.}
        \begin{tabular}{c|c|c|c|c}
            \hline
             & $\varphi_{s,\bar{s}}$ & $\phi_{s,\bar{s}}$ & $\psi_{s,\bar{s}}$ & $\bar{\psi}_{s,\bar{s}}$\\\hline\hline
            $M_{r,\bar{r}}$
                & $\phi_{r+s,\bar{r}+\bar{s}}$
                & $0$
                & $0$
                & $0$ \\\hline
            $Q^{(\frac{1}{2},0)}_{r,\bar{r}}$
                & $-\frac{i}{\sqrt{2}}\psi_{r+s,\bar{r}+\bar{s}}$
                & $0$
                & $0$
                & $\frac{i}{\sqrt{2}}\phi_{r+s,\bar{r}+\bar{s}}$ \\\hline
            $Q^{(0,\frac{1}{2})}_{r,\bar{r}}$
                & $-\frac{i}{\sqrt{2}}\bar{\psi}_{r+s,\bar{r}+\bar{s}}$
                & $0$
                & $\frac{i}{\sqrt{2}}\phi_{r+s,\bar{r}+\bar{s}}$ 
                & $0$ \\\hline
        \end{tabular}
        \label{tab:ModesTransformationElectricFull}
    \end{table}

    The corresponding field theory can be written as
    \begin{equation}\label{eq:ElectricTheory}
        \mathcal{L} = \frac{1}{2}\partial_0\phi\partial_0\phi - \psi\partial_0\bar{\psi} - \bar{\psi} \partial_0\psi,
    \end{equation}
    where the bosonic part is exactly \eqref{eq:BosonicFieldTheory}, and $\psi$ and $\bar{\psi}$ are fermionic fields with spin $\pm \frac{1}{2}$ respectively. In terms of Carrollian theories, $\phi$ is also referred as electric scalar, and $\psi$ are referred as electric spinor. The conformal dimensions are $\Delta_\phi=\frac{1}{2}$ and $\Delta_\psi=\Delta_{\bar{\psi}}=1$. The fermionic fields satisfy the equations of motion
    \begin{equation}
        \partial_0\psi = 0, \quad \partial_0\bar{\psi}=0,
    \end{equation}
    with mode expansions
    \begin{equation}
        \begin{aligned}
            &\psi(t,z,\bar{z}) = \sum_{r,\bar{r}} \psi_{r,\bar{r}}  z^{-r-\frac{3}{4}}\bar{z}^{-\bar{r}-\frac{1}{4}}, \quad \bar{\psi}(t,z,\bar{z}) = \sum_{r,\bar{r}} \bar{\psi}_{r,\bar{r}}  z^{-r-\frac{1}{4}}\bar{z}^{-\bar{r}-\frac{3}{4}}. \\
        \end{aligned}
    \end{equation}
    The canonical anticommutation relation
    \begin{equation}
        \{\phi(t,z,\bar{z}),\bar{\psi}(t,z^\prime,\bar{z}^\prime )\} = i\delta^2(z-z^\prime)
    \end{equation}
    is consistent with the mode commutator \eqref{eq:ElectricFermionModesCommutator}. In this theory, the bosonic field $\phi$ has one degree of freedom, while $\psi$ and $\bar{\psi}$ together account for one fermionic degree of freedom, as they are canonical conjugates. Hence, the numbers of bosonic and fermionic degrees of freedom are equal. Similarly, in the other constructed theories, the total bosonic and fermionic degrees of freedom match, even if the number of individual fields differs. \par

    \subsubsection*{Realization of electric algebra \eqref{eq:ElectricBMS4Type12A}}
    
    The same bosonic modes \eqref{eq:ElectricBosonModesCommutator} and fermionic modes \eqref{eq:ElectricFermionModesCommutator} can be used to realize \eqref{eq:ElectricBMS4Type12A}. The construction of the BMS$_4$ algebra remains the same as in \eqref{eq:ElectricBMS4ModesConstruction}, but the supercharges are now defined differently:
    \begin{equation}
        Q^{(\frac{1}{2},1)}_{r,\bar{r}} = 2\sqrt{2}\pi^2 \sum_{s,\bar{s}}\left(\bar{s} - \frac{3}{4}\bar{r}\right)\phi_{s,\bar{s}} \psi_{r-s,\bar{r}-\bar{s}}, \quad Q^{(0,\frac{1}{2})}_{r,\bar{r}} = 2\sqrt{2}\pi^2 \sum_{s,\bar{s}}\phi_{s,\bar{s}} \bar{\psi}_{r-s,\bar{r}-\bar{s}}.
    \end{equation}
    The transformation of the field modes under the bosonic symmetries is the same as in Table \ref{tab:ModesTransformationElectricFull}, while their transformation under the supercharges is summarized in Table \ref{tab:ModesTransformationElectricType12A}. \par

    \begin{table}[htpb]
        \renewcommand\arraystretch{2}
        \centering
        \caption{\centering Field modes transformation under supercharges of \eqref{eq:ElectricBMS4Type12A}.}
        \begin{tabular}{c|c|c|c|c}
            \hline
             & $\varphi_{s,\bar{s}}$ & $\phi_{s,\bar{s}}$ & $\psi_{s,\bar{s}}$ & $\bar{\psi}_{s,\bar{s}}$\\\hline\hline
            $Q^{(\frac{1}{2},1)}_{r,\bar{r}}$
                & $-\frac{i}{\sqrt{2}}\left(-\frac{3}{4}\bar{r} - \bar{s}\right)\psi_{r+s,\bar{r}+\bar{s}}$
                & $0$
                & $0$
                & $-\frac{i}{\sqrt{2}}\left(-\frac{1}{4}\bar{r} - \bar{s}\right)\phi_{r+s,\bar{r}+\bar{s}}$ \\\hline
            $Q^{(0,\frac{1}{2})}_{r,\bar{r}}$
                & $-\frac{i}{\sqrt{2}}\bar{\psi}_{r+s,\bar{r}+\bar{s}}$
                & $0$
                & $\frac{i}{\sqrt{2}}\phi_{r+s,\bar{r}+\bar{s}}$ 
                & $0$ \\\hline
        \end{tabular}
        \label{tab:ModesTransformationElectricType12A}
    \end{table}

\subsubsection*{Realization of electric algebra \eqref{eq:ElectricBMS4Type12B}}
    
    The same bosonic and fermionic modes \eqref{eq:ElectricBosonModesCommutator} and \eqref{eq:ElectricFermionModesCommutator} can also be employed to realize \eqref{eq:ElectricBMS4Type12B}, with a distinct construction for the supercharges:
    \begin{equation}
        Q^{(\frac{1}{2},0)}_{r,\bar{r}} = 2\sqrt{2}\pi^2 \sum_{s,\bar{s}} \phi_{s,\bar{s}} \psi_{r-s,\bar{r}-\bar{s}}, \quad Q^{(0,\frac{3}{2})}_{r,\bar{r}} = 2\sqrt{2}\pi^2 \sum_{s,\bar{s}} \left(\frac{3}{2}\bar{s} - \frac{3}{4}\bar{r}\right) \phi_{s,\bar{s}} \bar{\psi}_{r-s,\bar{r}-\bar{s}}.
    \end{equation}
    The transformations of the field modes under these supercharges are summarized in Table \ref{tab:ModesTransformationElectricType12B}. \par

    \begin{table}[htpb]
        \renewcommand\arraystretch{2}
        \centering
        \caption{\centering Field modes transformation under supercharges of \eqref{eq:ElectricBMS4Type12B}.}
        \begin{tabular}{c|c|c|c|c}
            \hline
             & $\varphi_{s,\bar{s}}$ & $\phi_{s,\bar{s}}$ & $\psi_{s,\bar{s}}$ & $\bar{\psi}_{s,\bar{s}}$\\\hline\hline
            $Q^{(\frac{1}{2},0)}_{r,\bar{r}}$
                & $-\frac{i}{\sqrt{2}}\psi_{r+s,\bar{r}+\bar{s}}$
                & $0$
                & $0$
                & $\frac{i}{\sqrt{2}}\phi_{r+s,\bar{r}+\bar{s}}$ \\\hline
            $Q^{(0,\frac{3}{2})}_{r,\bar{r}}$
                & $-\frac{i}{\sqrt{2}}\left(-\frac{3}{4}\bar{r} - \frac{3}{2}\bar{s}\right)\bar{\psi}_{r+s,\bar{r}+\bar{s}}$
                & $0$
                & $-\frac{i}{\sqrt{2}}\left(-\frac{3}{4}\bar{r} - \frac{3}{2}\bar{s}\right)\phi_{r+s,\bar{r}+\bar{s}}$ 
                & $0$ \\\hline
        \end{tabular}
        \label{tab:ModesTransformationElectricType12B}
    \end{table}

    Interestingly, the theory \eqref{eq:ElectricTheory} is sufficiently simple that it accommodates a large number of supercharges. Explicitly counting the conjugation, the field modes in \eqref{eq:ElectricTheory} can generate the following supercharges:
    \begin{equation}
        Q^{(\frac{1}{2},0)}, \quad Q^{(0,\frac{1}{2})}, \quad Q^{(\frac{1}{2},1)}, \quad Q^{(1,\frac{1}{2})}, \quad Q^{(\frac{3}{2},0)}, \quad Q^{(0,\frac{3}{2})},
    \end{equation}
    where all except $Q^{(1,\frac{1}{2})}$ and $Q^{(\frac{3}{2},1)}$ were constructed previously. However, considering all supercharges simultaneously quickly becomes cumbersome. For instance, one finds that
    \begin{equation}
        \{Q^{(\frac{1}{2},1)}_{r,\bar{r}}, Q^{(0,\frac{3}{2})}_{s,\bar{s}}\} = -i(4\pi^2) \sum_{t,\bar{t}} -\frac{3}{16} (3\bar{r}^2 + 2\bar{r}(\bar{s}-4\bar{t}) + 2(\bar{s}-2\bar{t})^2) \phi_{t,\bar{t}}\phi_{r+s-t,\bar{r}+\bar{s}-\bar{t}}.
    \end{equation}
    This shows that the anticommutator introduces bosonic generators beyond $M$. Similar structures appear in other anticommutators as well. While these additional generators are indeed symmetries of the specific theory \eqref{eq:ElectricTheory}, they do not correspond to symmetries of generic BMS$_4$ theories. \par

\subsection{Magnetic super \texorpdfstring{BMS$_4$}{BMS4} algebra} \label{subsec:MagneticRealization}

    We now turn to the realization of the magnetic super BMS$_4$ algebras. The Type I algebra \eqref{eq:MagneticBMS4TypeI} can be constructed using the same modes as in \eqref{eq:ElectricBosonModesCommutator} and \eqref{eq:ElectricFermionModesCommutator}. However, the altered transformation rules under the symmetry imply that these modes correspond to a different physical theory. In contrast, realizing the Type II algebra \eqref{eq:MagneticBMS4TypeII} requires introducing a distinct set of modes. \par

    \subsubsection*{Realization of magnetic Type I algebra \eqref{eq:MagneticBMS4TypeI}}
    
    For convenience, we repeat the modes \eqref{eq:ElectricBosonModesCommutator} and \eqref{eq:ElectricFermionModesCommutator} here, with a renaming of one set of fermionic modes:
    \begin{equation}
        \begin{aligned}
            &\varphi^{(-\frac{3}{4},-\frac{3}{4})}, \quad \phi^{(-\frac{1}{4},-\frac{1}{4})}, \quad [\varphi^{(-\frac{3}{4},-\frac{3}{4})}_{r,\bar{r}},\phi^{(-\frac{1}{4},-\frac{1}{4})}_{s,\bar{s}}] = \frac{i}{4\pi^2}\delta_{r+s} \delta_{\bar{r}+\bar{s}}, \\
            &\psi^{(-\frac{1}{4},-\frac{3}{4})}, \quad \chi^{(-\frac{3}{4},-\frac{1}{4})}, \quad \{\psi^{(-\frac{1}{4},-\frac{3}{4})}_{r,\bar{r}},\chi^{(-\frac{3}{4},-\frac{1}{4})}_{s,\bar{s}}\} = \frac{i}{4\pi^2}\delta_{r+s} \delta_{\bar{r}+\bar{s}}.\\
        \end{aligned}
    \end{equation}
    The Type I algebra \eqref{eq:MagneticBMS4TypeI} is realized as
    \begin{equation}\label{eq:MagneticBMS4TypeIModesConstruction}
        \begin{aligned}
            & L_{m} = -4\pi^2i \sum_{s,\bar{s}}\left(s-\frac{1}{4}m\right)\varphi_{s,\bar{s}} \phi_{m-s,-\bar{s}} 
                - \left(s-\frac{3}{4}m\right)\psi_{s,\bar{s}} \chi_{m-s,-\bar{s}}, \\
            & \bar{L}_{\bar{m}} = -4\pi^2i \sum_{s,\bar{s}}\left(\bar{s}-\frac{1}{4}\bar{m}\right)\varphi_{s,\bar{s}} \phi_{-s,\bar{m}-\bar{s}}
                - \left(\bar{s}-\frac{1}{4}\bar{m}\right)\psi_{s,\bar{s}} \chi_{-s,\bar{m}-\bar{s}}, \\
            & M_{r,\bar{r}} = 2\pi^2i \sum_{s,\bar{s}}\phi_{s,\bar{s}} \phi_{r-s,\bar{r}-\bar{s}}
                +2\left(\frac{1}{2}s - \frac{1}{4}r\right)\chi_{s,\bar{s}}\chi_{r-s,\bar{r}-\bar{s}}, \\
        \end{aligned}
    \end{equation}
    and
    \begin{equation}
        \begin{aligned}
            & Q^{(\frac{1}{2},0)}_{r} = -2\sqrt{2}\pi^2 i \sum_{s,\bar{s}} 2 \left(\frac{1}{2}s - \frac{1}{4}r\right) \varphi_{s,\bar{s}} \chi_{r-s,-\bar{s}}
                +\phi_{s,\bar{s}}\psi_{r-s,-\bar{s}}, \\
            & Q^{(0,\frac{1}{2})}_{r,\bar{r}} = 2\sqrt{2}\pi^2 i \sum_{s,\bar{s}} \phi_{s,\bar{s}} \chi_{r-s,\bar{r}-\bar{s}}.\\
        \end{aligned}
    \end{equation}
    Notice that the construction of $M$ in \eqref{eq:MagneticBMS4TypeIModesConstruction} differs from that in \eqref{eq:ElectricBMS4ModesConstruction}. Under the algebra \eqref{eq:MagneticBMS4TypeI}, the modes transform as summarized in Table \ref{tab:ModesTransformationMagneticTypeI}. \par

    \begin{table}[htpb]
        \renewcommand\arraystretch{2}
        \centering
        \caption{\centering Field modes transformation under Type I BMS$_4$ algebra \eqref{eq:MagneticBMS4TypeI}.}
        \begin{tabular}{c|c|c|c|c}
            \hline
             & $\varphi_{s,\bar{s}}$ & $\phi_{s,\bar{s}}$ & $\psi_{s,\bar{s}}$ & $\chi_{s,\bar{s}}$\\\hline\hline
            $M_{r,\bar{r}}$
                & $\phi_{r+s,\bar{r}+\bar{s}}$
                & $0$
                & $2\left(\frac{1}{4}r+\frac{1}{2}s\right)\chi_{r+s,\bar{r}+\bar{s}}$
                & $0$ \\\hline
            $Q^{(\frac{1}{2},0)}_{r}$
                & $-\frac{1}{\sqrt{2}}\psi_{r+s,\bar{s}}$
                & $\sqrt{2}\left(-\frac{1}{4}r - \frac{1}{2}s\right)\chi_{r+s,\bar{s}}$
                & $-\sqrt{2}\left(-\frac{1}{4}r - \frac{1}{2}s\right)\varphi_{r+s,\bar{s}}$
                & $\frac{1}{\sqrt{2}}\phi_{r+s,\bar{s}}$ \\\hline
            $Q^{(0,\frac{1}{2})}_{r,\bar{r}}$
                & $\frac{1}{\sqrt{2}}\chi_{r+s,\bar{r}+\bar{s}}$
                & $0$
                & $-\frac{1}{\sqrt{2}}\phi_{r+s,\bar{r}+\bar{s}}$ 
                & $0$ \\\hline
        \end{tabular}
        \label{tab:ModesTransformationMagneticTypeI}
    \end{table}

    The corresponding field theory for this construction is
    \begin{equation}\label{eq:MagneticTypeITheory}
        \mathcal{L} = \frac{1}{2}\partial_0\phi\partial_0\phi - \psi_{+}\partial_0\psi_{-} - \psi_{-}\partial_0\psi_{+} + \psi_{-}\partial_{z}\psi_{-}.
    \end{equation}
    In this theory, there is a single scalar field $\phi$ with conformal dimension $\Delta_\phi = \frac{1}{2}$, and two fermionic fields $\psi_{\pm}$ with spin $\pm \frac{1}{2}$ and conformal dimensions $\Delta_{\psi_{\pm}} = 1$. The equations of motion derived from the Lagrangian read
    \begin{equation}
        \partial_0^2\phi = 0, \quad \partial_0\psi_{+} = -\partial_{z}\psi_{-}, \quad \partial_0\psi_{-} = 0,
    \end{equation}
    and the corresponding mode expansions consistent with these EOMs are
    \begin{align}
        &\begin{aligned}
            & \phi(t,z,\bar{z}) = \sum_{r,\bar{r}} \varphi_{r,\bar{r}} z^{-r-\frac{1}{4}}\bar{z}^{-\bar{r}-\frac{1}{4}} 
                + t \phi_{r,\bar{r}} z^{-r-\frac{3}{4}}\bar{z}^{-\bar{r}-\frac{3}{4}}, \\
        \end{aligned}\\
        &\begin{aligned}
            & \psi_{+}(t,z,\bar{z}) = \sum_{r,\bar{r}} \psi_{r,\bar{r}} z^{-r-\frac{3}{4}}\bar{z}^{-\bar{r}-\frac{1}{4}} 
                + \left(r+\frac{1}{4}\right)\frac{t}{z} ~ \chi_{r,\bar{r}} z^{-r-\frac{1}{4}}\bar{z}^{-\bar{r}-\frac{3}{4}}, \\
            & \psi_{-}(t,z,\bar{z}) = \sum_{r,\bar{r}} \chi_{r,\bar{r}} z^{-r-\frac{1}{4}}\bar{z}^{-\bar{r}-\frac{3}{4}} . \\
        \end{aligned}
    \end{align}
    \par

    \subsubsection*{Realization of magnetic Type II algebra \eqref{eq:MagneticBMS4TypeII}}

    The construction of the magnetic Type II algebra employs slightly different modes. The bosonic modes remain the same as in the Type I case, while the fermionic modes carry different (anti-)holomorphic spins:
    \begin{equation}
        \begin{aligned}
            & \varphi^{(-\frac{3}{4},-\frac{3}{4})}, \quad \phi^{(-\frac{1}{4},-\frac{1}{4})}, \quad [\varphi^{(-\frac{3}{4},-\frac{3}{4})}_{r,\bar{r}},\phi^{(-\frac{1}{4},-\frac{1}{4})}_{s,\bar{s}}] = \frac{i}{4\pi^2}\delta_{r+s} \delta_{\bar{r}+\bar{s}}, \\
            & \psi^{(-\frac{5}{4},-\frac{3}{4})}, \quad \chi^{(\frac{1}{4},-\frac{1}{4})}, \quad \{\psi^{(-\frac{5}{4},-\frac{3}{4})}_{r,\bar{r}},\chi^{(\frac{1}{4},-\frac{1}{4})}_{s,\bar{s}}\} = \frac{i}{4\pi^2}\delta_{r+s} \delta_{\bar{r}+\bar{s}}.\\
        \end{aligned}
    \end{equation}
    The Type II algebra \eqref{eq:MagneticBMS4TypeII} is realized through the following mode constructions:
    \begin{equation}\label{eq:MagneticBMS4TypeIIModesConstruction}
        \begin{aligned}
            & L_{m} = -4\pi^2i \sum_{s,\bar{s}}\left(s-\frac{1}{4}m\right)\varphi_{s,\bar{s}} \phi_{m-s,-\bar{s}} 
                - \left(s+\frac{1}{4}m\right)\psi_{s,\bar{s}} \chi_{m-s,-\bar{s}}, \\
            & \bar{L}_{\bar{m}} = -4\pi^2i \sum_{s,\bar{s}}\left(\bar{s}-\frac{1}{4}\bar{m}\right)\varphi_{s,\bar{s}} \phi_{-s,\bar{m}-\bar{s}}
                - \left(\bar{s}-\frac{1}{4}\bar{m}\right)\psi_{s,\bar{s}} \chi_{-s,\bar{m}-\bar{s}}, \\
            & M_{r,\bar{r}} = 2\pi^2i \sum_{s,\bar{s}}\phi_{s,\bar{s}} \phi_{r-s,\bar{r}-\bar{s}}, \\
        \end{aligned}
    \end{equation}
    \begin{equation}
        \begin{aligned}
            & Q^{(\frac{1}{2},0)}_{r} = -4\pi^2 i \sum_{s,\bar{s}} \varphi_{s,\bar{s}} \chi_{r-s,-\bar{s}}
                -\left(\frac{1}{2}s-\frac{3}{4}r\right)\phi_{s,\bar{s}} \psi_{r-s,-\bar{s}}, \\
            & Q^{(1,\frac{1}{2})}_{r,\bar{r}} = -4\pi^2 i \sum_{s,\bar{s}}\phi_{s,\bar{s}} \chi_{r-s,\bar{r}-\bar{s}}.\\
        \end{aligned}
    \end{equation}
    The transformations of these modes under the Type II algebra are summarized in Table \ref{tab:ModesTransformationMagneticTypeII}. \par

    \begin{table}[htpb]
        \renewcommand\arraystretch{2}
        \centering
        \caption{\centering Field modes transformation under Type II BMS$_4$ algebra \eqref{eq:MagneticBMS4TypeII}.}
        \begin{tabular}{c|c|c|c|c}
            \hline
             & $\varphi_{s,\bar{s}}$ & $\phi_{s,\bar{s}}$ & $\psi_{s,\bar{s}}$ & $\chi_{s,\bar{s}}$\\\hline\hline
            $M_{r,\bar{r}}$
                & $\phi_{r+s,\bar{r}+\bar{s}}$
                & $0$
                & $0$
                & $0$ \\\hline
            $Q^{(\frac{1}{2},0)}_{r}$
                & $\left(-\frac{3}{4}r - \frac{1}{2}s\right)\psi_{r+s,\bar{s}}$
                & $\chi_{r+s,\bar{s}}$
                & $\varphi_{r+s,\bar{s}}$
                & $\left(\frac{1}{4}r - \frac{1}{2}s\right)\phi_{r+s,\bar{s}}$ \\\hline
            $Q^{(0,\frac{1}{2})}_{r,\bar{r}}$
                & $-\chi_{r+s,\bar{r}+\bar{s}}$
                & $0$
                & $\phi_{r+s,\bar{r}+\bar{s}}$ 
                & $0$ \\\hline
        \end{tabular}
        \label{tab:ModesTransformationMagneticTypeII}
    \end{table}

    The field theory corresponding to this construction is given by
    \begin{equation}\label{eq:MagneticTypeIITheory}
        \mathcal{L} = \frac{1}{2}\partial_0\phi\partial_0\phi - \psi_{+}\partial_0\psi_{-} - \psi_{-}\partial_0\psi_{+}.
    \end{equation}
    This theory consists of a single scalar field $\phi$ and two fermionic fields $\psi_{\pm}$ with spins $\pm \frac{1}{2}$. Although the Lagrangian \eqref{eq:MagneticTypeIITheory} appears similar to the electric super BMS$_4$ field theory \eqref{eq:ElectricTheory}, it describes a different theory because the conformal dimensions of the fermionic fields differ: $\Delta_{\psi_{+}} = 2$ and $\Delta_{\psi_{-}} = 0$. The equations of motion read
    \begin{equation}
        \partial_0^2\phi = 0, \quad \partial_0\psi_{+} = \partial_0\psi_{-} = 0,
    \end{equation}
    and the consistent mode expansions for these fields are
    \begin{align}
        &\begin{aligned}
            & \phi(t,z,\bar{z}) = \sum_{r,\bar{r}} \varphi_{r,\bar{r}} z^{-r-\frac{1}{4}}\bar{z}^{-\bar{r}-\frac{1}{4}} 
                + t \phi_{r,\bar{r}} z^{-r-\frac{3}{4}}\bar{z}^{-\bar{r}-\frac{3}{4}}, \\
        \end{aligned}\\
        &\begin{aligned}
            & \psi_{+}(t,z,\bar{z}) = \sum_{r,\bar{r}} \chi_{r,\bar{r}} z^{-r-\frac{5}{4}}\bar{z}^{-\bar{r}-\frac{3}{4}} , \quad \psi_{-}(t,z,\bar{z}) = \sum_{r,\bar{r}} \psi_{r,\bar{r}} z^{-r+\frac{1}{4}}\bar{z}^{-\bar{r}-\frac{1}{4}} . \\
        \end{aligned}
    \end{align}
    \par

\subsubsection*{Realization of magnetic Type I-I algebra \eqref{eq:MagneticBMS4TypeI-I}}

    We now consider a theory that possesses both left- and right-handed super BMS$_4$ symmetries. The mode content of this theory is specified as follows:
    \begin{equation}
        \begin{aligned}
            & \varphi^{(-\frac{3}{4},-\frac{3}{4})}, \quad \phi^{(-\frac{1}{4},-\frac{1}{4})}, \quad [\varphi^{(-\frac{3}{4},-\frac{3}{4})}_{r,\bar{r}},\phi^{(-\frac{1}{4},-\frac{1}{4})}_{s,\bar{s}}] = \frac{i}{4\pi^2}\delta_{r+s} \delta_{\bar{r}+\bar{s}}, \\
            & \psi^{(-\frac{1}{4},-\frac{3}{4})}, \quad \chi^{(-\frac{3}{4},-\frac{1}{4})}, \quad \{\psi^{(-\frac{1}{4},-\frac{3}{4})}_{r,\bar{r}},\chi^{(-\frac{3}{4},-\frac{1}{4})}_{s,\bar{s}}\} = \frac{i}{4\pi^2}\delta_{r+s} \delta_{\bar{r}+\bar{s}},\\
            & \bar{\psi}^{(-\frac{3}{4},-\frac{1}{4})}, \quad \bar{\chi}^{(-\frac{1}{4},-\frac{3}{4})}, \quad \{\bar{\psi}^{(-\frac{3}{4},-\frac{1}{4})}_{r,\bar{r}},\bar{\chi}^{(-\frac{1}{4},-\frac{3}{4})}_{s,\bar{s}}\} = \frac{i}{4\pi^2}\delta_{r+s} \delta_{\bar{r}+\bar{s}},\\
            & \bar{\varphi}^{(-\frac{1}{4},-\frac{1}{4})}, \quad \bar{\phi}^{(-\frac{3}{4},-\frac{3}{4})}, \quad [\bar{\varphi}^{(-\frac{1}{4},-\frac{1}{4})}_{r,\bar{r}}, \bar{\phi}^{(-\frac{3}{4},-\frac{3}{4})}_{s,\bar{s}}] = \frac{i}{4\pi^2}\delta_{r+s} \delta_{\bar{r}+\bar{s}}. \\
        \end{aligned}
    \end{equation}
    The Type I-I algebra \eqref{eq:MagneticBMS4TypeI-I} features an additional bosonic generator $R$. The BMS$_4$ part of the algebra can be realized as
    \begin{equation}\label{eq:MagneticBMS4TypeI-IModesConstruction}
        \begin{aligned}
            & L_{m} = -4\pi^2i \sum_{s,\bar{s}} \left(s-\frac{1}{4}m\right) \varphi_{s,\bar{s}} \phi_{m-s,-\bar{s}} 
                - \left(s-\frac{3}{4}m\right) \psi_{s,\bar{s}} \chi_{m-s,-\bar{s}} \\[0.5em]
                &\qquad\qquad\qquad\qquad 
                    - \left(s-\frac{1}{4}m\right) \bar{\psi}_{s,\bar{s}} \bar{\chi}_{m-s,-\bar{s}} 
                    + \left(s-\frac{3}{4}m\right) \bar{\varphi}_{s,\bar{s}} \bar{\phi}_{m-s,-\bar{s}}, \\[1em]
            & \bar{L}_{\bar{m}} = -4\pi^2i \sum_{s,\bar{s}} \left(\bar{s}-\frac{1}{4}\bar{m}\right) \varphi_{s,\bar{s}} \phi_{-s,\bar{m}-\bar{s}}
                - \left(\bar{s}-\frac{1}{4}\bar{m}\right) \psi_{s,\bar{s}} \chi_{-s,\bar{m}-\bar{s}} \\[0.5em]
                &\qquad\qquad\qquad\qquad 
                    - \left(\bar{s}-\frac{3}{4}\bar{m}\right) \bar{\psi}_{s,\bar{s}} \bar{\chi}_{-s,\bar{m}-\bar{s}} 
                    + \left(\bar{s}-\frac{3}{4}\bar{m}\right) \bar{\varphi}_{s,\bar{s}} \bar{\phi}_{-s,\bar{m}-\bar{s}}, \\[1em]
            & M_{r,\bar{r}} = 2\pi^2i \sum_{s,\bar{s}} \phi_{s,\bar{s}} \phi_{r-s,\bar{r}-\bar{s}}
                + 2 \left(\frac{1}{2}\bar{s}-\frac{1}{4}\bar{m}\right) \chi_{s,\bar{s}} \chi_{r-s,\bar{m}-\bar{s}} \\[0.5em]
                &\qquad\qquad 
                    + 2 \left(\frac{1}{2}\bar{s} - \frac{1}{4}\bar{r}\right) \bar{\chi}_{s,\bar{s}} \bar{\chi}_{m-s,\bar{r}-\bar{s}} 
                    - 2 \left(\frac{1}{2}\bar{s}-\frac{1}{4}\bar{m}\right) \left(\frac{1}{2}\bar{s} - \frac{1}{4}\bar{r}\right) \bar{\phi}_{s,\bar{s}} \bar{\phi}_{m-s,\bar{r}-\bar{s}}, \\
        \end{aligned}
    \end{equation}
    the fermionic generators are constructed as
    \begin{equation}
        \begin{aligned}
            & Q^{(\frac{1}{2},0)}_{r} = -2\pi^2 i \sum_{s,\bar{s}} 2\sqrt{2} \left(\frac{1}{2}s - \frac{1}{4}r\right) \varphi_{s,\bar{s}} \chi_{r-s,-\bar{s}}
                + \sqrt{2} ~ \phi_{s,\bar{s}} \psi_{r-s,-\bar{s}} \\[0.5em]
                &\qquad\qquad\qquad\qquad 
                    + 2 \left(\frac{1}{2}s - \frac{1}{4}r\right) \bar{\psi}_{s,\bar{s}} \bar{\phi}_{r-s,-\bar{s}}
                    + 2 ~ \bar{\chi}_{s,\bar{s}} \bar{\varphi}_{r-s,-\bar{s}}\\[1em]
            & Q^{(0,\frac{1}{2})}_{r,\bar{r}} = 2\pi^2 i \sum_{s,\bar{s}} \sqrt{2} ~ \phi_{s,\bar{s}} \chi_{r-s,\bar{r}-\bar{s}}
                + 2 \left(\frac{1}{2}\bar{s} - \frac{1}{4}\bar{r}\right) \bar{\chi}_{s,\bar{s}} \bar{\phi}_{r-s,\bar{r}-\bar{s}},\\[1em]
            & \bar{Q}^{(0,\frac{1}{2})}_{\bar{r}} = -2\pi^2 i \sum_{s,\bar{s}} 2\sqrt{2} \left(\frac{1}{2}\bar{s} - \frac{1}{4}\bar{r}\right) \varphi_{s,\bar{s}} \bar{\chi}_{-s,\bar{r}-\bar{s}}
                + \sqrt{2} ~ \phi_{s,\bar{s}} \bar{\psi}_{-s,\bar{r}-\bar{s}} \\[0.5em]
                &\qquad\qquad\qquad\qquad 
                    - 2 \left(\frac{1}{2}\bar{s} - \frac{1}{4}\bar{r}\right) \psi_{s,\bar{s}} \bar{\phi}_{-s,\bar{r}-\bar{s}}
                    - 2 ~ \chi_{s,\bar{s}} \bar{\varphi}_{-s,\bar{r}-\bar{s}}\\[1em]
            & \bar{Q}^{(\frac{1}{2},0)}_{r,\bar{r}} = 2\pi^2 i \sum_{s,\bar{s}} \sqrt{2} ~ \phi_{s,\bar{s}} \bar{\chi}_{r-s,\bar{r}-\bar{s}}
                - 2 \left(\frac{1}{2}s - \frac{1}{4}r\right) \chi_{s,\bar{s}} \bar{\phi}_{r-s,\bar{r}-\bar{s}},\\
        \end{aligned}
    \end{equation}
    and the $R$ generator is defined as
    \begin{equation}
        \begin{aligned}
            & R^{(0,0)}_{r,\bar{r}} = \pi^2 i \sum_{s,\bar{s}} \sqrt{2} ~ \phi_{s,\bar{s}} \bar{\phi}_{r-s,\bar{r}-\bar{s}}
                + 2 ~ \chi_{s,\bar{s}} \bar{\chi}_{r-s,\bar{r}-\bar{s}}. \\
        \end{aligned}
    \end{equation}
    The modes transformation is shown in Table \ref{tab:ModesTransformationMagneticTypeI-I}. \par

    \begin{table}[htpb]
        \renewcommand\arraystretch{2}
        \centering
        \caption{\centering Field modes transformation under Type I-I BMS$_4$ algebra \eqref{eq:MagneticBMS4TypeI-I}.}
        \resizebox{\linewidth}{!}{\begin{tabular}{c|c|c|c|c}
            \hline
             & $\varphi_{s,\bar{s}}$ & $\phi_{s,\bar{s}}$ & $\psi_{s,\bar{s}}$ & $\chi_{s,\bar{s}}$ \\\hline\hline
            $M_{r,\bar{r}}$
                & $\phi_{r+s,\bar{r}+\bar{s}}$
                & $0$
                & $-2\left(-\frac{1}{4}r - \frac{1}{2}s\right)\chi_{r+s,\bar{s}}$
                & $0$ \\\hline
            $Q^{(\frac{1}{2},0)}_{r}$
                & $-\frac{1}{\sqrt{2}} \psi_{r+s,\bar{s}}$
                & $\sqrt{2} \left(-\frac{1}{4}r - \frac{1}{2}s\right) \chi_{r+s,\bar{s}}$
                & $-\sqrt{2} \left(-\frac{1}{4}r - \frac{1}{2}s\right) \varphi_{r+s,\bar{s}}$
                & $\frac{1}{\sqrt{2}} \phi_{r+s,\bar{s}}$ \\\hline
            $Q^{(0,\frac{1}{2})}_{r,\bar{r}}$
                & $\frac{1}{\sqrt{2}} \chi_{r+s,\bar{r}+\bar{s}}$
                & $0$
                & $-\frac{1}{\sqrt{2}} \phi_{r+s,\bar{r}+\bar{s}}$ 
                & $0$ \\\hline
            $\bar{Q}^{(0,\frac{1}{2})}_{\bar{r}}$
                & $-\frac{1}{\sqrt{2}} \bar{\psi}_{s,\bar{r}+\bar{s}}$
                & $\sqrt{2} \left(-\frac{1}{4}\bar{r} - \frac{1}{2}\bar{s}\right) \bar{\chi}_{s,\bar{r}+\bar{s}}$
                & $-\bar{\varphi}_{s,\bar{r}+\bar{s}}$
                & $-\left(-\frac{1}{4}\bar{r} - \frac{1}{2}\bar{s}\right) \bar{\phi}_{s,\bar{r}+\bar{s}}$ \\\hline
            $\bar{Q}^{(\frac{1}{2},0)}_{r,\bar{r}}$
                & $\frac{1}{\sqrt{2}} \bar{\chi}_{r+s,\bar{r}+\bar{s}}$
                & $0$
                & $\left(-\frac{1}{4}r - \frac{1}{2}s\right)\phi_{r+s,\bar{r}+\bar{s}}$ 
                & $0$ \\\hline
            $R^{(0,0)}_{r,\bar{r}}$
                & $\frac{1}{2\sqrt{2}} \bar{\phi}_{r+s,\bar{r}+\bar{s}}$
                & $0$
                & $\frac{1}{2} \bar{\chi}_{r+s,\bar{r}+\bar{s}}$
                & $0$ \\\hline
        \end{tabular}}\\[1em]
        \resizebox{\linewidth}{!}{\begin{tabular}{c|c|c|c|c}
            \hline
             & $\bar{\psi}_{s,\bar{s}}$ & $\bar{\chi}_{s,\bar{s}}$ & $\bar{\varphi}_{s,\bar{s}}$ & $\bar{\phi}_{s,\bar{s}}$ \\\hline\hline
            $M_{r,\bar{r}}$
                & $-2\left(-\frac{1}{4}\bar{r} - \frac{1}{2}\bar{s}\right) \bar{\chi}_{r+s,\bar{r}+\bar{s}}$
                & $0$
                & $-2\left(-\frac{1}{4}r - \frac{1}{2}s\right) \left(-\frac{1}{4}\bar{r} - \frac{1}{2}\bar{s}\right) \bar{\phi}_{r+s,\bar{r}+\bar{s}}$
                & $0$ \\\hline
            $Q^{(\frac{1}{2},0)}_{r}$
                & $\bar{\varphi}_{r+s,\bar{s}}$
                & $\left(-\frac{1}{4}r - \frac{1}{2}s\right) \bar{\phi}_{r+s,\bar{s}}$
                & $\left(-\frac{1}{4}r - \frac{1}{2}s\right) \bar{\psi}_{r+s,\bar{s}}$
                & $\bar{\chi}_{r+s,\bar{s}}$ \\\hline
            $Q^{(0,\frac{1}{2})}_{r,\bar{r}}$
                & $-\left(-\frac{1}{4}\bar{r} - \frac{1}{2}\bar{s}\right) \bar{\phi}_{r+s,\bar{r}+\bar{s}}$
                & $0$
                & $-\left(-\frac{1}{4}\bar{r} - \frac{1}{2}\bar{s}\right) \bar{\chi}_{r+s,\bar{r}+\bar{s}}$ 
                & $0$ \\\hline
            $\bar{Q}^{(0,\frac{1}{2})}_{\bar{r}}$
                & $-\sqrt{2} \left(-\frac{1}{4}\bar{r} - \frac{1}{2}\bar{s}\right) \varphi_{s,\bar{r}+\bar{s}}$
                & $\frac{1}{\sqrt{2}} \phi_{s,\bar{r}+\bar{s}}$
                & $-\left(-\frac{1}{4}\bar{r} - \frac{1}{2}\bar{s}\right) \psi_{s,\bar{r}+\bar{s}}$
                & $-\chi_{s,\bar{r}+\bar{s}}$ \\\hline
            $\bar{Q}^{(\frac{1}{2},0)}_{r,\bar{r}}$
                & $-\frac{1}{\sqrt{2}} \phi_{r+s,\bar{r}+\bar{s}}$
                & $0$
                & $\left(-\frac{1}{4}r - \frac{1}{2}s\right) \chi_{r+s,\bar{r}+\bar{s}}$ 
                & $0$ \\\hline
            $R^{(0,0)}_{r,\bar{r}}$
                & $-\frac{1}{2} \chi_{r+s,\bar{r}+\bar{s}}$
                & $0$
                & $\frac{1}{2\sqrt{2}} \phi_{r+s,\bar{r}+\bar{s}}$
                & $0$ \\\hline
        \end{tabular}}
        \label{tab:ModesTransformationMagneticTypeI-I}
    \end{table}

    The field theory corresponding to these modes is
    \begin{equation}\label{eq:MagneticTypeI-ITheory}
        \begin{aligned}
            \mathcal{L} &= \frac{1}{2}\partial_0\phi\partial_0\phi - \psi_{+}\partial_0\psi_{-} - \psi_{-}\partial_0\psi_{+} + \psi_{-}\partial_{z}\psi_{-} \\
                &\qquad - \bar{\psi}_{-}\partial_0\bar{\psi}_{+} - \bar{\psi}_{+}\partial_0\bar{\psi}_{-} + \bar{\psi}_{+}\partial_{\bar{z}}\bar{\psi}_{+} 
                + \bar{\pi}\partial_0\bar{\phi} + \partial_{z}\bar{\phi}\partial_{\bar{z}}\bar{\phi}.
        \end{aligned}
    \end{equation}
    Interestingly, this theory incorporates both the electric Carrollian scalar $\phi$ and the magnetic Carrollian scalars $\bar{\pi}$ and $\bar{\phi}$ \cite{Chen:2023pqf}, providing a total of two bosonic degrees of freedom. The spin and conformal dimensions of these fields are listed as follows
    \begin{equation}
        \begin{aligned}
            &\phi: l_{\phi} = 0, \quad \Delta_{\phi} = \frac{1}{2}, \\
            &\psi_{+}: l_{\psi_{+}} = \frac{1}{2}, \quad \Delta_{\psi_{+}} = 1, 
                &&\quad\psi_{-}: l_{\psi_{-}} = -\frac{1}{2}, \quad \Delta_{\psi_{-}} = 1, \\
            &\bar{\psi}_{+}: l_{\bar{\psi}_{+}} = \frac{1}{2}, \quad \Delta_{\bar{\psi}_{+}} = 1, 
                &&\quad\bar{\psi}_{-}: l_{\bar{\psi}_{-}} = -\frac{1}{2}, \quad \Delta_{\bar{\psi}_{-}} = 1, \\
            &\bar{\pi}: l_{\bar{\pi}} = 0, \quad \Delta_{\bar{\pi}} = \frac{3}{2}, 
                &&\quad\bar{\phi}: l_{\bar{\phi}} = 0, \quad \Delta_{\bar{\phi}} = \frac{1}{2}, \\
        \end{aligned}
    \end{equation}
    The EOMs and mode expansions are respectively 
    \begin{equation}
        \begin{aligned}
            & \partial_0^2\phi = 0, \quad \partial_0\psi_{+} = -\partial_{z}\psi_{-}, \quad \partial_0\psi_{-} = 0, \\
            & \partial_0\bar{\psi}_{-} = -\partial_{\bar{z}}\bar{\psi}_{+}, \quad \partial_0\bar{\psi}_{+} = 0, \quad \partial_0\bar{\pi} = -2\partial_{z}\partial_{\bar{z}} \bar{\phi}, \quad \partial_0\bar{\phi} = 0,
        \end{aligned}
    \end{equation}
    \begin{align}
        &\begin{aligned}
            & \phi(t,z,\bar{z}) = \sum_{r,\bar{r}} \varphi_{r,\bar{r}} z^{-r-\frac{1}{4}}\bar{z}^{-\bar{r}-\frac{1}{4}} 
                + t \phi_{r,\bar{r}} z^{-r-\frac{3}{4}}\bar{z}^{-\bar{r}-\frac{3}{4}}, \\
        \end{aligned}\\
        &\begin{aligned}
            & \psi_{+}(t,z,\bar{z}) = \sum_{r,\bar{r}} \psi_{r,\bar{r}} z^{-r-\frac{3}{4}}\bar{z}^{-\bar{r}-\frac{1}{4}} 
                + \left(r+\frac{1}{4}\right)\frac{t}{z} ~ \chi_{r,\bar{r}} z^{-r-\frac{1}{4}}\bar{z}^{-\bar{r}-\frac{3}{4}}, \\
            & \psi_{-}(t,z,\bar{z}) = \sum_{r,\bar{r}} \chi_{r,\bar{r}} z^{-r-\frac{1}{4}}\bar{z}^{-\bar{r}-\frac{3}{4}}, \\
        \end{aligned}\\
        &\begin{aligned}
            & \bar{\psi}_{-}(t,z,\bar{z}) = \sum_{r,\bar{r}} \bar{\psi}_{r,\bar{r}} z^{-r-\frac{1}{4}}\bar{z}^{-\bar{r}-\frac{3}{4}} 
                + \left(\bar{r}+\frac{1}{4}\right)\frac{t}{\bar{z}} ~ \bar{\chi}_{r,\bar{r}} z^{-r-\frac{3}{4}} \bar{z}^{-\bar{r}-\frac{1}{4}}, \\
            & \bar{\psi}_{+}(t,z,\bar{z}) = \sum_{r,\bar{r}} \bar{\chi}_{r,\bar{r}} z^{-r-\frac{3}{4}}\bar{z}^{-\bar{r}-\frac{1}{4}}, \\
        \end{aligned}\\
        &\begin{aligned}
            & \bar{\pi}(t,z,\bar{z}) = \sum_{r,\bar{r}} \bar{\varphi}_{r,\bar{r}} z^{-r-\frac{3}{4}} \bar{z}^{-\bar{r}-\frac{3}{4}} 
                - 2 \left(r+\frac{1}{4}\right) \left(r+\frac{1}{4}\right) \bar{\phi}_{r,\bar{r}} z^{-r-\frac{1}{4}} \bar{z}^{-\bar{r}-\frac{1}{4}}, \\
            & \bar{\phi}(t,z,\bar{z}) = \sum_{r,\bar{r}} \bar{\phi}_{r,\bar{r}} z^{-r-\frac{1}{4}}\bar{z}^{-\bar{r}-\frac{1}{4}}.
        \end{aligned}
    \end{align}
    \par

\subsubsection*{Realization of magnetic Type II-II algebra \eqref{eq:MagneticBMS4TypeII-II}}

    The modes required to realize the Type II-II algebra are:
    \begin{equation}
        \begin{aligned}
            & \varphi^{(-\frac{3}{4},-\frac{3}{4})}, \quad \phi^{(-\frac{1}{4},-\frac{1}{4})}, \quad [\varphi^{(-\frac{3}{4},-\frac{3}{4})}_{r,\bar{r}},\phi^{(-\frac{1}{4},-\frac{1}{4})}_{s,\bar{s}}] = \frac{i}{4\pi^2}\delta_{r+s} \delta_{\bar{r}+\bar{s}}, \\
            & \psi^{(-\frac{5}{4},-\frac{3}{4})}, \quad \chi^{(\frac{}{4},-\frac{1}{4})}, \quad \{\psi^{(-\frac{5}{4},-\frac{3}{4})}_{r,\bar{r}},\chi^{(\frac{1}{4},-\frac{1}{4})}_{s,\bar{s}}\} = \frac{i}{4\pi^2}\delta_{r+s} \delta_{\bar{r}+\bar{s}},\\
            & \bar{\psi}^{(-\frac{3}{4},-\frac{5}{4})}, \quad \bar{\chi}^{(-\frac{1}{4},\frac{1}{4})}, \quad \{\bar{\psi}^{(-\frac{3}{4},-\frac{5}{4})}_{r,\bar{r}},\bar{\chi}^{(-\frac{1}{4},\frac{1}{4})}_{s,\bar{s}}\} = \frac{i}{4\pi^2}\delta_{r+s} \delta_{\bar{r}+\bar{s}},\\
            & \bar{\varphi}^{(-\frac{5}{4},-\frac{5}{4})}, \quad \bar{\phi}^{(\frac{1}{4},\frac{1}{4})}, \quad [\bar{\varphi}^{(-\frac{5}{4},-\frac{5}{4})}_{r,\bar{r}}, \bar{\phi}^{(\frac{1}{4},\frac{1}{4})}_{s,\bar{s}}] = \frac{i}{4\pi^2}\delta_{r+s} \delta_{\bar{r}+\bar{s}}. \\
        \end{aligned}
    \end{equation}
    The algebra is constructed as
    \begin{equation}
        \begin{aligned}
            & L_{m} = -4\pi^2i \sum_{s,\bar{s}} \left(s-\frac{1}{4}m\right) \varphi_{s,\bar{s}} \phi_{m-s,-\bar{s}} 
                - \left(s+\frac{1}{4}m\right)\psi_{s,\bar{s}} \chi_{m-s,-\bar{s}} \\[0.5em]
                &\qquad\qquad\qquad\qquad 
                    - \left(s-\frac{1}{4}m\right) \bar{\psi}_{s,\bar{s}} \bar{\chi}_{m-s,-\bar{s}} 
                    + \left(s+\frac{1}{4}m\right) \bar{\varphi}_{s,\bar{s}} \bar{\phi}_{m-s,-\bar{s}}, \\[1em]
            & \bar{L}_{\bar{m}} = -4\pi^2i \sum_{s,\bar{s}} \left(\bar{s}-\frac{1}{4}\bar{m}\right) \varphi_{s,\bar{s}} \phi_{-s,\bar{m}-\bar{s}}
                - \left(\bar{s}-\frac{1}{4}\bar{m}\right)\psi_{s,\bar{s}} \chi_{-s,\bar{m}-\bar{s}} \\[0.5em]
                &\qquad\qquad\qquad\qquad 
                    - \left(\bar{s}+\frac{1}{4}\bar{m}\right) \bar{\psi}_{s,\bar{s}} \bar{\chi}_{-s,\bar{m}-\bar{s}} 
                    + \left(\bar{s}+\frac{1}{4}\bar{m}\right) \bar{\varphi}_{s,\bar{s}} \bar{\phi}_{-s,\bar{m}-\bar{s}}, \\[1em]
            & M_{r,\bar{r}} = 2\pi^2i \sum_{s,\bar{s}} \phi_{s,\bar{s}} \phi_{r-s,\bar{r}-\bar{s}}, \\
        \end{aligned}
    \end{equation}
    \begin{equation}
        \begin{aligned}
            & Q^{(\frac{1}{2},0)}_{r} = -4\pi^2 i \sum_{s,\bar{s}} \varphi_{s,\bar{s}} \chi_{r-s,-\bar{s}}
                - \left(\frac{1}{2}s - \frac{3}{4}r\right) \phi_{s,\bar{s}} \psi_{r-s,-\bar{s}} \\[0.5em]
                &\qquad\qquad\qquad\qquad 
                    + \bar{\psi}_{s,\bar{s}} \bar{\phi}_{r-s,-\bar{s}}
                    - \left(\frac{1}{2}s - \frac{3}{4}r\right) \bar{\varphi}_{r-s,-\bar{s}}\\[1em]
            & Q^{(1,\frac{1}{2})}_{r,\bar{r}} = -4\pi^2 i \sum_{s,\bar{s}} \phi_{s,\bar{s}} \chi_{r-s,\bar{r}-\bar{s}},\\[1em]
            & \bar{Q}^{(0,\frac{1}{2})}_{\bar{r}} = -4\pi^2 i \sum_{s,\bar{s}} \varphi_{s,\bar{s}} \bar{\chi}_{-s,\bar{r}-\bar{s}}
                - \left(\frac{1}{2}\bar{s} - \frac{3}{4}\bar{r}\right) \phi_{s,\bar{s}} \bar{\psi}_{-s,\bar{r}-\bar{s}} \\[0.5em]
                &\qquad\qquad\qquad\qquad 
                    - \psi_{s,\bar{s}} \bar{\phi}_{-s,\bar{r}-\bar{s}}
                    + \left(\frac{1}{2}\bar{s} - \frac{3}{4}\bar{r}\right) \chi_{s,\bar{s}} \bar{\varphi}_{-s,\bar{r}-\bar{s}}\\[1em]
            & \bar{Q}^{(\frac{1}{2},1)}_{r,\bar{r}} = 2\pi^2 i \sum_{s,\bar{s}} \sqrt{2} ~ \phi_{s,\bar{s}} \bar{\chi}_{r-s,\bar{r}-\bar{s}},\\
        \end{aligned}
    \end{equation}
    and
    \begin{equation}
        \begin{aligned}
            & R^{(1,1)}_{r,\bar{r}} = -4\pi^2 i \sum_{s,\bar{s}} \phi_{s,\bar{s}} \bar{\phi}_{r-s,\bar{r}-\bar{s}}
                + \chi_{s,\bar{s}} \bar{\chi}_{r-s,\bar{r}-\bar{s}}. \\
        \end{aligned}
    \end{equation}
    The transformation of these modes under the algebra is summarized in Table \ref{tab:ModesTransformationMagneticTypeII-II}. \par

    \begin{table}[htpb]
        \renewcommand\arraystretch{2}
        \centering
        \caption{\centering Field modes transformation under Type II-II BMS$_4$ algebra \eqref{eq:MagneticBMS4TypeII-II}.}
        \begin{tabular}{c|c|c|c|c}
            \hline
             & $\varphi_{s,\bar{s}}$ & $\phi_{s,\bar{s}}$ & $\psi_{s,\bar{s}}$ & $\chi_{s,\bar{s}}$ \\\hline\hline
            $M_{r,\bar{r}}$
                & $\phi_{r+s,\bar{r}+\bar{s}}$
                & $0$
                & $0$
                & $0$ \\\hline
            $Q^{(\frac{1}{2},0)}_{r}$
                & $\left(-\frac{3}{4}r - \frac{1}{2}s\right) \psi_{r+s,\bar{s}}$
                & $\chi_{r+s,\bar{s}}$
                & $\varphi_{r+s,\bar{s}}$
                & $-\left(-\frac{1}{4}r - \frac{1}{2}s\right)\phi_{r+s,\bar{s}}$ \\\hline
            $Q^{(1,\frac{1}{2})}_{r,\bar{r}}$
                & $-\chi_{r+s,\bar{r}+\bar{s}}$
                & $0$
                & $\phi_{r+s,\bar{r}+\bar{s}}$ 
                & $0$ \\\hline
            $\bar{Q}^{(0,\frac{1}{2})}_{\bar{r}}$
                & $\left(-\frac{3}{4}\bar{r} - \frac{1}{2}\bar{s}\right) \bar{\psi}_{s,\bar{r}+\bar{s}}$
                & $\bar{\chi}_{s,\bar{r}+\bar{s}}$
                & $\left(-\frac{3}{4}\bar{r} - \frac{1}{2}\bar{s}\right) \bar{\varphi}_{s,\bar{r}+\bar{s}}$
                & $-\bar{\phi}_{s,\bar{r}+\bar{s}}$ \\\hline
            $\bar{Q}^{(\frac{1}{2},1)}_{r,\bar{r}}$
                & $-\bar{\chi}_{r+s,\bar{r}+\bar{s}}$
                & $0$
                & $0$ 
                & $0$ \\\hline
            $R^{(1,1)}_{r,\bar{r}}$
                & $-\bar{\phi}_{r+s,\bar{r}+\bar{s}}$
                & $0$
                & $-\bar{\chi}_{r+s,\bar{r}+\bar{s}}$
                & $0$ \\\hline
        \end{tabular}\\[1em]
        \begin{tabular}{c|c|c|c|c}
            \hline
             & $\bar{\psi}_{s,\bar{s}}$ & $\bar{\chi}_{s,\bar{s}}$ & $\bar{\varphi}_{s,\bar{s}}$ & $\bar{\phi}_{s,\bar{s}}$ \\\hline\hline
            $M_{r,\bar{r}}$
                & $0$
                & $0$
                & $0$
                & $0$ \\\hline
            $Q^{(\frac{1}{2},0)}_{r}$
                & $-\left(-\frac{3}{4}r - \frac{1}{2}s\right) \bar{\varphi}_{r+s,\bar{s}}$
                & $\bar{\phi}_{r+s,\bar{s}}$
                & $-\bar{\psi}_{r+s,\bar{s}}$
                & $\left(\frac{1}{4}r - \frac{1}{2}s\right) \bar{\chi}_{r+s,\bar{s}}$ \\\hline
            $Q^{(1,\frac{1}{2})}_{r,\bar{r}}$
                & $0$
                & $0$
                & $0$ 
                & $0$ \\\hline
            $\bar{Q}^{(0,\frac{1}{2})}_{\bar{r}}$
                & $\varphi_{s,\bar{r}+\bar{s}}$
                & $\left(\frac{1}{4}\bar{r} - \frac{1}{2}\bar{s}\right) \phi_{s,\bar{r}+\bar{s}}$
                & $\psi_{s,\bar{r}+\bar{s}}$
                & $-\left(\frac{1}{4}\bar{r} - \frac{1}{2}\bar{s}\right) \chi_{s,\bar{r}+\bar{s}}$ \\\hline
            $\bar{Q}^{(\frac{1}{2},1)}_{r,\bar{r}}$
                & $\phi_{r+s,\bar{r}+\bar{s}}$
                & $0$
                & $0$ 
                & $0$ \\\hline
            $R^{(1,1)}_{r,\bar{r}}$
                & $\chi_{r+s,\bar{r}+\bar{s}}$
                & $0$
                & $- \phi_{r+s,\bar{r}+\bar{s}}$
                & $0$ \\\hline
        \end{tabular}
        \label{tab:ModesTransformationMagneticTypeII-II}
    \end{table}

    The constructed theory is
    \begin{equation}\label{eq:MagneticTypeII-IITheory}
        \begin{aligned}
            \mathcal{L} &= \frac{1}{2}\partial_0\phi\partial_0\phi - \psi_{+}\partial_0\psi_{-} - \psi_{-}\partial_0\psi_{+} - \bar{\psi}_{-}\partial_0\bar{\psi}_{+} - \bar{\psi}_{+}\partial_0\bar{\psi}_{-} + \bar{\varphi}\partial_0 \bar{\phi}.
        \end{aligned}
    \end{equation}
    The spin and conformal dimension of the fields are given by
    \begin{equation}
        \begin{aligned}
            &\phi: l_{\phi} = 0, \quad \Delta_{\phi} = \frac{1}{2}, \\
            &\psi_{+}: l_{\psi_{+}} = \frac{1}{2}, \quad \Delta_{\psi_{+}} = 2, 
                &&\quad\psi_{-}: l_{\psi_{-}} = -\frac{1}{2}, \quad \Delta_{\psi_{-}} = 0, \\
            &\bar{\psi}_{+}: l_{\bar{\psi}_{+}} = \frac{1}{2}, \quad \Delta_{\bar{\psi}_{+}} = 0, 
                &&\quad\bar{\psi}_{-}: l_{\bar{\psi}_{-}} = -\frac{1}{2}, \quad \Delta_{\bar{\psi}_{-}} = 2, \\
            &\bar{\varphi}: l_{\bar{\varphi}} = 0, \quad \Delta_{\bar{\varphi}} = -\frac{1}{2}, 
                &&\quad \bar{\phi}: l_{\bar{\phi}} = 0, \quad \Delta_{\bar{\phi}} = \frac{5}{2}, \\
        \end{aligned}
    \end{equation}
    The EOMs and mode expansions are respectively 
    \begin{equation}
        \begin{aligned}
            & \partial_0^2\phi = 0, \quad \partial_0\psi_{+} = \partial_0\psi_{-} = 0, \quad \partial_0\bar{\psi}_{+} = \partial_0\bar{\psi}_{-} = 0, \quad \partial_0 \bar{\varphi} = \partial_0 \bar{\phi} = 0,
        \end{aligned}
    \end{equation}
    \begin{align}
        &\begin{aligned}
            & \phi(t,z,\bar{z}) = \sum_{r,\bar{r}} \varphi_{r,\bar{r}} z^{-r-\frac{1}{4}}\bar{z}^{-\bar{r}-\frac{1}{4}} 
                + t \phi_{r,\bar{r}} z^{-r-\frac{3}{4}}\bar{z}^{-\bar{r}-\frac{3}{4}}, \\
        \end{aligned}\\
        &\begin{aligned}
            & \psi_{+}(t,z,\bar{z}) = \sum_{r,\bar{r}} \chi_{r,\bar{r}} z^{-r-\frac{5}{4}}\bar{z}^{-\bar{r}-\frac{3}{4}}, \quad \psi_{-}(t,z,\bar{z}) = \sum_{r,\bar{r}} \chi_{r,\bar{r}} z^{-r+\frac{1}{4}}\bar{z}^{-\bar{r}-\frac{1}{4}}, \\
        \end{aligned}\\
        &\begin{aligned}
            & \bar{\psi}_{+}(t,z,\bar{z}) = \sum_{r,\bar{r}} \bar{\psi}_{r,\bar{r}} z^{-r-\frac{1}{4}}\bar{z}^{-\bar{r}+\frac{1}{4}} , \quad \bar{\psi}_{-}(t,z,\bar{z}) = \sum_{r,\bar{r}} \bar{\chi}_{r,\bar{r}} z^{-r-\frac{3}{4}}\bar{z}^{-\bar{r}-\frac{5}{4}}, \\
        \end{aligned}\\
        &\begin{aligned}
            & \bar{\varphi}(t,z,\bar{z}) = \sum_{r,\bar{r}} \bar{\varphi}_{r,\bar{r}} z^{-r+\frac{1}{4}} \bar{z}^{-\bar{r}+\frac{1}{4}}, \quad \bar{\phi}(t,z,\bar{z}) = \sum_{r,\bar{r}} \bar{\phi}_{r,\bar{r}} z^{-r-\frac{5}{4}}\bar{z}^{-\bar{r}-\frac{5}{4}}.
        \end{aligned}
    \end{align}
    \par

\subsubsection*{Realization of magnetic Type I-II algebra \eqref{eq:MagneticBMS4TypeI-II}}

    Finally, we construct the theory exhibiting Type I-II symmetry. The corresponding modes are given by:
    \begin{equation}
        \begin{aligned}
            & \varphi^{(-\frac{3}{4},-\frac{3}{4})}, \quad \phi^{(-\frac{1}{4},-\frac{1}{4})}, \quad [\varphi^{(-\frac{3}{4},-\frac{3}{4})}_{r,\bar{r}},\phi^{(-\frac{1}{4},-\frac{1}{4})}_{s,\bar{s}}] = \frac{i}{4\pi^2}\delta_{r+s} \delta_{\bar{r}+\bar{s}}, \\
            & \psi^{(-\frac{1}{4},-\frac{3}{4})}, \quad \chi^{(-\frac{3}{4},-\frac{1}{4})}, \quad \{\psi^{(-\frac{1}{4},-\frac{3}{4})}_{r,\bar{r}},\chi^{(-\frac{3}{4},-\frac{1}{4})}_{s,\bar{s}}\} = \frac{i}{4\pi^2}\delta_{r+s} \delta_{\bar{r}+\bar{s}},\\
            & \bar{\psi}^{(-\frac{3}{4},-\frac{5}{4})}, \quad \bar{\chi}^{(-\frac{1}{4},\frac{1}{4})}, \quad \{\bar{\psi}^{(-\frac{3}{4},-\frac{5}{4})}_{r,\bar{r}},\bar{\chi}^{(-\frac{1}{4},\frac{1}{4})}_{s,\bar{s}}\} = \frac{i}{4\pi^2}\delta_{r+s} \delta_{\bar{r}+\bar{s}},\\
            & \bar{\varphi}^{(-\frac{1}{4},-\frac{5}{4})}, \quad \bar{\phi}^{(-\frac{3}{4},\frac{1}{4})}, \quad [\bar{\varphi}^{(-\frac{1}{4},-\frac{5}{4})}_{r,\bar{r}}, \bar{\phi}^{(-\frac{3}{4},\frac{1}{4})}_{s,\bar{s}}] = \frac{i}{4\pi^2}\delta_{r+s} \delta_{\bar{r}+\bar{s}}. \\
        \end{aligned}
    \end{equation} 
    The algebra is constructed as:
    \begin{equation}
        \begin{aligned}
            & L_{m} = -4\pi^2i \sum_{s,\bar{s}} \left(s-\frac{1}{4}m\right) \varphi_{s,\bar{s}} \phi_{m-s,-\bar{s}} 
                - \left(s-\frac{3}{4}m\right) \psi_{s,\bar{s}} \chi_{m-s,-\bar{s}} \\[0.5em]
                &\qquad\qquad\qquad\qquad 
                    - \left(s-\frac{1}{4}m\right) \bar{\psi}_{s,\bar{s}} \bar{\chi}_{m-s,-\bar{s}} 
                    + \left(s-\frac{3}{4}m\right) \bar{\varphi}_{s,\bar{s}} \bar{\phi}_{m-s,-\bar{s}}, \\[1em]
            & \bar{L}_{\bar{m}} = -4\pi^2i \sum_{s,\bar{s}} \left(\bar{s}-\frac{1}{4}\bar{m}\right) \varphi_{s,\bar{s}} \phi_{-s,\bar{m}-\bar{s}}
                - \left(\bar{s}-\frac{1}{4}\bar{m}\right) \psi_{s,\bar{s}} \chi_{-s,\bar{m}-\bar{s}} \\[0.5em]
                &\qquad\qquad\qquad\qquad 
                    - \left(\bar{s}+\frac{1}{4}\bar{m}\right) \bar{\psi}_{s,\bar{s}} \bar{\chi}_{-s,\bar{m}-\bar{s}} 
                    + \left(\bar{s}+\frac{1}{4}\bar{m}\right) \bar{\varphi}_{s,\bar{s}} \bar{\phi}_{-s,\bar{m}-\bar{s}}, \\[1em]
            & M_{r,\bar{r}} = 2\pi^2i \sum_{s,\bar{s}} \phi_{s,\bar{s}} \phi_{r-s,\bar{r}-\bar{s}}
                + 2 \left(\frac{1}{2}\bar{s}-\frac{1}{4}\bar{m}\right) \chi_{s,\bar{s}} \chi_{r-s,\bar{m}-\bar{s}}, \\
        \end{aligned}
    \end{equation}
    the fermionic generators are
    \begin{equation}
        \begin{aligned}
            & Q^{(\frac{1}{2},0)}_{r} = -2\sqrt{2}\pi^2 i \sum_{s,\bar{s}} 2 \left(\frac{1}{2}s - \frac{1}{4}r\right) \varphi_{s,\bar{s}} \chi_{r-s,-\bar{s}}
                + \phi_{s,\bar{s}} \psi_{r-s,-\bar{s}} \\[0.5em]
                &\qquad\qquad\qquad\qquad 
                    - 2 \left(\frac{1}{2}s - \frac{1}{4}r\right) \bar{\psi}_{s,\bar{s}} \bar{\phi}_{r-s,-\bar{s}}
                    - \bar{\chi}_{s,\bar{s}} \bar{\varphi}_{r-s,-\bar{s}}\\[1em]
            & Q^{(0,\frac{1}{2})}_{r,\bar{r}} = 2\sqrt{2}\pi^2 i \sum_{s,\bar{s}} \phi_{s,\bar{s}} \chi_{r-s,\bar{r}-\bar{s}},\\[1em]
            & \bar{Q}^{(0,\frac{1}{2})}_{\bar{r}} = -4\pi^2 i \sum_{s,\bar{s}} \varphi_{s,\bar{s}} \bar{\chi}_{-s,\bar{r}-\bar{s}}
                - \left(\frac{1}{2}\bar{s} - \frac{3}{4}\bar{r}\right) \phi_{s,\bar{s}} \bar{\psi}_{-s,\bar{r}-\bar{s}} \\[0.5em]
                &\qquad\qquad\qquad\qquad 
                    + \psi_{s,\bar{s}} \bar{\phi}_{-s,\bar{r}-\bar{s}}
                    - \left(\frac{1}{2}\bar{s} - \frac{3}{4}\bar{r}\right)  \chi_{s,\bar{s}} \bar{\varphi}_{-s,\bar{r}-\bar{s}}\\[1em]
            & \bar{Q}^{(\frac{1}{2},1)}_{r,\bar{r}} = -4\pi^2 i \sum_{s,\bar{s}} \phi_{s,\bar{s}} \bar{\chi}_{r-s,\bar{r}-\bar{s}}
                + 2 \left(\frac{1}{2}s - \frac{1}{4}r\right) \chi_{s,\bar{s}} \bar{\phi}_{r-s,\bar{r}-\bar{s}},\\
        \end{aligned}
    \end{equation}
    and the bosonic $R$ generators are
    \begin{equation}
        \begin{aligned}
            & R^{(0,1)}_{r,\bar{r}} = 2\sqrt{2}\pi^2 i \sum_{s,\bar{s}} \phi_{s,\bar{s}} \bar{\phi}_{r-s,\bar{r}-\bar{s}}
               - \chi_{s,\bar{s}} \bar{\chi}_{r-s,\bar{r}-\bar{s}}. \\
        \end{aligned}
    \end{equation}
    The modes transformation is shown in Table \ref{tab:ModesTransformationMagneticTypeI-II}. \par

    \begin{table}[htpb]
        \renewcommand\arraystretch{2}
        \centering
        \caption{\centering Field modes transformation under Type I-II BMS$_4$ algebra \eqref{eq:MagneticBMS4TypeI-II}.}
        \resizebox{\linewidth}{!}{\begin{tabular}{c|c|c|c|c}
            \hline
             & $\varphi_{s,\bar{s}}$ & $\phi_{s,\bar{s}}$ & $\psi_{s,\bar{s}}$ & $\chi_{s,\bar{s}}$ \\\hline\hline
            $M_{r,\bar{r}}$
                & $\phi_{r+s,\bar{r}+\bar{s}}$
                & $0$
                & $-2\left(-\frac{1}{4}r - \frac{1}{2}s\right)\chi_{r+s,\bar{s}}$
                & $0$ \\\hline
            $Q^{(\frac{1}{2},0)}_{r}$
                & $-\frac{1}{\sqrt{2}} \psi_{r+s,\bar{s}}$
                & $\sqrt{2} \left(-\frac{1}{4}r - \frac{1}{2}s\right) \chi_{r+s,\bar{s}}$
                & $-\sqrt{2} \left(-\frac{1}{4}r - \frac{1}{2}s\right) \varphi_{r+s,\bar{s}}$
                & $\frac{1}{\sqrt{2}} \phi_{r+s,\bar{s}}$ \\\hline
            $Q^{(0,\frac{1}{2})}_{r,\bar{r}}$
                & $\frac{1}{\sqrt{2}} \chi_{r+s,\bar{r}+\bar{s}}$
                & $0$
                & $-\frac{1}{\sqrt{2}} \phi_{r+s,\bar{r}+\bar{s}}$ 
                & $0$ \\\hline
            $\bar{Q}^{(0,\frac{1}{2})}_{\bar{r}}$
                & $\left(-\frac{3}{4}\bar{r} - \frac{1}{2}\bar{s}\right) \bar{\psi}_{s,\bar{r}+\bar{s}}$
                & $\bar{\chi}_{s,\bar{r}+\bar{s}}$
                & $-\left(-\frac{3}{4}\bar{r} - \frac{1}{2}\bar{s}\right) \bar{\varphi}_{s,\bar{r}+\bar{s}}$
                & $\bar{\phi}_{s,\bar{r}+\bar{s}}$ \\\hline
            $\bar{Q}^{(\frac{1}{2},1)}_{r,\bar{r}}$
                & $- \bar{\chi}_{r+s,\bar{r}+\bar{s}}$
                & $0$
                & $2\left(-\frac{1}{4}r - \frac{1}{2}s\right)\phi_{r+s,\bar{r}+\bar{s}}$ 
                & $0$ \\\hline
            $R^{(0,1)}_{r,\bar{r}}$
                & $\frac{1}{\sqrt{2}} \bar{\phi}_{r+s,\bar{r}+\bar{s}}$
                & $0$
                & $-\frac{1}{\sqrt{2}} \bar{\chi}_{r+s,\bar{r}+\bar{s}}$
                & $0$ \\\hline
        \end{tabular}}\\[1em]
        \resizebox{\linewidth}{!}{\begin{tabular}{c|c|c|c|c}
            \hline
             & $\bar{\psi}_{s,\bar{s}}$ & $\bar{\chi}_{s,\bar{s}}$ & $\bar{\varphi}_{s,\bar{s}}$ & $\bar{\phi}_{s,\bar{s}}$ \\\hline\hline
            $M_{r,\bar{r}}$
                & $0$
                & $0$
                & $0$
                & $0$ \\\hline
            $Q^{(\frac{1}{2},0)}_{r}$
                & $-\frac{1}{\sqrt{2}} \bar{\varphi}_{r+s,\bar{s}}$
                & $-\sqrt{2}\left(-\frac{1}{4}r - \frac{1}{2}s\right) \bar{\phi}_{r+s,\bar{s}}$
                & $-\sqrt{2}\left(-\frac{1}{4}r - \frac{1}{2}s\right) \bar{\psi}_{r+s,\bar{s}}$
                & $-\frac{1}{\sqrt{2}} \bar{\chi}_{r+s,\bar{s}}$ \\\hline
            $Q^{(0,\frac{1}{2})}_{r,\bar{r}}$
                & $0$
                & $0$
                & $0$ 
                & $0$ \\\hline
            $\bar{Q}^{(0,\frac{1}{2})}_{\bar{r}}$
                & $\varphi_{s,\bar{r}+\bar{s}}$
                & $\left(\frac{1}{4}\bar{r} - \frac{1}{2}\bar{s}\right)  \phi_{s,\bar{r}+\bar{s}}$
                & $-\psi_{s,\bar{r}+\bar{s}}$
                & $\left(\frac{1}{4}\bar{r} - \frac{1}{2}\bar{s}\right) \chi_{s,\bar{r}+\bar{s}}$ \\\hline
            $\bar{Q}^{(\frac{1}{2},1)}_{r,\bar{r}}$
                & $\phi_{r+s,\bar{r}+\bar{s}}$
                & $0$
                & $\left(-\frac{1}{4}r - \frac{1}{2}s\right) \chi_{r+s,\bar{r}+\bar{s}}$ 
                & $0$ \\\hline
            $R^{(0,1)}_{r,\bar{r}}$
                & $\frac{1}{\sqrt{2}} \chi_{r+s,\bar{r}+\bar{s}}$
                & $0$
                & $\frac{1}{\sqrt{2}} \phi_{r+s,\bar{r}+\bar{s}}$
                & $0$ \\\hline
        \end{tabular}}
        \label{tab:ModesTransformationMagneticTypeI-II}
    \end{table}

    The field theory incorporating these modes is
    \begin{equation}\label{eq:MagneticTypeI-IITheory}
        \begin{aligned}
            \mathcal{L} &= \frac{1}{2}\partial_0\phi\partial_0\phi - \psi_{+}\partial_0\psi_{-} - \psi_{-}\partial_0\psi_{+} + \psi_{-}\partial_{z}\psi_{-} - \bar{\psi}_{-}\partial_0\bar{\psi}_{+} - \bar{\psi}_{+}\partial_0\bar{\psi}_{-} 
                + A_{+}\partial_0 A_{-}.
        \end{aligned}
    \end{equation}
    It is worth noting that this nonchiral symmetric theory includes spin-$\pm 1$ bosons $A_{\pm}$. Especially, the $A_{\pm}$ and $\bar{\psi}_{\pm}$ part of this theory is single copy of the theory studied in \cite{Adamo:2014yya} as perturbative supergravity.  The spin and conformal dimensions of the fields are summarized as
    \begin{equation}
        \begin{aligned}
            &\phi: l_{\phi} = 0, \quad \Delta_{\phi} = \frac{1}{2}, \\
            &\psi_{+}: l_{\psi_{+}} = \frac{1}{2}, \quad \Delta_{\psi_{+}} = 1, 
                &&\quad\psi_{-}: l_{\psi_{-}} = -\frac{1}{2}, \quad \Delta_{\psi_{-}} = 1, \\
            &\bar{\psi}_{+}: l_{\bar{\psi}_{+}} = \frac{1}{2}, \quad \Delta_{\bar{\psi}_{+}} = 0, 
                &&\quad\bar{\psi}_{-}: l_{\bar{\psi}_{-}} = -\frac{1}{2}, \quad \Delta_{\bar{\psi}_{-}} = 2, \\
            &A_{+}: l_{A_{+}} = 1, \quad \Delta_{A_{+}} = \frac{1}{2}, 
                &&\quad A_{-}: l_{A_{-}} = -1, \quad \Delta_{A_{-}} = \frac{3}{2}, \\
        \end{aligned}
    \end{equation}
    The EOMs and mode expansions are respectively 
    \begin{equation}
        \begin{aligned}
            & \partial_0^2\phi = 0, \quad \partial_0\psi_{+} = -\partial_{z}\psi_{-}, \quad \partial_0\psi_{-} = 0, \\
            & \partial_0\bar{\psi}_{-} = \partial_0\bar{\psi}_{+} = 0, \quad \partial_0 A_{+} = \partial_0 A_{-} = 0,
        \end{aligned}
    \end{equation}
    \begin{align}
        &\begin{aligned}
            & \phi(t,z,\bar{z}) = \sum_{r,\bar{r}} \varphi_{r,\bar{r}} z^{-r-\frac{1}{4}}\bar{z}^{-\bar{r}-\frac{1}{4}} 
                + t \phi_{r,\bar{r}} z^{-r-\frac{3}{4}}\bar{z}^{-\bar{r}-\frac{3}{4}}, \\
        \end{aligned}\\
        &\begin{aligned}
            & \psi_{+}(t,z,\bar{z}) = \sum_{r,\bar{r}} \psi_{r,\bar{r}} z^{-r-\frac{3}{4}}\bar{z}^{-\bar{r}-\frac{1}{4}} 
                + \left(r+\frac{1}{4}\right)\frac{t}{z} ~ \chi_{r,\bar{r}} z^{-r-\frac{1}{4}}\bar{z}^{-\bar{r}-\frac{3}{4}}, \\
            & \psi_{-}(t,z,\bar{z}) = \sum_{r,\bar{r}} \chi_{r,\bar{r}} z^{-r-\frac{1}{4}}\bar{z}^{-\bar{r}-\frac{3}{4}}, \\
        \end{aligned}\\
        &\begin{aligned}
            & \bar{\psi}_{+}(t,z,\bar{z}) = \sum_{r,\bar{r}} \bar{\psi}_{r,\bar{r}} z^{-r-\frac{1}{4}}\bar{z}^{-\bar{r}+\frac{1}{4}}, \quad \bar{\psi}_{-}(t,z,\bar{z}) = \sum_{r,\bar{r}} \bar{\chi}_{r,\bar{r}} z^{-r-\frac{3}{4}}\bar{z}^{-\bar{r}-\frac{5}{4}}, \\
        \end{aligned}\\
        &\begin{aligned}
            & A_{+}(t,z,\bar{z}) = \sum_{r,\bar{r}} \bar{\varphi}_{r,\bar{r}} z^{-r-\frac{3}{4}} \bar{z}^{-\bar{r}+\frac{1}{4}}, \quad A_{-}(t,z,\bar{z}) = \sum_{r,\bar{r}} \bar{\phi}_{r,\bar{r}} z^{-r-\frac{1}{4}}\bar{z}^{-\bar{r}-\frac{5}{4}}.
        \end{aligned}
    \end{align}
    \par

    In the preceding constructions, we have realized the magnetic super BMS$_4$ algebras through the theories \eqref{eq:MagneticTypeITheory}, \eqref{eq:MagneticTypeIITheory}, \eqref{eq:MagneticTypeI-ITheory}, \eqref{eq:MagneticTypeII-IITheory}, and \eqref{eq:MagneticTypeI-IITheory}. Notably, there is no single simple theory that accommodates all types of magnetic algebras, highlighting their nontrivial structure compared to the electric algebras. Theories \eqref{eq:MagneticTypeI-ITheory}, \eqref{eq:MagneticTypeII-IITheory}, and \eqref{eq:MagneticTypeI-IITheory} are not standard extended supersymmetric theories; nevertheless, the inclusion of $R$-symmetry is essential for the closure of the algebra. From this perspective, the left-hand and right-hand fermionic generators can be viewed as independent symmetry algebras, and their combinations should be regarded as extensions of the superalgebra. In each case, the $R$-symmetry maps bosons to bosons and fermions to fermions. It is also noteworthy that in \eqref{eq:MagneticTypeI-ITheory}, the presence of the Carrollian magnetic scalar demonstrates that, for supersymmetric BMS$_4$ theories, the electric and magnetic sectors are equally significant and are related via $R$-symmetry. In the case of \eqref{eq:MagneticTypeI-IITheory}, the emergence of spin-$1$ fields corresponds to massless vector fields in the bulk $4$D theory, rendering this nontrivial theory particularly relevant for future discussions in flat holography. \par

\section{Discussion} \label{sec:Discussion}

    In this paper, we conducted a systematic study of the supersymmetric extensions of the BMS$_4$ algebra. In Section \ref{sec:Algebra}, we explored all possible electric and magnetic super BMS$_4$ algebras that admit finite-dimensional subalgebras. For the electric case, we imposed the condition that the anticommutator of supercharges contains only supertranslations $M$, whereas for the magnetic case, the inclusion of superrotations $L$ and $\bar{L}$ is required. Our analysis revealed ten distinct electric super BMS$_4$ algebras, two types of chiral magnetic super BMS$_4$ algebras, and four nonchiral magnetic superalgebras formed from combinations of two chiral superalgebras. \par

    In Section \ref{sec:Realization}, we implemented an inverse construction to realize these superalgebras in free field theories. Notably, we identified a simple theory \eqref{eq:ElectricTheory} that realizes \eqref{eq:ElectricBMS4FromPoincareFull}, \eqref{eq:ElectricBMS4Type12A}, and \eqref{eq:ElectricBMS4Type12B}, consisting of one electric Carrollian scalar and two electric Carrollian spinors. For magnetic supersymmetry, there is a unique realization for each type. Although the of theory \eqref{eq:MagneticTypeIITheory} resembles the electric theory \eqref{eq:ElectricTheory}, the conformal dimensions of the fermions differ, highlighting the distinct nature of magnetic constructions. The significance of magnetic Carrollian scalars is particularly evident in theory \eqref{eq:MagneticTypeI-ITheory}, while theory \eqref{eq:MagneticTypeI-IITheory} plays a key role in the study of theories with bulk electromagnetic fields. Especially, part of \eqref{eq:MagneticTypeI-IITheory} was studied in \cite{Adamo:2014yya} as model of perturbative supergravity. The constructions presented in Section \ref{sec:Realization} focus on the simplest scalar supermultiplets, but the framework allows straightforward study of more general multiplets for further applications. \par

    The global part of the BMS$_4$ algebra is isomorphic to both the $4$D Poincar\'e algebra and the $3$D Carrollian conformal algebra. Similar relations hold for their supersymmetric extensions and the corresponding field theory realizations. As a supersymmetric extension of the $4$D Poincar\'e algebra, it is imposed strict constraints by Haag--Łopuszański--Sohnius theorem. Indeed, the super Poincar\'e algebra coincides with the global part of the electric super BMS$_4$ algebra \eqref{eq:ElectricBMS4FromPoincareFull}. The remaining electric superalgebras discussed in Section \ref{subsec:ElectricAlgebra} involve supercharges with spin greater than $\tfrac{1}{2}$, and therefore their global parts do not correspond to physically realized superalgebras. Similarly, the magnetic superalgebras, whose supercharge anticommutators also produce superrotations, lack a direct physical superalgebra interpretation for their global sectors. Nevertheless, this does not diminish their relevance in the context of flat holography. On the contrary, these algebras highlight how supersymmetric extensions of BMS$_4$ can go beyond the strict confines of super Poincar\'e, potentially capturing structures that only emerge in asymptotic or holographic regimes. In particular, such algebras may arise from large-radius limits of extended super asymptotic symmetries, and the appearance of $R$-symmetry offers a novel perspective on how internal symmetries interplay with asymptotic supersymmetry. \par

    The BMS$_4$ algebra is closely related to the $3$D Carrollian conformal algebra in both algebraic and field-theoretic aspects. By requiring that supercharges admit finite-dimensional subsectors, the global part of all superalgebras discussed in this work can be naturally reinterpreted as $3$D super Carrollian conformal algebras. This dual perspective not only unifies the classification but also connects the free field realizations to a broader Carrollian framework. Indeed, all the free theories constructed here can be viewed as super Carrollian conformal theories containing fields of diverse spins and conformal dimensions. The interplay between electric and magnetic Carrollian scalar theories, related by the $R$-symmetry of the Type I-I magnetic superalgebra, further illustrates how Carrollian holography accommodates supersymmetric structures beyond conventional relativistic situations. In this way, our results underline the role of Carrollian methods as a powerful organizing language for flat space holography and pave the way for future explorations of supersymmetric Carrollian conformal field theories \cite{Zheng:2025cuw}. \par

    There are several promising directions for extending this study:
    \par\noindent \textbf{Super BMS$_4$ algebra from asymptotic symmetry.} The origin of the BMS$_4$ algebra lies in the study of the symmetries of asymptotically flat gravity. In a similar manner, the super BMS$_4$ algebra can also be derived from this perspective \cite{Barnich:2014cwa, Flanagan:2015pxa, Banerjee:2017gzj, Fuentealba:2017fck, Henneaux:2020ekh}. It would be interesting to extend such an asymptotic analysis to the new super BMS$_4$ algebras discussed in this work, and further investigate the physical interpretation of these new superalgebras. \par

    \par\noindent \textbf{Extended supersymmetry.} Extended super BMS$_3$ and BMS$_4$ algebras with $\mathcal{N}$ copy of electric super charges was well studied in the literature \cite{Banerjee:2016nio, Lodato:2016alv, Banerjee:2017gzj, Fuentealba:2017fck, Banerjee:2018hbl, Caroca:2018obf, Banerjee:2019lrv, Banerjee:2022lnz}. The presence of $R$-symmetry in the magnetic Type I-I, Type II-II, and Type I-II algebras indicates that these should also be regarded as extended superalgebras. Interestingly, in general superalgebras, the $R$-symmetry can carry nonzero spin. It is straightforward to consider the case of conventional extended supersymmetry, i.e., multiple sets of the same supercharges, including both electric and magnetic types. In such scenarios, $R$-symmetries with generic spin are naturally expected in general extended super BMS$_4$ algebras. Furthermore, the physical interpretation of these $R$-symmetries, particularly those with spin, is an intriguing question. From a holographic perspective, such novel $R$-symmetries may correspond to symmetries associated with compactified dimensions, thereby providing a bridge between asymptotic flat space physics and string/M-theory constructions. 

    \par\noindent \textbf{More general supersymmetry.} In this work, we focused on supercharges in the charge representation \eqref{eq:ChargeRepresentation}. A more general supersymmetric extension may admit charges in arbitrary representations, including structures reminiscent of $w$-algebras \cite{Banerjee:2015kcx,Banerjee:2022abf}. It is expected that super BMS$_4$ algebras with higher-spin generators are naturally accommodated within a $w$-algebra framework, which could have direct applications in the study of soft theorems and memory effects. Extending the present framework to encompass higher-spin symmetries would not only generalize the classification but also provide algebraic tools for organizing scattering data in flat space, in close analogy with the role of $W$-symmetry in $2$D CFT. \par

    \par\noindent \textbf{Holography with supersymmetry.} This paper has classified all possible super BMS$_4$ algebras in the charge representation with finite subalgebras, providing a firm foundation for understanding supersymmetry in flat holography. These results serve as fundamental building blocks for constructing Fock spaces, boundary field theories, and analyzing constraints imposed by supersymmetry in holographic setups. In particular, the structure of $R$-symmetry offers guidance for constructing full $10$D spaces of string theories in flat holography, potentially shedding light on the embedding of flat space holography into string theory. Additionally, the results presented here suggest new directions for celestial holography: free superfield theories allow for explicit computation of correlation functions, thereby establishing bridges between supersymmetric scattering amplitudes and celestial correlators. Notably, the theory \eqref{eq:MagneticTypeI-IITheory}, which contains a massless spin-$1$ particle, is directly relevant for the study of soft photon theorems and may provide a testing ground for exploring how supersymmetry constrains celestial amplitudes. \par

\section*{Acknowledgments}
   We are deeply grateful to Bin Chen and Sergio Cecotti for valuable discussions.\par

\vspace{2cm}

\appendix
\renewcommand{\appendixname}{Appendix~\Alph{section}}

\section{Tensor Product of Virasoro Algebra Representations}\label{app:TensorProductOfVirReprs}

    In this appendix we show the solutions of equation \eqref{eq:JacobiEquation} and \eqref{eq:CombinedModesOfVirReprsSolution}. \par

    For convenience, we repeat equation \eqref{eq:JacobiEquation} here:
    \begin{equation}
        (l_k n - (r_i+r_j)) f_{l_i,l_j}^{l_k}(r_i,r_j) - (l_j n - r_j) f_{l_i,l_j}^{l_k}(r_i,r_j + n) - (l_i n - r_i) f_{l_i,l_j}^{l_k}(r_i+n,r_j) = 0.
    \end{equation} 
    This is a difference equation, and solving it analytically for generic values of the parameters $l_i$, $l_j$, and $l_k$ is challenging. Instead, we proceed by obtaining numerical solutions for specific parameter choices. The setup is as follows:
    \begin{equation}
        li, lj, l_k \in \mathbb{Z}/2, \quad -4\le li, lj \le 4, \quad \abs{l_k- (l_i + l_j)}\le 3.
    \end{equation}
    A solution is referred as continuous if $f$ is a continuous function, and denoted as discrete if $f$ involves Kronecker $\delta$ function. The results are summarized as follows
    \par\noindent\textbf{continuous}
    \begin{subequations}
        \begin{align}
            & f_{l_i,l_j}^{l_k}(r_i,r_j) = 1, && l_k=l_i+l_j, \label{eq:AppdixJacobiSolution1}\\
            & f_{l_i,l_j}^{l_k}(r_i,r_j) = l_j r_i - l_i r_j, && l_k=l_i+l_j-1, \label{eq:AppdixJacobiSolution2}\\
            & f_{l_i,l_j}^{l_k}(r_i,r_j) = r_i(l_j r_i + r_j), && l_i = 0, \quad l_k=l_j-2, \label{eq:AppdixJacobiSolution3}\\
            & f_{l_i,l_j}^{l_k}(r_i,r_j) = (r_i + r_j)(l_j r_i - l_i r_j), && l_i +l_j = 1, \quad l_k=-1, \\
            & f_{l_i,l_j}^{l_k}(r_i,r_j) = r_i, && l_i = l_j = 0, \quad l_k=-1, \\
            & f_{l_i,l_j}^{l_k}(r_i,r_j) = l_j r_i - l_i r_j, && l_i = l_j = 0, \quad l_k=-3, \\
            & f_{l_i,l_j}^{l_k}(r_i,r_j) = r_i(r_i + r_j)(2r_i + r_j), && l_i = 0, \quad l_j = 2, \quad l_k=-1.
        \end{align}
    \end{subequations}
    \par\noindent\textbf{discrete}
    \begin{subequations}
        \begin{align}
            & f_{l_i,l_j}^{l_k}(r_i,r_j) = \delta_{r_i + r_j}, && l_i+l_j = -1, \quad l_k = 0, \label{eq:AppdixJacobiSolution4}\\
            & f_{l_i,l_j}^{l_k}(r_i,r_j) = \delta_{r_i}, && l_i = -1, \quad l_j=l_k, \\
            & f_{l_i,l_j}^{l_k}(r_i,r_j) = r_i\delta(r_i + r_j), && l_i = l_j = l_k = 0, \\
            & f_{l_i,l_j}^{l_k}(r_i,r_j) = r_i \delta_{r_j}, && l_i = 0, \quad l_k=l_j=-1, \\
            & f_{l_i,l_j}^{l_k}(r_i,r_j) = \delta_{r_i}\delta_{r_j}, && l_i = l_j = -1, \quad l_k = 0, \\
            & f_{l_i,l_j}^{l_k}(r_i,r_j) = \frac{1}{r_i}\delta_{r_i + r_j} + \frac{1}{r_j}\delta_{r_i} - \frac{1}{r_i}\delta_{r_j}, && l_i = l_j = -1, \quad l_k = 0. \label{eq:AppdixJacobiSolution5}
        \end{align}
    \end{subequations}
    These solutions remain valid under the exchange of $i$ and $j$.  Among them, \eqref{eq:AppdixJacobiSolution1}, \eqref{eq:AppdixJacobiSolution2}, and \eqref{eq:AppdixJacobiSolution3} coincide with the solutions obtained in \eqref{eq:TensorProductOfVirReprsSolution}. Solution \eqref{eq:AppdixJacobiSolution5} does not play a role in the construction of the superalgebras. And the other solutions are only consistent for negative parameter values, so they are not in our interest for constructing the superalgebras. \par

    It is worth emphasizing that \eqref{eq:AppdixJacobiSolution4} is the solution used in \eqref{eq:DistreteJacobiSolution}, which automatically satisfied by the canonical commutation relations of the field modes. For instance, \eqref{eq:BosonicCommutationCondition} requires $l_\varphi + l_\phi = -1$ and $\bar{l}_\varphi + \bar{l}_\phi = -1$, while the spin indices of the identity operator are $l_I = \bar{l}_I = 0$. \par

    The difference equation for the field modes \eqref{eq:ModeCombinationEquation} is repeated here:
    \begin{equation}
        ((l_i+1)m-r)c(r-m,t)+(l_j m-t+r)c(r,t)-(l_Q m-t)c(r,t+m)=0.
    \end{equation}
    The numerical method we used are similar, but the setup is
    \begin{equation}
        li, lj, l_k \in \mathbb{Z}/2, \quad -3\le li, lj \le 3, \quad \abs{l_k- (l_i + l_j)}\le 3.
    \end{equation}
    Using this setup, we find the following solutions
    \par\noindent\textbf{continuous}
    \begin{subequations}
        \begin{align}
            & c_{l_i,l_j}^{l_k}(r,n) = 1, && l_k=l_i+l_j+1, \label{eq:AppdixModesSolution1}\\
            & c_{l_i,l_j}^{l_k}(r,n) = l_k r - (l_i+1) n, && l_k=l_i+l_j+2, \label{eq:AppdixModesSolution2}\\
            & c_{l_i,l_j}^{l_k}(r,n) = r((2 + l_j) r - n), && l_i = -1, \quad l_k=l_j+2, \\
            & c_{l_i,l_j}^{l_k}(r,n) = n( r + (l_i+1) n), && l_i+l_j = -3, \quad l_k=0,
        \end{align}
    \end{subequations}
    \par\noindent\textbf{discrete}
    \begin{subequations}
        \begin{align}
            & c_{l_i,l_j}^{l_k}(r,n) = \delta_{r}, && l_i = 0, \quad l_k=l_j, \\
            & c_{l_i,l_j}^{l_k}(r,n) = n\delta_{r}, && l_i = l_k=0, \quad l_j = -1, \\
            & c_{l_i,l_j}^{l_k}(r,n) = \frac{1}{n}\delta_{n-r}-\frac{1}{r}\delta_{n}-\frac{1}{n}\delta_{r}, && l_i = l_j=0, \quad l_k = -1, \\
            & c_{l_i,l_j}^{l_k}(r,n) = \delta_{r}\delta_{n}, && l_i = l_j=0, \quad l_k = -1, \\
            & c_{l_i,l_j}^{l_k}(r,n) = \delta_{n}, && l_i + l_j = l_k=-1, \\
            & c_{l_i,l_j}^{l_k}(r,n) = r\delta_{n}, && l_i = l_j = l_k=-1.
        \end{align}
    \end{subequations}
    These solutions remain valid under the exchange $l_i \leftrightarrow l_j$ accompanied by the shift $n \to n-r$. Among them, only \eqref{eq:AppdixModesSolution1} and \eqref{eq:AppdixModesSolution2}, which correspond to \eqref{eq:CombinedModesOfVirReprsSolution}, are relevant for constructing the field modes.  \par





\bibliographystyle{JHEP}
\bibliography{refs.bib}

\providecommand{\href}[2]{#2}\begingroup\raggedright\begin{thebibliography}{10}

\bibitem{Bondi:1962px}
H.~Bondi, M.G.J.~van~der Burg and A.W.K.~Metzner, \emph{{Gravitational waves in general relativity. 7. Waves from axisymmetric isolated systems}}, \href{https://doi.org/10.1098/rspa.1962.0161}{\emph{Proc. Roy. Soc. Lond. A} {\bfseries 269} (1962) 21}.

\bibitem{Sachs:1962wk}
R.K.~Sachs, \emph{{Gravitational waves in general relativity. 8. Waves in asymptotically flat space-times}}, \href{https://doi.org/10.1098/rspa.1962.0206}{\emph{Proc. Roy. Soc. Lond. A} {\bfseries 270} (1962) 103}.

\bibitem{Sachs:1962zza}
R.~Sachs, \emph{{Asymptotic symmetries in gravitational theory}}, \href{https://doi.org/10.1103/PhysRev.128.2851}{\emph{Phys. Rev.} {\bfseries 128} (1962) 2851}.

\bibitem{Penrose:1962ij}
R.~Penrose, \emph{{Asymptotic properties of fields and space-times}}, \href{https://doi.org/10.1103/PhysRevLett.10.66}{\emph{Phys. Rev. Lett.} {\bfseries 10} (1963) 66}.

\bibitem{Banks:2003vp}
T.~Banks, \emph{{A Critique of pure string theory: Heterodox opinions of diverse dimensions}},  \href{https://arxiv.org/abs/hep-th/0306074}{{\ttfamily hep-th/0306074}}.

\bibitem{Barnich:2009se}
G.~Barnich and C.~Troessaert, \emph{{Symmetries of asymptotically flat 4 dimensional spacetimes at null infinity revisited}}, \href{https://doi.org/10.1103/PhysRevLett.105.111103}{\emph{Phys. Rev. Lett.} {\bfseries 105} (2010) 111103} [\href{https://arxiv.org/abs/0909.2617}{{\ttfamily 0909.2617}}].

\bibitem{Barnich:2010eb}
G.~Barnich and C.~Troessaert, \emph{{Aspects of the BMS/CFT correspondence}}, \href{https://doi.org/10.1007/JHEP05(2010)062}{\emph{JHEP} {\bfseries 05} (2010) 062} [\href{https://arxiv.org/abs/1001.1541}{{\ttfamily 1001.1541}}].

\bibitem{Barnich:2010ojg}
G.~Barnich and C.~Troessaert, \emph{{Supertranslations call for superrotations}}, \href{https://doi.org/10.22323/1.127.0010}{\emph{PoS} {\bfseries CNCFG2010} (2010) 010} [\href{https://arxiv.org/abs/1102.4632}{{\ttfamily 1102.4632}}].

\bibitem{Kapec:2014opa}
D.~Kapec, V.~Lysov, S.~Pasterski and A.~Strominger, \emph{{Semiclassical Virasoro symmetry of the quantum gravity $ \mathcal{S}$-matrix}}, \href{https://doi.org/10.1007/JHEP08(2014)058}{\emph{JHEP} {\bfseries 08} (2014) 058} [\href{https://arxiv.org/abs/1406.3312}{{\ttfamily 1406.3312}}].

\bibitem{Campiglia:2014yka}
M.~Campiglia and A.~Laddha, \emph{{Asymptotic symmetries and subleading soft graviton theorem}}, \href{https://doi.org/10.1103/PhysRevD.90.124028}{\emph{Phys. Rev. D} {\bfseries 90} (2014) 124028} [\href{https://arxiv.org/abs/1408.2228}{{\ttfamily 1408.2228}}].

\bibitem{Alessio:2017lps}
F.~Alessio and G.~Esposito, \emph{{On the structure and applications of the Bondi{\textendash}Metzner{\textendash}Sachs group}}, \href{https://doi.org/10.1142/S0219887818300027}{\emph{Int. J. Geom. Meth. Mod. Phys.} {\bfseries 15} (2018) 1830002} [\href{https://arxiv.org/abs/1709.05134}{{\ttfamily 1709.05134}}].

\bibitem{Ashtekar:2018lor}
A.~Ashtekar, M.~Campiglia and A.~Laddha, \emph{{Null infinity, the BMS group and infrared issues}}, \href{https://doi.org/10.1007/s10714-018-2464-3}{\emph{Gen. Rel. Grav.} {\bfseries 50} (2018) 140} [\href{https://arxiv.org/abs/1808.07093}{{\ttfamily 1808.07093}}].

\bibitem{Strominger:2017zoo}
A.~Strominger, \emph{{Lectures on the Infrared Structure of Gravity and Gauge Theory}} (3, 2017), [\href{https://arxiv.org/abs/1703.05448}{{\ttfamily 1703.05448}}].

\bibitem{Strominger:2013jfa}
A.~Strominger, \emph{{On BMS Invariance of Gravitational Scattering}}, \href{https://doi.org/10.1007/JHEP07(2014)152}{\emph{JHEP} {\bfseries 07} (2014) 152} [\href{https://arxiv.org/abs/1312.2229}{{\ttfamily 1312.2229}}].

\bibitem{He:2014laa}
T.~He, V.~Lysov, P.~Mitra and A.~Strominger, \emph{{BMS supertranslations and Weinberg{\textquoteright}s soft graviton theorem}}, \href{https://doi.org/10.1007/JHEP05(2015)151}{\emph{JHEP} {\bfseries 05} (2015) 151} [\href{https://arxiv.org/abs/1401.7026}{{\ttfamily 1401.7026}}].

\bibitem{He:2014cra}
T.~He, P.~Mitra, A.P.~Porfyriadis and A.~Strominger, \emph{{New Symmetries of Massless QED}}, \href{https://doi.org/10.1007/JHEP10(2014)112}{\emph{JHEP} {\bfseries 10} (2014) 112} [\href{https://arxiv.org/abs/1407.3789}{{\ttfamily 1407.3789}}].

\bibitem{Strominger:2014pwa}
A.~Strominger and A.~Zhiboedov, \emph{{Gravitational Memory, BMS Supertranslations and Soft Theorems}}, \href{https://doi.org/10.1007/JHEP01(2016)086}{\emph{JHEP} {\bfseries 01} (2016) 086} [\href{https://arxiv.org/abs/1411.5745}{{\ttfamily 1411.5745}}].

\bibitem{Pasterski:2015tva}
S.~Pasterski, A.~Strominger and A.~Zhiboedov, \emph{{New Gravitational Memories}}, \href{https://doi.org/10.1007/JHEP12(2016)053}{\emph{JHEP} {\bfseries 12} (2016) 053} [\href{https://arxiv.org/abs/1502.06120}{{\ttfamily 1502.06120}}].

\bibitem{Conde:2016rom}
E.~Conde and P.~Mao, \emph{{BMS Supertranslations and Not So Soft Gravitons}}, \href{https://doi.org/10.1007/JHEP05(2017)060}{\emph{JHEP} {\bfseries 05} (2017) 060} [\href{https://arxiv.org/abs/1612.08294}{{\ttfamily 1612.08294}}].

\bibitem{Nande:2017dba}
A.~Nande, M.~Pate and A.~Strominger, \emph{{Soft Factorization in QED from 2D Kac-Moody Symmetry}}, \href{https://doi.org/10.1007/JHEP02(2018)079}{\emph{JHEP} {\bfseries 02} (2018) 079} [\href{https://arxiv.org/abs/1705.00608}{{\ttfamily 1705.00608}}].

\bibitem{Adamo:2014yya}
T.~Adamo, E.~Casali and D.~Skinner, \emph{{Perturbative gravity at null infinity}}, \href{https://doi.org/10.1088/0264-9381/31/22/225008}{\emph{Class. Quant. Grav.} {\bfseries 31} (2014) 225008} [\href{https://arxiv.org/abs/1405.5122}{{\ttfamily 1405.5122}}].

\bibitem{Pasterski:2016qvg}
S.~Pasterski, S.-H.~Shao and A.~Strominger, \emph{{Flat Space Amplitudes and Conformal Symmetry of the Celestial Sphere}}, \href{https://doi.org/10.1103/PhysRevD.96.065026}{\emph{Phys. Rev. D} {\bfseries 96} (2017) 065026} [\href{https://arxiv.org/abs/1701.00049}{{\ttfamily 1701.00049}}].

\bibitem{Fotopoulos:2020bqj}
A.~Fotopoulos, S.~Stieberger, T.R.~Taylor and B.~Zhu, \emph{{Extended Super BMS Algebra of Celestial CFT}}, \href{https://doi.org/10.1007/JHEP09(2020)198}{\emph{JHEP} {\bfseries 09} (2020) 198} [\href{https://arxiv.org/abs/2007.03785}{{\ttfamily 2007.03785}}].

\bibitem{Raclariu:2021zjz}
A.-M.~Raclariu, \emph{{Lectures on Celestial Holography}},  \href{https://arxiv.org/abs/2107.02075}{{\ttfamily 2107.02075}}.

\bibitem{Pasterski:2021rjz}
S.~Pasterski, \emph{{Lectures on celestial amplitudes}}, \href{https://doi.org/10.1140/epjc/s10052-021-09846-7}{\emph{Eur. Phys. J. C} {\bfseries 81} (2021) 1062} [\href{https://arxiv.org/abs/2108.04801}{{\ttfamily 2108.04801}}].

\bibitem{Prema:2021sjp}
A.B.~Prema, G.~Comp{\`e}re, L.~Pipolo~de Gioia, I.~Mol and B.~Swidler, \emph{{Celestial holography: Lectures on asymptotic symmetries}}, \href{https://doi.org/10.21468/SciPostPhysLectNotes.47}{\emph{SciPost Phys. Lect. Notes} {\bfseries 47} (2022) 1} [\href{https://arxiv.org/abs/2109.00997}{{\ttfamily 2109.00997}}].

\bibitem{Donnay:2023mrd}
L.~Donnay, \emph{{Celestial holography: An asymptotic symmetry perspective}}, \href{https://doi.org/10.1016/j.physrep.2024.04.003}{\emph{Phys. Rept.} {\bfseries 1073} (2024) 1} [\href{https://arxiv.org/abs/2310.12922}{{\ttfamily 2310.12922}}].

\bibitem{Barnich:2011mi}
G.~Barnich and C.~Troessaert, \emph{{BMS charge algebra}}, \href{https://doi.org/10.1007/JHEP12(2011)105}{\emph{JHEP} {\bfseries 12} (2011) 105} [\href{https://arxiv.org/abs/1106.0213}{{\ttfamily 1106.0213}}].

\bibitem{Barnich:2014kra}
G.~Barnich and B.~Oblak, \emph{{Notes on the BMS group in three dimensions: I. Induced representations}}, \href{https://doi.org/10.1007/JHEP06(2014)129}{\emph{JHEP} {\bfseries 06} (2014) 129} [\href{https://arxiv.org/abs/1403.5803}{{\ttfamily 1403.5803}}].

\bibitem{Barnich:2015uva}
G.~Barnich and B.~Oblak, \emph{{Notes on the BMS group in three dimensions: II. Coadjoint representation}}, \href{https://doi.org/10.1007/JHEP03(2015)033}{\emph{JHEP} {\bfseries 03} (2015) 033} [\href{https://arxiv.org/abs/1502.00010}{{\ttfamily 1502.00010}}].

\bibitem{Compere:2020lrt}
G.~Comp{\`e}re, A.~Fiorucci and R.~Ruzziconi, \emph{{The $\Lambda$-BMS$_4$ charge algebra}}, \href{https://doi.org/10.1007/JHEP10(2020)205}{\emph{JHEP} {\bfseries 10} (2020) 205} [\href{https://arxiv.org/abs/2004.10769}{{\ttfamily 2004.10769}}].

\bibitem{Donnay:2020fof}
L.~Donnay, G.~Giribet and F.~Rosso, \emph{{Quantum BMS transformations in conformally flat space-times and holography}}, \href{https://doi.org/10.1007/JHEP12(2020)102}{\emph{JHEP} {\bfseries 12} (2020) 102} [\href{https://arxiv.org/abs/2008.05483}{{\ttfamily 2008.05483}}].

\bibitem{Chen:2025gaz}
B.~Chen and Z.~Hu, \emph{{Carrollian superstring in the flipped vacuum}}, \href{https://doi.org/10.1103/b2jy-223t}{\emph{Phys. Rev. D} {\bfseries 112} (2025) 046005} [\href{https://arxiv.org/abs/2501.11011}{{\ttfamily 2501.11011}}].

\bibitem{Donnay:2015abr}
L.~Donnay, G.~Giribet, H.A.~Gonzalez and M.~Pino, \emph{{Supertranslations and Superrotations at the Black Hole Horizon}}, \href{https://doi.org/10.1103/PhysRevLett.116.091101}{\emph{Phys. Rev. Lett.} {\bfseries 116} (2016) 091101} [\href{https://arxiv.org/abs/1511.08687}{{\ttfamily 1511.08687}}].

\bibitem{Hawking:2016msc}
S.W.~Hawking, M.J.~Perry and A.~Strominger, \emph{{Soft Hair on Black Holes}}, \href{https://doi.org/10.1103/PhysRevLett.116.231301}{\emph{Phys. Rev. Lett.} {\bfseries 116} (2016) 231301} [\href{https://arxiv.org/abs/1601.00921}{{\ttfamily 1601.00921}}].

\bibitem{Hawking:2016sgy}
S.W.~Hawking, M.J.~Perry and A.~Strominger, \emph{{Superrotation Charge and Supertranslation Hair on Black Holes}}, \href{https://doi.org/10.1007/JHEP05(2017)161}{\emph{JHEP} {\bfseries 05} (2017) 161} [\href{https://arxiv.org/abs/1611.09175}{{\ttfamily 1611.09175}}].

\bibitem{Awada:1985by}
M.A.~Awada, G.W.~Gibbons and W.T.~Shaw, \emph{{CONFORMAL SUPERGRAVITY, TWISTORS AND THE SUPER BMS GROUP}}, \href{https://doi.org/10.1016/S0003-4916(86)80023-9}{\emph{Annals Phys.} {\bfseries 171} (1986) 52}.

\bibitem{Barnich:2015sca}
G.~Barnich, L.~Donnay, J.~Matulich and R.~Troncoso, \emph{{Super-BMS$_{3}$ invariant boundary theory from three-dimensional flat supergravity}}, \href{https://doi.org/10.1007/JHEP01(2017)029}{\emph{JHEP} {\bfseries 01} (2017) 029} [\href{https://arxiv.org/abs/1510.08824}{{\ttfamily 1510.08824}}].

\bibitem{Fuentealba:2020aax}
O.~Fuentealba, M.~Henneaux, S.~Majumdar, J.~Matulich and T.~Neogi, \emph{{Asymptotic structure of the Rarita-Schwinger theory in four spacetime dimensions at spatial infinity}}, \href{https://doi.org/10.1007/JHEP02(2021)031}{\emph{JHEP} {\bfseries 02} (2021) 031} [\href{https://arxiv.org/abs/2011.04669}{{\ttfamily 2011.04669}}].

\bibitem{Henneaux:2020ekh}
M.~Henneaux, J.~Matulich and T.~Neogi, \emph{{Asymptotic realization of the super-BMS algebra at spatial infinity}}, \href{https://doi.org/10.1103/PhysRevD.101.126016}{\emph{Phys. Rev. D} {\bfseries 101} (2020) 126016} [\href{https://arxiv.org/abs/2004.07299}{{\ttfamily 2004.07299}}].

\bibitem{Narayanan:2020amh}
S.A.~Narayanan, \emph{{Massive Celestial Fermions}}, \href{https://doi.org/10.1007/JHEP12(2020)074}{\emph{JHEP} {\bfseries 12} (2020) 074} [\href{https://arxiv.org/abs/2009.03883}{{\ttfamily 2009.03883}}].

\bibitem{Pano:2021ewd}
Y.~Pano, S.~Pasterski and A.~Puhm, \emph{{Conformally soft fermions}}, \href{https://doi.org/10.1007/JHEP12(2021)166}{\emph{JHEP} {\bfseries 12} (2021) 166} [\href{https://arxiv.org/abs/2108.11422}{{\ttfamily 2108.11422}}].

\bibitem{Fuentealba:2023hzq}
O.~Fuentealba and M.~Henneaux, \emph{{Simplifying (super-)BMS algebras}}, \href{https://doi.org/10.1007/JHEP11(2023)108}{\emph{JHEP} {\bfseries 11} (2023) 108} [\href{https://arxiv.org/abs/2309.07600}{{\ttfamily 2309.07600}}].

\bibitem{Tropper:2024evi}
A.~Tropper, \emph{{Symmetries of the Celestial Supersphere}},  \href{https://arxiv.org/abs/2412.13113}{{\ttfamily 2412.13113}}.

\bibitem{Teitelboim:1977fs}
C.~Teitelboim, \emph{{Supergravity and Square Roots of Constraints}}, \href{https://doi.org/10.1103/PhysRevLett.38.1106}{\emph{Phys. Rev. Lett.} {\bfseries 38} (1977) 1106}.

\bibitem{Tabensky:1977ic}
R.~Tabensky and C.~Teitelboim, \emph{{The Square Root of General Relativity}}, \href{https://doi.org/10.1016/0370-2693(77)90843-7}{\emph{Phys. Lett. B} {\bfseries 69} (1977) 453}.

\bibitem{Avery:2015iix}
S.G.~Avery and B.U.W.~Schwab, \emph{{Residual Local Supersymmetry and the Soft Gravitino}}, \href{https://doi.org/10.1103/PhysRevLett.116.171601}{\emph{Phys. Rev. Lett.} {\bfseries 116} (2016) 171601} [\href{https://arxiv.org/abs/1512.02657}{{\ttfamily 1512.02657}}].

\bibitem{Banerjee:2015kcx}
N.~Banerjee, D.P.~Jatkar, S.~Mukhi and T.~Neogi, \emph{{Free-field realisations of the BMS$_{3}$ algebra and its extensions}}, \href{https://doi.org/10.1007/JHEP06(2016)024}{\emph{JHEP} {\bfseries 06} (2016) 024} [\href{https://arxiv.org/abs/1512.06240}{{\ttfamily 1512.06240}}].

\bibitem{Fuentealba:2021xhn}
O.~Fuentealba, M.~Henneaux, S.~Majumdar, J.~Matulich and T.~Neogi, \emph{{Local supersymmetry and the square roots of Bondi-Metzner-Sachs supertranslations}}, \href{https://doi.org/10.1103/PhysRevD.104.L121702}{\emph{Phys. Rev. D} {\bfseries 104} (2021) L121702} [\href{https://arxiv.org/abs/2108.07825}{{\ttfamily 2108.07825}}].

\bibitem{Bagchi:2022owq}
A.~Bagchi, D.~Grumiller and P.~Nandi, \emph{{Carrollian superconformal theories and super BMS}}, \href{https://doi.org/10.1007/JHEP05(2022)044}{\emph{JHEP} {\bfseries 05} (2022) 044} [\href{https://arxiv.org/abs/2202.01172}{{\ttfamily 2202.01172}}].

\bibitem{Lodato:2016alv}
I.~Lodato and W.~Merbis, \emph{{Super-BMS$_{3}$ algebras from $ \mathcal{N}=2 $ flat supergravities}}, \href{https://doi.org/10.1007/JHEP11(2016)150}{\emph{JHEP} {\bfseries 11} (2016) 150} [\href{https://arxiv.org/abs/1610.07506}{{\ttfamily 1610.07506}}].

\bibitem{Bagchi:2017cte}
A.~Bagchi, A.~Banerjee, S.~Chakrabortty and P.~Parekh, \emph{{Inhomogeneous Tensionless Superstrings}}, \href{https://doi.org/10.1007/JHEP02(2018)065}{\emph{JHEP} {\bfseries 02} (2018) 065} [\href{https://arxiv.org/abs/1710.03482}{{\ttfamily 1710.03482}}].

\bibitem{Chen:2023esw}
B.~Chen, Z.~Hu, Z.-f.~Yu and Y.-f.~Zheng, \emph{{Path-integral quantization of tensionless (super) string}}, \href{https://doi.org/10.1007/JHEP08(2023)133}{\emph{JHEP} {\bfseries 08} (2023) 133} [\href{https://arxiv.org/abs/2302.05975}{{\ttfamily 2302.05975}}].

\bibitem{Banerjee:2019lrv}
N.~Banerjee, A.~Bhattacharjee, Neetu and T.~Neogi, \emph{{New $ \mathcal{N} $ = 2 SuperBMS$_{3}$ algebra and invariant dual theory for 3D supergravity}}, \href{https://doi.org/10.1007/JHEP11(2019)122}{\emph{JHEP} {\bfseries 11} (2019) 122} [\href{https://arxiv.org/abs/1905.10239}{{\ttfamily 1905.10239}}].

\bibitem{Prabhu:2021bod}
K.~Prabhu, \emph{{Novel supersymmetric extension of BMS symmetries at null infinity}}, \href{https://doi.org/10.1103/PhysRevD.105.064054}{\emph{Phys. Rev. D} {\bfseries 105} (2022) 064054} [\href{https://arxiv.org/abs/2112.07186}{{\ttfamily 2112.07186}}].

\bibitem{Banks:2014iha}
T.~Banks, \emph{{The Super BMS Algebra, Scattering and Holography}},  \href{https://arxiv.org/abs/1403.3420}{{\ttfamily 1403.3420}}.

\bibitem{Adamo:2014wea}
T.~Adamo, E.~Casali and D.~Skinner, \emph{{A Worldsheet Theory for Supergravity}}, \href{https://doi.org/10.1007/JHEP02(2015)116}{\emph{JHEP} {\bfseries 02} (2015) 116} [\href{https://arxiv.org/abs/1409.5656}{{\ttfamily 1409.5656}}].

\bibitem{Banerjee:2022lnz}
N.~Banerjee, T.~Rahnuma and R.K.~Singh, \emph{{Asymptotic symmetry algebra of N=8 supergravity}}, \href{https://doi.org/10.1103/PhysRevD.109.046010}{\emph{Phys. Rev. D} {\bfseries 109} (2024) 046010} [\href{https://arxiv.org/abs/2212.12133}{{\ttfamily 2212.12133}}].

\bibitem{Taylor:2023bzj}
T.R.~Taylor and B.~Zhu, \emph{{Celestial Supersymmetry}}, \href{https://doi.org/10.1007/JHEP06(2023)210}{\emph{JHEP} {\bfseries 06} (2023) 210} [\href{https://arxiv.org/abs/2302.12830}{{\ttfamily 2302.12830}}].

\bibitem{Duval:2014uva}
C.~Duval, G.W.~Gibbons and P.A.~Horvathy, \emph{{Conformal Carroll groups and BMS symmetry}}, \href{https://doi.org/10.1088/0264-9381/31/9/092001}{\emph{Class. Quant. Grav.} {\bfseries 31} (2014) 092001} [\href{https://arxiv.org/abs/1402.5894}{{\ttfamily 1402.5894}}].

\bibitem{Duval:2014uoa}
C.~Duval, G.W.~Gibbons, P.A.~Horvathy and P.M.~Zhang, \emph{{Carroll versus Newton and Galilei: two dual non-Einsteinian concepts of time}}, \href{https://doi.org/10.1088/0264-9381/31/8/085016}{\emph{Class. Quant. Grav.} {\bfseries 31} (2014) 085016} [\href{https://arxiv.org/abs/1402.0657}{{\ttfamily 1402.0657}}].

\bibitem{Duval:2014lpa}
C.~Duval, G.W.~Gibbons and P.A.~Horvathy, \emph{{Conformal Carroll groups}}, \href{https://doi.org/10.1088/1751-8113/47/33/335204}{\emph{J. Phys. A} {\bfseries 47} (2014) 335204} [\href{https://arxiv.org/abs/1403.4213}{{\ttfamily 1403.4213}}].

\bibitem{Haag:1974qh}
R.~Haag, J.T.~Lopuszanski and M.~Sohnius, \emph{{All Possible Generators of Supersymmetries of the s Matrix}}, \href{https://doi.org/10.1016/0550-3213(75)90279-5}{\emph{Nucl. Phys. B} {\bfseries 88} (1975) 257}.

\bibitem{DiFrancesco:1997nk}
P.~Di~Francesco, P.~Mathieu and D.~Senechal, \emph{{Conformal Field Theory}}, Graduate Texts in Contemporary Physics, Springer-Verlag, New York (1997), \href{https://doi.org/10.1007/978-1-4612-2256-9}{10.1007/978-1-4612-2256-9}.

\bibitem{Green:1987sp}
M.B.~Green, J.H.~Schwarz and E.~Witten, \emph{{SUPERSTRING THEORY. VOL. 1: INTRODUCTION}}, Cambridge Monographs on Mathematical Physics, Cambridge University Press (7, 1988).

\bibitem{Polchinski:1998rq}
J.~Polchinski, \emph{{String theory. Vol. 1: An introduction to the bosonic string}}, Cambridge Monographs on Mathematical Physics, Cambridge University Press (12, 2007), \href{https://doi.org/10.1017/CBO9780511816079}{10.1017/CBO9780511816079}.

\bibitem{Bagchi:2022eav}
A.~Bagchi, A.~Banerjee, S.~Dutta, K.S.~Kolekar and P.~Sharma, \emph{{Carroll covariant scalar fields in two dimensions}}, \href{https://doi.org/10.1007/JHEP01(2023)072}{\emph{JHEP} {\bfseries 01} (2023) 072} [\href{https://arxiv.org/abs/2203.13197}{{\ttfamily 2203.13197}}].

\bibitem{Hao:2022xhq}
P.-X.~Hao, W.~Song, Z.~Xiao and X.~Xie, \emph{{BMS-invariant free fermion models}}, \href{https://doi.org/10.1103/PhysRevD.109.025002}{\emph{Phys. Rev. D} {\bfseries 109} (2024) 025002} [\href{https://arxiv.org/abs/2211.06927}{{\ttfamily 2211.06927}}].

\bibitem{Yu:2022bcp}
Z.-f.~Yu and B.~Chen, \emph{{Free field realization of the BMS Ising model}}, \href{https://doi.org/10.1007/JHEP08(2023)116}{\emph{JHEP} {\bfseries 08} (2023) 116} [\href{https://arxiv.org/abs/2211.06926}{{\ttfamily 2211.06926}}].

\bibitem{Bagchi:2024qsb}
A.~Bagchi, P.~Chakraborty, S.~Chakrabortty, S.~Fredenhagen, D.~Grumiller and P.~Pandit, \emph{{Boundary Carrollian Conformal Field Theories and Open Null Strings}}, \href{https://doi.org/10.1103/PhysRevLett.134.071604}{\emph{Phys. Rev. Lett.} {\bfseries 134} (2025) 071604} [\href{https://arxiv.org/abs/2409.01094}{{\ttfamily 2409.01094}}].

\bibitem{Chen:2024voz}
B.~Chen, H.~Sun and Y.-f.~Zheng, \emph{{Quantization of Carrollian conformal scalar theories}}, \href{https://doi.org/10.1103/PhysRevD.110.125010}{\emph{Phys. Rev. D} {\bfseries 110} (2024) 125010} [\href{https://arxiv.org/abs/2406.17451}{{\ttfamily 2406.17451}}].

\bibitem{Chen:2021xkw}
B.~Chen, R.~Liu and Y.-f.~Zheng, \emph{{On Higher-dimensional Carrollian and Galilean Conformal Field Theories}}, \href{https://doi.org/10.21468/SciPostPhys.14.5.088}{\emph{SciPost Phys.} {\bfseries 14} (2023) 088} [\href{https://arxiv.org/abs/2112.10514}{{\ttfamily 2112.10514}}].

\bibitem{Chen:2023pqf}
B.~Chen, R.~Liu, H.~Sun and Y.-f.~Zheng, \emph{{Constructing Carrollian field theories from null reduction}}, \href{https://doi.org/10.1007/JHEP11(2023)170}{\emph{JHEP} {\bfseries 11} (2023) 170} [\href{https://arxiv.org/abs/2301.06011}{{\ttfamily 2301.06011}}].

\bibitem{Zheng:2025cuw}
Y.-f.~Zheng and B.~Chen, \emph{{Structure of Carrollian (conformal) superalgebra}}, \href{https://doi.org/10.1007/JHEP08(2025)111}{\emph{JHEP} {\bfseries 08} (2025) 111} [\href{https://arxiv.org/abs/2503.22160}{{\ttfamily 2503.22160}}].

\bibitem{Barnich:2014cwa}
G.~Barnich, L.~Donnay, J.~Matulich and R.~Troncoso, \emph{{Asymptotic symmetries and dynamics of three-dimensional flat supergravity}}, \href{https://doi.org/10.1007/JHEP08(2014)071}{\emph{JHEP} {\bfseries 08} (2014) 071} [\href{https://arxiv.org/abs/1407.4275}{{\ttfamily 1407.4275}}].

\bibitem{Flanagan:2015pxa}
{\'E}.{\'E}.~Flanagan and D.A.~Nichols, \emph{{Conserved charges of the extended Bondi-Metzner-Sachs algebra}}, \href{https://doi.org/10.1103/PhysRevD.95.044002}{\emph{Phys. Rev. D} {\bfseries 95} (2017) 044002} [\href{https://arxiv.org/abs/1510.03386}{{\ttfamily 1510.03386}}].

\bibitem{Banerjee:2017gzj}
N.~Banerjee, I.~Lodato and T.~Neogi, \emph{{N=4 Supersymmetric BMS3 algebras from asymptotic symmetry analysis}}, \href{https://doi.org/10.1103/PhysRevD.96.066029}{\emph{Phys. Rev. D} {\bfseries 96} (2017) 066029} [\href{https://arxiv.org/abs/1706.02922}{{\ttfamily 1706.02922}}].

\bibitem{Fuentealba:2017fck}
O.~Fuentealba, J.~Matulich and R.~Troncoso, \emph{{Asymptotic structure of $\mathcal{N}=2$ supergravity in 3D: extended super-BMS$_3$ and nonlinear energy bounds}}, \href{https://doi.org/10.1007/JHEP09(2017)030}{\emph{JHEP} {\bfseries 09} (2017) 030} [\href{https://arxiv.org/abs/1706.07542}{{\ttfamily 1706.07542}}].

\bibitem{Banerjee:2016nio}
N.~Banerjee, D.P.~Jatkar, I.~Lodato, S.~Mukhi and T.~Neogi, \emph{{Extended Supersymmetric BMS$_3$ algebras and Their Free Field Realisations}}, \href{https://doi.org/10.1007/JHEP11(2016)059}{\emph{JHEP} {\bfseries 11} (2016) 059} [\href{https://arxiv.org/abs/1609.09210}{{\ttfamily 1609.09210}}].

\bibitem{Banerjee:2018hbl}
N.~Banerjee, A.~Bhattacharjee, I.~Lodato and T.~Neogi, \emph{{Maximally $ \mathcal{N} $ -extended super-BMS$_{3}$ algebras and generalized 3D gravity solutions}}, \href{https://doi.org/10.1007/JHEP01(2019)115}{\emph{JHEP} {\bfseries 01} (2019) 115} [\href{https://arxiv.org/abs/1807.06768}{{\ttfamily 1807.06768}}].

\bibitem{Caroca:2018obf}
R.~Caroca, P.~Concha, O.~Fierro and E.~Rodr{\'\i}guez, \emph{{Three-dimensional Poincar{\'e} supergravity and $N$-extended supersymmetric $BMS_3$ algebra}}, \href{https://doi.org/10.1016/j.physletb.2019.02.049}{\emph{Phys. Lett. B} {\bfseries 792} (2019) 93} [\href{https://arxiv.org/abs/1812.05065}{{\ttfamily 1812.05065}}].

\bibitem{Banerjee:2022abf}
N.~Banerjee, A.~Mitra, D.~Mukherjee and H.R.~Safari, \emph{{Supersymmetrization of deformed BMS algebras}}, \href{https://doi.org/10.1140/epjc/s10052-022-11036-y}{\emph{Eur. Phys. J. C} {\bfseries 83} (2023) 3} [\href{https://arxiv.org/abs/2201.09853}{{\ttfamily 2201.09853}}].

\end{thebibliography}\endgroup
\end{document}